\documentclass[preprint]{aastex}
\pdfoutput=1 
\usepackage{multirow}
\usepackage{rotating}

\newcommand{\myemail}{wiclarks@indiana.edu}

\newcommand{\muk}{\bar{\vec{\mu}_k}}
\newcommand{\sigk}{{\bf \Sigma}_k}
\newcommand{\pik}{\vec{\pi}_k}
\newcommand{\sigki}{{\bf \Sigma}_{ki}}
\newcommand{\zk}{{\bf Z}_k}
\newcommand{\sigi}{{\bf S}_i}
\newcommand{\sigimean}{\bar{\bf S}}

\newcommand{\MProjObs}{\ensuremath{M(R < 0.4~{\rm pc}) } }
\newcommand{\MKineObs}{\ensuremath{M(r < 1.0~{\rm pc}) } }
\newcommand{\MProjRad}{\ensuremath{M(< R)}}
\newcommand{\MCl}{\ensuremath{M_{Cl}}}
\newcommand{\SurfdensN}{\ensuremath{\Sigma_N(R) }}
\newcommand{\SurfdensM}{\ensuremath{\Sigma_{mass}(R) }}
\newcommand{\Surfdensfac}{\ensuremath{\Sigma_{N,0}/\rho_0}}

\newcommand{\vvec}{\vec{v_i}}

\newcommand{\mukm}{$\muk$~}
\newcommand{\sigkm}{$\sigk$~}
\newcommand{\pikm}{$\pik$~}

\newcommand{\IMF}{\Gamma_0}
\newcommand{\PDMF}{\Gamma}

\newcommand{\tmeanm}{\bar{t}}
\newcommand{\tmean}{$\tmeanm$~}

\newcommand{\treference}{$t_{ref}$~}
\newcommand{\tzero}{$t_0$}

\newcommand{\bestmass}{\ensuremath{\MProjObs = 0.90^{+0.40}_{-0.35} \times 10^4 M_{\odot}}}
\newcommand{\besttot}{\ensuremath{\MKineObs = 1.5^{+0.74}_{-0.60} \times 10^4 M_{\odot}}}
\newcommand{\bestmcl}{\ensuremath{M_{cl} = 3.16^{+2.46}_{-2.09} \times 10^4 M_{\odot}}}

\newcommand{\adde}{\delta}
\newcommand{\addx}{\delta_x}
\newcommand{\addy}{\delta_y}

\newcommand{\addxbest}{\overline{\delta_x}}
\newcommand{\addybest}{\overline{\delta_y}}

\newcommand{\Mrad}{\ensuremath{M(<r)}}
\newcommand{\ntracer}{\ensuremath{n}}
\newcommand{\sigr}{\ensuremath{\sigma^2_r}}
\newcommand{\sigt}{\ensuremath{\sigma^2_t}}

\shorttitle{Keck LGS-AO motions of the Arches cluster}
\shortauthors{Clarkson et al.}

\begin{document}


\title{Proper motions of the Arches cluster with Keck LGS-Adaptive optics: the first kinematic mass measurement of the Arches}


\author{W. I. Clarkson\altaffilmark{1,2}, A. M. Ghez \altaffilmark{2}, M. R. Morris \altaffilmark{2}, J. R. Lu \altaffilmark{3,4}, A. Stolte\altaffilmark{5}, N. McCrady\altaffilmark{6}, Tuan Do\altaffilmark{7} \& Sylvana Yelda\altaffilmark{2}}




\altaffiltext{1}{Department of Astronomy, Indiana University, Bloomington, 727 East 3rd Street, Swain West 319, Bloomington, IN 47405-7105, USA; \myemail}
\altaffiltext{2}{Division of Astronomy \& Astrophysics, University of California, Los Angeles, Physics and Astronomy Building, 430 Portola Plaza, Box 951547, Los Angeles, CA 90095-1547, USA}
\altaffiltext{3}{Institute for Astronomy, 2680 Woodlawn Drive, Honolulu, HI 96822-1839, USA}
\altaffiltext{4}{Department of Astronomy, California Institute of Technology, 1200 East California Blvd, Pasadena CA 91125, USA}
\altaffiltext{5}{Argelander Institut f\"{u}r Astronomie, Universit\"{a}t Bonn, Auf dem H\"{u}gel 71, 53121 Bonn, Germany}
\altaffiltext{6}{Department of Physics and Astronomy, University of Montana, 32 Campus Drive, \#1080, Missoula, MT 59812, USA}
\altaffiltext{7}{Department of Physics \& Astronomy, University of California, Irvine, 4129 Frederick Reines Hall, Irvine, CA 92697-4575, USA}


\begin{abstract}
We report the first detection of the intrinsic velocity dispersion of
the Arches cluster - a young ($\sim$~2 Myr),
massive ($10^4 M_{\odot}$) starburst cluster located only 26 pc in
projection from the Galactic center. This was accomplished using
proper motion measurements within the central $10'' \times 10''$~of
the cluster, obtained with the laser guide star adaptive optics system
at Keck Observatory over a 3 year time baseline (2006-2009).  This
uniform dataset results in proper motion measurements that are
improved by a factor $\sim$~5 over previous measurements from
heterogeneous instruments.  By careful, simultaneous accounting of the
cluster and field contaminant distributions as well as the possible
sources of measurement uncertainties, we estimate the internal
velocity dispersion to be $0.15 \pm 0.01$~mas yr$^{-1}$, which
corresponds to $5.4 \pm 0.4$~km s$^{-1}$~at a distance of 8.4 kpc.

Projecting a simple model for the cluster onto the sky to compare with
our proper motion dataset, in conjunction with surface density data,
we estimate the total {\it present-day}~mass of the cluster to be
\besttot. The mass in stars observed within a cylinder of radius
$R$~(for comparison to photometric estimates) is found to be
\bestmass~at formal $3\sigma$~confidence. This mass measurement is
free from assumptions about the mass function of the cluster, and thus
may be used to check mass estimates from photometry and
simulation. Photometric mass estimates assuming an initially Salpeter
mass function ($\IMF = 1.35$, or $\PDMF\sim 1.0$~at present, where
$dN/d(\log M) \propto M^{\PDMF}$)~suggest a total cluster mass $M_{cl}
\sim (4-6) \times 10^4 M_{\odot}$~and projected mass ($\sim 2 \le
M(R<0.4~{\rm pc}) \le 3$)~$\times 10^4 M_{\odot}$. Photometric mass
estimates assuming a globally top-heavy or strongly truncated
present-day mass function (PDMF, with $\PDMF \sim 0.6$) yield mass
estimates closer to $\MProjObs \sim 1-1.2 \times 10^4
M_{\odot}$. Consequently, our results support a PDMF that is either
top-heavy or truncated at low-mass, or both.

Collateral benefits of our data and analysis include: (i) cluster
membership probabilities, which may be used to extract a clean cluster
sample for future photometric work; (ii) a refined estimate of the
bulk motion of the Arches cluster with respect to the field, which we
find to be 172 $\pm$~15 km s$^{-1}$, which is slightly slower than
suggested by previous VLT-Keck measurements; and (iii) a velocity
dispersion estimate for the field itself, which is likely dominated by
the inner galactic bulge and the nuclear disk.

\end{abstract}


\keywords{astrometry --- techniques: high angular resolution --- open clusters and associations: individual (Arches) --- Galaxies: clusters: individual (Arches) --- Stars: kinematics and dynamics --- Galaxy: kinematics and dynamics}


\section{Introduction}\label{ss_intro}


The spectrum of masses produced during the star formation process (the
Initial Mass Function, or IMF) is a key prediction of the star
formation process as it indirectly links to the observable Present-Day
Mass Function (PDMF) of the population \citep[for example, see][for
  review]{millerscalo79, mckee07, bastian10}. Because star formation
depends on collapse by self-gravity out of a turbulent medium threaded
with a magnetic field, there is some expectation that the physical
conditions in the parent cloud should affect the slope of the IMF, its
minimum mass, or both \citep[e.g.][]{morris93}. Numerical modeling
provides some support for environment-dependent IMF variations,
particularly in the unusual environment of the Galactic center
\citep[e.g.][]{bonnell04, klessen07, krumholz08, bonnell08}. However,
the resulting IMF variations may be so small as to be observable only
in extreme environments \citep[e.g. ][]{elmegreen08}. There is some
observational support for a varying IMF and turn-over mass in the
extreme environments of the young ($\sim$~few Myr), massive ($\sim
10^4 M_{\odot}$) starburst clusters NGC 3603 \citep{harayama08} and
for the stellar cluster at the Galactic center itself
\citep{bartko10}.

The young, massive cluster (YMC) known as the Arches Cluster
\citep[e.g.][]{nagata95, cotera96}\footnote{Throughout this communication, ``the
  Arches'' refers to the star cluster, not the arched radio filaments
  \citep{yusef84, morris89}, against which the cluster is
  projected, and with which it is physically associated
  \citep{lang04}.} is a particularly well-studied example of an extreme
environment for star formation. It is massive \citep[Total mass
  $M_{cl}\sim (2-7) \times 10^{4} M_{\odot}$;][]{figer99,figer02}
dense \citep[$\rho_c \sim 10^5 M_{\odot}$~pc$^{-3}$][]{espinoza09} and
young \citep[$\sim 2-2.5$~My;][]{najarro04, martins08}. It contains a
substantial number of massive stars \citep{serabyn98} which both
contribute to and heat the surrounding medium \citep[e.g.][]{figer02,
  yusef02, lang04}.

The Arches cluster is located only 26 pc in projection from the
Galactic center~(hereafter the GC). It therefore likely formed in
  an environment characterised by high gas pressure and velocity
  dispersion in the parent cloud, and high ambient temperature,
  particularly when compared to the relatively more benign environment
  of NGC 3603. As these parameters are thought to impact the IMF
  \citep{morris93, klessen07}, the Arches cluster is expected to be
an excellent candidate for observing a non-canonical IMF, whether in
its mass function exponent, low-mass turnover, a low-mass cut-off, or
all three \citep[e.g.][]{stolte02, klessen07}. It is also young enough
that the most massive main-sequence stars are still present, making it
one of the few clusters in which the upper mass-limit to the star
formation process may be observationally tested \citep{figer05,
  crowther10}. It has thus received substantial observational
attention, with efforts focused particularly on estimates of its
  IMF. Indeed, the Arches was originally the prototypical object for
a non-standard IMF \citep{figer99}, with an observed present-day
luminosity function indicating an overabundance of massive stars
compared to the canonical Salpeter IMF (parameterized as $dN/d(\log M)
\propto M^{-\Gamma}$, with $\Gamma = 1.35$; see \citealt{bastian10}
for a review).

However, a number of effects conspire to obscure the true IMF from
observation, complicating the interpretation of the PDMF, and indeed
the present consensus seems to be that the Arches began with an IMF
that is consistent with the canonical Salpeter IMF found in most
environments. Photometric efforts to chart the present-day luminosity
function of the cluster suffer from two important
  limitations. Firstly, the observations are difficult; strong
crowding and high, spatially-variable extinction are observed
  across the field of view, so that the photometric completeness is
challenging to estimate for masses lower than a few $M_{\odot}$. There
is evidence for mass segregation in the cluster \citep{figer99,
  stolte05}, seen as a steepening of the present-day luminosity
function towards the cluster center, implying strong spatial
selection effects when attempting to constrain the IMF. 

Secondly, the relationship of the PDMF to the IMF is not trivial
  to evaluate. Stellar evolution must be taken into account when
  relating the PDMF to the IMF, requiring a prescription for mass-loss
  from high-mass objects \citep[e.g.,][]{espinoza09}. In addition,
  mass segregation and tidal stripping are both likely to have been
  important for the evolution of the Arches. Present-day mass
segregation need not be primordial, since the Arches is likely
  already in a post-collapse phase \citep[e.g.][]{pz07,
    allison09}. The Arches cluster sits in a strong tidal field, such
  that as much as half of its stars may already have been stripped
  into the field over the $\sim 2.5$~My of its history \citep{kim00,
    pz02}. Mass segregation and tidal stripping together imply that
  the true IMF of the cluster may differ from the IMF drawn from the
  subset of stars that have remained within the Arches cluster to the
  present day. We review the literature mass determinations in
Section \ref{ss_masscomp}.

A {\it kinematic} mass measurement provides a direct test of the PDMF
of the Arches cluster, because its selection effects are somewhat less
stringent. Stars below the typical photometric completeness limit of
$\sim 1-2 M_{\odot}$~are observable through their contribution to the
total cluster mass. \citet{figer02}~were the first to attempt this, by
estimating the radial velocity dispersion of a sample of emission-line
stars and assuming the cluster is spherically symmetric and in virial
equilibrium. However, the estimate is complicated by the difficulty of
resolving the blended lines, their high width, and intrinsic
line-profile variation among the sample, so that the resulting mass
estimate is strongly dependent on atmosphere models. Mass estimates
using the velocity dispersion derived from proper motions are
independent of the details of the atmospheres of the tracer stars, and
in principle allow for the mass distribution to be derived in a more
assumption-free manner \citep{lm89}.

The advent of adaptive optics on large telescopes in the near-infrared
has enabled the measurement of precise proper motions of a large
number of stars in the Galactic center clusters. In a pioneering
proper motion study of the Arches, \citet{stolte08} used one epoch
each of VLT/NACO and Keck/NIRC2 separated by 4.3 years to measure the
motion of the cluster. However, differential distortion between the
cameras limited the proper motion precision to $\sim 0.7$~mas
yr$^{-1}$, somewhat too coarse to measure the internal velocity
dispersion\footnote{We use the term ``velocity dispersion'' to refer
  to both the dispersion in mas yr$^{-1}$~and km
  s$^{-1}$~throughout.}, for which the expected order of magnitude is
about $\sim 0.2$~mas yr$^{-1}$.

We have observed the central $10'' \times 10''$~of the Arches across
five epochs in three years (2006-2009) with a uniform observational
setup (PI Morris). Using NIRC2 on Keck-2, behind the LGS Adaptive
Optics facility \citep{ghez05, wizinowich06}, these cross-instrument
systematics encountered by \citet{stolte08} are not present in our
observations, and we are able to attain proper motion measurements
with error lower than the expected velocity dispersion. We report here
on our results, which provide the first kinematic mass estimate of the
Arches cluster from proper motions.

This communication is organised as follows. Section \ref{s_obs}
describes the observations and positional measurement technique, while
Section \ref{s_analyse} describes the process of proper motion
measurement and error assignment. Section \ref{s_res} describes the
techniques used to fit the cluster membership probabilities and
kinematic parameters. Section \ref{s_discuss} provides our mass
measurement and new bulk motion measurement for the Arches, and
briefly discusses the implications.

\section{Observations \& Measurements}\label{s_obs}

\begin{figure}
\begin{center}
\includegraphics[width=14cm]{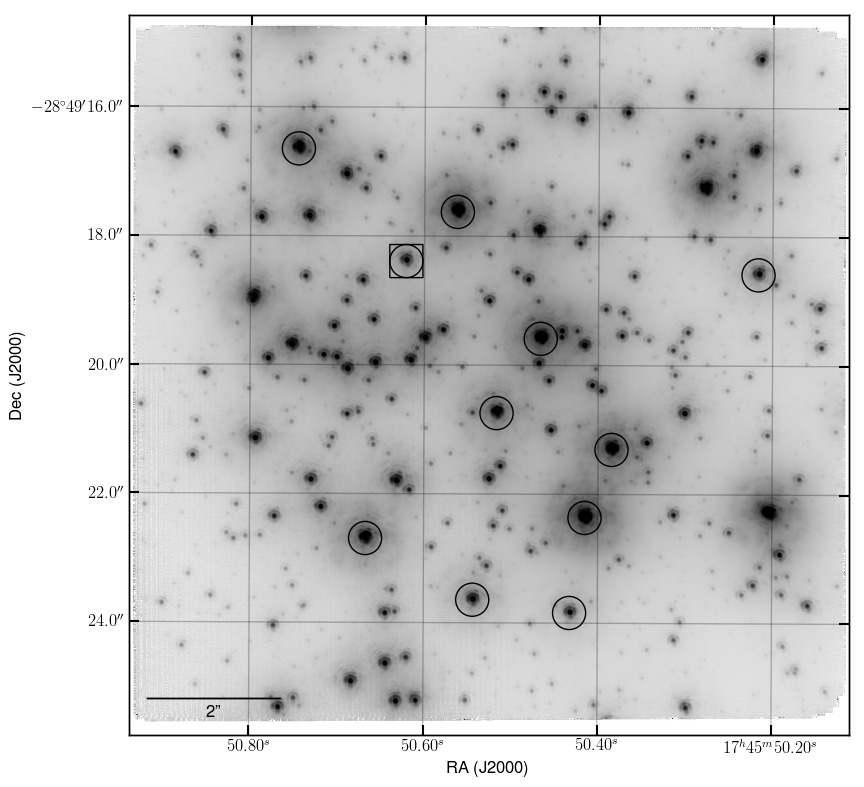}
\caption{NIRC2 $K'$~mosaic of the core field of the Arches in May
  2009. This is our best map in terms of both resolution (51 mas) and
  sensitivity ($K'_{lim} = 20.59$~mag; Table 1). All stars on which we
  report in this paper fall within the field of view indicated
  here. The scale-bar is two arcseconds in length. Stars used as PSF
  reference-stars are indicated by circles. When stellar membership
  probabilities are reported, positions are reported as offsets from
  the reference star indicated by the square in this figure.}
\label{fig_findingchart}
\end{center}
\end{figure}

Observations of the central $10'' \times 10''$~of the Arches cluster
were obtained between May 2006 and May 2009 with the Keck
near-infrared camera (NIRC2: PI K. Matthews), behind the Laser Guide
Star Adaptive Optics \citep[LGSAO;][] {vandam06, wizinowich06} system
on the W. M. Keck II 10-meter telescope. All observations were
obtained with the narrow-field mode of NIRC2 (field of view $10.2''
\times 10.2''$), which has a pixel scale of 9.952 $\pm 0.003$~mas
pix$^{-1}$~\citep[][hereafter Y10]{yelda10}. Observations were taken
in the $K'$~filter ($\Delta \lambda = 0.35\mu m$, $\lambda_0 = 2.12
\mu m$). Five epochs of the central field in $K'$~have now been taken
(Table \ref{tab_obsns}), the second of which (2006 July 18) was first
reported in \citet{stolte08}. Observations were designed to be as
uniform as possible across the epochs, with detector-Y commanded to
align with the S-N direction at each epoch, with the same pseudorandom
dither pattern within a $0.7''\times 0.7''$~box applied at each epoch
\citep{ghez08}, and with observations taken at pointings with as
uniform a range of zenith angles as practical. Three images were taken
at each position within the dither pattern. Figure
\ref{fig_findingchart}~shows the mean image constructed from the May
2009 dataset. This is our best map in terms of both angular resolution
(FWHM 51 mas)~and sensitivity ($K'_{lim} = 20.59$~mag; Table 1). Table
\ref{tab_psfstars}~shows the PSF stars used in the analysis.

\section{Analysis}\label{s_analyse}

Our goal is to extract relative proper motions of Arches stars against
the field, which is mostly composed of bulge\footnote{We refer to the
  bulge/bar system simply as the ``bulge'' throughout. See Section
  \ref{ss_fieldprops} for a brief discussion.} stars. The analysis
proceeds in the following stages: (1). Positions are estimated from a
master-image at each epoch (Section \ref{ss_epochs}). (2) The
extracted positions are transformed into a reference frame common to
all epochs using likely cluster members, and proper motions extracted
from the positional time-series in this frame, using statistical
uncertainties (Section \ref{ss_meas}). (3) Possible sources of
additional proper motion uncertainties are explored (Section
\ref{ss_errors}) and, when such additional error sources are
identified, motions are re-extracted incorporating the updated
errors. Section \ref{ss_errors} briefly discusses the proper
motion precision attained.

\subsection{Production of star-lists from each epoch}\label{ss_epochs}

The procedures used by our group to produce lists of stellar positions
and fluxes from the image-sets at each epoch have been fully described
elsewhere \citep{ghez08, lu09, yelda10}; here we recapitulate briefly
the aspects relevant for the present work. Images are calibrated and
corrected for differential atmospheric refraction, and corrected for
static distortion using the most accurate distortion characterization
currently available (Y10). Within an epoch, the corrected
images are combined into a mean image using positional shifts only,
weighting by the Strehl ratio estimate for each image. Images are
combined using the Drizzle algorithm \citep{fruchterhook02}, and the
mean frame is not supersampled since the pixels already provide
2.5$\times$~Nyquist sampling. This is in contrast to common practice
when using Drizzle with Hubble data, which is typically only barely
Nyquist sampled. The shifts to use are estimated using
cross-correlation of the scene between images. By combining using
shifts only, we average through any rapid variations in instantaneous
distortion, and average over slow drifts in image orientation, which,
based on transformations between ccommon stars across a set of images
within the night, appear to be only $\sim 1'$~over the course of a
night and are therefore negligible. A modified version of the IDL
routine {\it Starfinder} \citep{diolaiti00} is then used to measure
star positions in the mean frame (Y10 \& refs. therein) by
simultaneous fitting of the point-spread function (herein PSF) to many
stars. In each epoch, 500-900 stars are identified, depending on the
performance of the AO system and the number of frames collected (Table
1).

Within an epoch, the positional estimate for each star is associated
with random measurement uncertainty on the mean of all images within
the night that passed quality control (we call this random uncertainty
the ``centroiding uncertainty;'' see Section \ref{ss_centroiding} for
details on its measurement).

\subsection{Transformations to common reference frame and proper motion measurement}\label{ss_meas}

The Arches cluster moves rapidly with respect to the field ($\sim
5$~mas yr$^{-1}$), shows low velocity dispersion compared to that of
the field, and accounts for most of the stars in the field of view
(Stolte et al. 2008). We therefore measure motions in the reference
frame in which the cluster is at rest\footnote{Throughout this
  discussion the term ``frame'' refers to reference frames of a given
  epoch or constructed from the positions, not to individual
  images.}. Details of this process are given in Appendix A; here we
outline the important considerations.

First, stars are matched across epochs to produce a master catalogue
containing all the original position measurements of each star. Each
star-list is transformed to the frame of a single star-list at a
chosen epoch \tzero.\footnote{In the following discussion, we use
  "\tzero'' (or ``\treference'')~as short-hands for ``the reference
  frame at time \tzero~(or \treference).''} Motions are estimated from
straight-line fits to the transformed measurements in \tzero. These
measurements are then used to construct a refined cluster reference
frame at a chosen time \treference and the original star-lists are
then re-mapped onto this refined frame and proper motions re-evaluated
in this frame.

The choice of epoch \tzero~is determined by the data quality and by
the epochs of observation. Proper motions are determined from
straight-line fits to positions in the chosen reference frame;
choosing a reference frame near the pivot point of these straight line
fits will minimize the error when mapping star-lists. Of the three
deepest epochs (Table 1), epoch 2008.5 is closest to this pivot point,
and is adopted as \tzero. Once a first estimate of proper motions has
been produced in this reference frame, the distribution of pivot
points of the first pass at straight-line fits is assessed; its mode
is \treference=2008.0. The reference frame for motions is then
constructed by evaluating the fits to the positional time-series from
the first pass, at time \treference.

Because the field and cluster populations show significantly different
motion, field stars are removed from consideration when evaluating the
transformation parameters that map reference frames onto each
other. This is achieved by clipping outliers in the vector point
diagram; after a few iterations of clipping and re-fitting, the
centroid of the cluster population is at zero motion in the vector
point diagram.

When mapping star-lists between epochs, the transformation parameters
are estimated by $\chi^2$~minimization using the positional
differences in each coordinate separately. The appropriate order of
transformation - a second-order polynomial - was determined by
evaluating the positional residuals as a function of order (Appendix
C). Transformation parameters are given in Table \ref{tab_transform}.

Measurements are inverse-variance weighted using the error estimates
in each coordinate for each star. As part of the fitting, three passes
of sigma-clipping (with 4$\sigma$~bounds) are used to reduce
sensitivity to measurement-outliers, mismatches or misidentification
of cluster members among the reference stars. This typically removes a
few stars from the reference list used for the mapping and can be
regarded as a fine-tuning of the reference star list for a given
mapping. 

As the analysis proceeds, additional estimates of positional error
become available to use as weights (see Section \ref{ss_errors}). When
mapping star-lists onto each other, errors used in the weighting are
the positional errors associated with each star-list. When mapping
star-lists onto the reference frame \tzero, these errors are just the
centroiding error in each list; when mapping onto \treference the
errors associated with the target frame are the errors in the
predicted position in \treference based on the first pass of motion
estimation. When evaluating velocities in a given frame, the
positional error and error associated with the mapping into this frame
(the alignment error; Section \ref{ss_align}) are added in quadrature.
Measured velocities are then used to investigate any additional error
not taken into account. Upon discovery of an error source missing from
the analysis, the entire analysis is repeated with the missing error
term included. Additional random error determined from the velocity
fits is associated with random variations in position-measurement of a
given source between epochs, and so is added in quadrature to the
centroiding error in the frame mapping and subsequent mapping. The
size of this additional error is examined in Section
\ref{ss_additive}.

\subsection{Positional and Motion errors}\label{ss_errors}

We now describe the error sources that are included in our
analysis in some detail. 

\subsubsection{Centroiding error}\label{ss_centroiding}

The centroiding uncertainty (random error in position-measurement on
the mean image from an epoch) is estimated empirically. The stack of
images from each epoch is divided into three subsets of equal length,
yielding three ``submaps'' - mean images of each of the three
subsets. Images are sorted in decreasing order of Strehl ratio and
each submap constructed from every third image surviving quality
control in the resulting list. Each submap has therefore been formed
from images spanning the same range of observationally relevant
conditions (such as Strehl ratio and pointing). Because each submap
describes a similar path through auxiliary parameter-space both in
relation to each other and to the mean image from the night, images
are mapped onto the same reference frame using simple translations
before averaging into a submap in the same manner as the production of
the mean image. Positions are also measured on each of the three
submaps in the same way as for the mean image, and these positions
then mapped onto the reference frame of the mean image using shifts
only. This produces three position measurements from the night, each
using one third of the information from the night and taking the same
path through auxiliary parameter-space. The rms of stellar position
measurements across the three submaps is evaluated for each star to
estimate the random error on the mean of one third of the images
within the night. This must be scaled by~$1/\sqrt{3}$~to estimate the
centroiding error from the mean image for the star in question. For
stars brighter than $K'=16$, this centroiding uncertainty is typically
0.1 mas (Table \ref{tab_budget} and Figure 2).

\subsubsection{Alignment error}\label{ss_align}

Error in predicted positions due to the mapping between frames was
estimated through Monte Carlo resampling: sets of half the reference
stars were randomly drawn and the frame-mapping re-fit and
re-evaluated for each trial set to produce a trial set of positions as
transformed into the target frame. The rms of the differences between
these predicted positions and those predicted from the full list is
then adopted as the contribution to random positional error due to the
fitting process. This error is always included when positional errors
after transformation are needed (steps 4 and onward in Appendix
A). Figure \ref{fig_additive_det}~shows the typical magnitude of
alignment errors.

\subsubsection{Additional random errors}\label{ss_additive}

When velocities were extracted, the distribution of $\chi^2$~values
from the velocity fits is quite different from that expected if all
random errors had correctly been included (Appendix B); clearly
additional positional variation is present between epochs that is not
accounted for by centroiding and alignment error alone. To properly
represent random error along each positional time-series, an
additional temporally-random error $\adde$~(denoted here as ``additive
error'')~must be added in quadrature to the random error sources
estimated thus far. The size of additive errors $\addx,
\addy$~required ({\it after} accounting for higher-than-linear
frame-to-frame transformations; see below) are determined by Maximum
Likelihood, as detailed in Appendix B1. We find that a flat
distribution of additive error with magnitude produces a velocity
$\chi^2$~distribution significantly more discrepant from statistical
expectation than a magnitude-dependent additive error (Appendix
B2). We therefore adopt estimates of $\addx, \addy$~that vary with
magnitude (see Table \ref{tab_budget}; specifically, for $K' < 16$~the
values are $\addx, \addy = 0.16, 0.15$~mas).

While the balance of important terms varies across different
facilities, the major causes of additional error are discussed in some
detail by \citet{fritz10}; we give three example causes here that
cannot practically be overcome by experimental design. (i) The
AO-corrected PSF core sits on top of a broad halo with radius similar
to the seeing radius ($\sim$1/20$^{\rm th}$ the size of the entire
field of view), resulting in significant background spatial
structure. Because the spatial scale of this structure is a
significant fraction of the field of view, and the field itself is
highly crowded, astrometric error due to seeing halo bias is difficult
to model with high accuracy. As the seeing halo depends on seeing
conditions during the observation, it varies between epochs and
therefore manifests itself as an additional time-random error
component \footnote{This is {\it not} the ``Halo Noise'' of Fritz et
  al. 2010); they include PSF cores, seeing-halos and all other
  spatially-varying background light under this term.}. (ii) Related
to this is source confusion, where the PSF of an object of interest
overlaps that of another object (resolved or unresolved); the
magnitude and direction of the bias depends on the relative brightness
of the two objects and on the spatial structure of the PSF. Note that
this astrometric bias may vary systematically with time (due to
time-varying object separation due to object relative motion;
e.g. Ghez et al. 2008), or randomly with time due to variation in the
PSF structure between epochs (which we assume to be temporally random
on timescales of months-years). This confusion-error may be spatially
correlated if the variation of the PSF between epochs is
spatially-correlated. (iii) The distortion suffered along the path
from source to pixels may vary between epochs in a number of ways,
producing spatially correlated but temporally-random spurious apparent
motions between epochs. For example: the mean reference frame within
an epoch is constructed from a strehl-weighted sum of instantaneous
images, and therefore depends on the time-variation of observing
conditions throught the night, which varies between epochs. Thus, a
variation in distortion may be expected between epochs even in the
limit of a perfectly stable instrument and telescope. Of the three
error-sources above, source confusion (source ii) is expected to vary
the most strongly with target object magnitude, and thus is a strong
candidate for the additive error $\addx$, $\addy$.

\subsection{Resulting Proper motion precision}\label{ss_resultingprecision}

We have achieved proper motion precision sufficiently high to measure
intrinsic dispersion. Positional time-series for a selection of
objects, along with the motions fitted to the time-series of the
objects, are provided in Figure \ref{f_example_curves}. The proper
motion precision adopted is shown in Figure
\ref{f_motion_errors}. Table \ref{tab_budget} summarises the error
budget of our astrometric measurements.

\section{Results}\label{s_res}

The cluster shares the field of view with a significant field
population. To estimate membership probabilities, we fit the kinematic
parameters of the cluster and field components simultaneously with the
relative contribution each component makes to the population in the
image (Section \ref{ss_fit}). Armed with membership probabilities, we
also estimate the velocity dispersion of the cluster and subsamples
within the cluster by direct Maximum Likelihood fitting of the motions
of likely cluster members (Section \ref{ss_direct_calc}). Section
\ref{ss_founddisp}~discusses the velocity dispersions resulting from
each approach. Since the two approaches produce similar estimates for
the dispersion, we re-express velocities in terms of radial and
tangential components in order to calculate the proper motion
dispersion profile for use in mass modeling (Section \ref{ss_projvels}).

\subsection{Membership Probabilities from kinematic fitting}\label{ss_membprob}

Table \ref{table_memprob}~provides a complete catalogue of formal
membership probabilities for cluster and field for all 432 objects
surviving the culls in Appendix A and for which five epochs of
position-measurement are available. Given best-fit kinematic
parameters, the probability that a given star is a member of the
$k$'th kinematic component, is then the usual
\begin{equation}
  P(k)_i = \frac{\phi_{ki}}{\sum^K_j \phi_{ji}}
\label{eq_prob}
\end{equation}
\noindent where $\phi_{ki}$~gives the likelihood that the $i$`th star
belongs to the $k$`th component, and depends on the fit to the
kinematic parameters of the field and cluster. We describe the process
of obtaining $\phi_{ki}$~below. 

\subsubsection{Kinematic fitting}\label{ss_fit}

Too few field objects are present in our sample to decompose the field
population by distance based on our measurements, so we appeal to the
literature. The field population is likely dominated by stars in the
inner region of the Milky Way bulge and may contain some population
from the outer regions of the nuclear stellar disk \citep[hereafter
  NSD;][]{launhardt02}. Both the bulge and NSD should show some degree
of central concentration along our line of sight (however, not
necessarily centered on the distance of the Arches; see Section
\ref{ss_meanmotion}). For the bulge we expect to preferentially sample
field stars within a few hundred parsecs of the Arches population
itself along the line of sight \citep[e.g.,][]{cabreralavers07}. The
velocity signature of the field component should thus be a sum of
differential rotation along the line of sight and intrinsic velocity
dispersion, sampled from the bulge and NSD. The contributions of the
NSD and galactic bulge to the field of the arches may be comparable in
size (e.g. Figure 2 of \citealt{launhardt02}), however the uncertainty
in the mass model in the inner regions is still rather large. For the
purposes of this work, we parameterize the sum of bulge + NSD as a
single two-dimensional Gaussian within the VPD.

Unlike many cluster studies \citep[e.g.,][]{platais03}, our field
component is expected to be significantly asymmetric in the vector
point diagram (hereafter VPD), as the bulge velocity dispersion is
larger along the galactic plane than perpendicular to it
\citep{kuijkenrich02}; at 350pc from the Galactic center, for example,
proper motion dispersions are of order $5$~and $3$~mas yr$^{-1}$~along
and perpendicular to the galactic plane, respectively
\citep{clarkson08}.

Because the cluster distribution in the vector point diagram is so
much tighter than the field distribution, fitting to the binned VPD is
not appropriate for this dataset as there is no binning factor that
allows simultaneous resolution of both cluster and field
components. Instead we maximize the likelihood $L$(data given the
component fractions, kinematic parameters, measurement errors) without
recourse to binning. The component fraction $\pi_k$~describes the
proportion of the sample of tracer stars in the image that belong to
the $k$'th kinematic component. Since our field fits entirely within
the flat core of the surface density radial profile (Espinoza et
al. 2009), we assume that $\pi_k$~is uniform within our field of view.

We follow standard practice \citep[e.g.,][]{sanders71,joneswalker88,
  kozplatais95} in using a two-dimensional gaussian profile to model
the intrinsic kinematic properties of the cluster in the VPD. The
field population is likely to be dominated by bulge objects which
occupy a rather narrow distance range compared to the distance to the
Galactic center; we therefore parameterize the field component with a
two-dimensional gaussian. Because the convolution of two bivariate
gaussians is another bivariate gaussian, whose covariance matrix is
the sum of those of the two components, this choice of model form
allows errors to be included naturally in the analysis. The likelihood
of finding a star at a given location in the VPD is thus given by the
sum of $K$~gaussian components:
\begin{eqnarray}
  L(\vvec) & = & \sum ^K _k \pi_k \frac{1}{2 \pi | \sigki |^{1/2}} 
  \exp{\left( -\frac{1}{2} (\vvec - \bar{\mu_k} )^T \sigki^{-1} (\vvec - \bar{\mu_k}) \right)}  \nonumber \\
  & \equiv & \sum^K_k \phi_{ki} 
\label{eq_like}
\end{eqnarray}
\noindent where $\phi_{ki}$~describes the likelihood of finding a
given star in a given component at its measured location in the VPD.
In (\ref{eq_like}), $\overline{\mu_k}$~is the centroid of the $k$'th
component in the VPD and $\sigki$~the covariance of the
$k'th$~component for the $i$'th star. Because the kinematic model and
error model are both gaussian, the covariances due to error and model
combine in the form $\sigki = \sigi + \zk$, where the diagonal
positional-error matrix $\sigi$~has components ($\sigma^2_{v,x},
\sigma^2_{v,y}$)~and the covariance matrix of the $k$'th~model
component is given by $\zk$.

Once the best-fit $\sigki$~has been estimated from the proper motion
data and errors, the intrinsic velocity dispersions $\sigma_a,
\sigma_b$~and the major-axis orientation $\theta$~of each component
are found from the eigenvalues and eigenvectors of $\sigki - \sigi$.

Best-fit parameters and component fractions are found by maximizing
$\sum^N_i \ln L(\vvec) = \sum^N_i \ln\left( \sum^K_k
\phi_{ik}\right)$~over the sample of tracer stars of interest, under
the constraint $\sum^K_k \pik = 1$. The maximum-likelhood $\pi_k$~and
the kinematic parameters are evaluated sequentially and iteratively
until convergence. If the errors are constant over the sample of
interest (so $\sigki \approx \zk + \sigimean$), then each update step
requires the evaluation of analytic expressions for the
maximum-likelihood estimate of the updated $\pi_k'$~given the current
estimate of the parameters (and vice versa). This is the well-known
Expectation Maximization (hereafter EM) algorithm. This technique is
well-established outside astronomy (Chapter 9 of \citealt{bishop06}
provides a thorough explanation, and the method has appeared in the
most recent edition of \citealt{press02}) and is becoming more
commonly employed to mixture problems in astronomy in which binning is
undesirable and/or a low number of reference objects is available
\citep{bovy09}.

Strongly-varying error on a star-by-star basis is a significant
complication, as the parameter-update step no longer has an analytic
form, and instead must be solved numerically. For the present
investigation, we use a single cluster component and choose instead a
variant of the technique of \citet[][and refs. therein]{kozplatais95}
in which membership probabilities for each star are estimated using
kinematic parameters fitted only from stars with roughly similar error
(so $\sigki \approx \zk + \sigimean = \sigk$~for the sub-sample). The
sample is broken into overlapping bins two magnitudes wide (so
$K'$=14.0-16.0, 15.0-17.0, 16.0-18.0 and 17.0-19.0) and the best-fit
parameters determined for each magnitude strip separately (Table
\ref{tab_kinparams} and Figure \ref{f_VPD_compilation}). Investigation
of remaining magnitude-dependent bias can be found in Appendix D. We
find that parameters do not become strongly affected by bias until
stars as faint as $K`=18$~are considered.

Between $15 \le K' < 18$, every star is at most half a magnitude from
the center of one of the magnitude-strips, and it is the parameters
corresponding to this nearest magnitude-strip that are used to
evaluate (\ref{eq_prob})~for these stars. Objects at $K' < 15.0$~use
the kinematic parameters evaluated for $K' < 16.0$~(see also Section
\ref{ss_direct_calc}), while objects at $K' > 18.0$~use the kinematic
parameters estimated from $17.0 \le K' < 19.0$. 

Errors on the parameters thus fit are estimated by Monte Carlo
bootstrap analysis. The observed vector point diagram is resampled by
random drawing of points from the full sample with replacement. The
full kinematic fitting process is re-applied to each trial and the
distribution of recovered values parameterized with its standard
deviation about the mean value using the full dataset. 

\subsubsection{Direct calculation of the velocity dispersion}\label{ss_direct_calc}

The process given in Section \ref{ss_fit} fails when the sample size
is small (i.e. $\la$~70 stars), or contains a negligible field
component (as occurs for $K \la 14.5$). Our end goal is to compute the
dispersion profile as a function of distance from the cluster center
for mass modeling, which may entail few stars per annulus. We
therefore compute the velocity dispersion from proper motions using a
direct Maximum Likelihood approach. Likely-cluster objects are
isolated using the membership probabilities estimated from the fits of
the previous section. Along each direction, we maximize
\begin{equation}
  L(\overline{v}, \sigma) = \prod^N_i (2\pi(\sigma^2 + e^2_i))^{-1/2} \exp\left\{ -(v_i - \overline{v})^2 / 2(\sigma^2 + e_i^2)\right\} 
\label{eq_direct}
\end{equation}
\noindent for $\overline{v}, \sigma$~iteratively, where $\sigma$~is
the intrinsic velocity dispersion and $e_i$~the measurement error. For
each $\sigma$, $\overline{v}$~is obtained by weighted averaging while
for each $\overline{v}$, $\sigma$~is found numerically by
bisection. Errors are estimated by Monte Carlo bootstrap analysis;
members of the sub-sample are drawn randomly with replacement and the
calculation repeated for a large number of trials. The standard
deviation of the fitted parameters across the set of trials is then
adopted as the fitting error from this process. Table
\ref{tab_dispersions}~shows the dispersions and their errors estimated
by this procedure.

\subsection{Cluster Velocity Dispersion}\label{ss_founddisp}

The kinematic fitting (Section \ref{ss_fit})~and direct (Section
\ref{ss_direct_calc})~approaches produce complementary estimates for
the intrinsic velocity dispersion of the Arches cluster. Both have
been corrected for measurement error using the procedures described
above. Tables \ref{tab_kinparams} \& \ref{tab_dispersions} show the
kinematic parameters fit to cluster and field, and the velocity
dispersion estimate for the cluster respectively. Figure
\ref{f_compare_erro}~shows the velocity dispersion graphically and its
comparison to the proper motion error.

Both estimates yield statistially significant measurements of the
cluster velocity dispersion. The mean cluster velocity dispersion and
its error is estimated by inverse variance-weighted average of the
determinations from non-overlapping magnitude bins, for $K' <
18.0$~(to avoid strong error and mass segregation effects). For the
dispersions estimated from the kinematic fitting (Section
\ref{ss_fit}; Table \ref{tab_kinparams}), magnitude bins $(14.0 \le K'
< 16.0)$~and $(16.0 \le K' < 18.0)$~are used. We do not use the
magnitude bin $K' > 18.0$~because this bin appears to suffer
misclassification bias under the fitting technique used; see Appendix
D.

This yields mean velocity dispersions along major and minor axes
respectively\footnote{{\it not}~galactic longitude and latitude}, of
($\overline{\sigma_a}, \overline{\sigma_b}) = (0.154 \pm 0.01, 0.136
\pm 0.008)$~mas yr$^{-1}$. Scaling to the distance of the Galactic
center ($8.4 \pm 0.4$~kpc; Ghez et al. 2008) this yields measured
velocity dispersion $(5.8 \pm 0.48, 5.2 \pm 0.40)$~km s$^{-1}$.

For the dispersions estimated directly from likely cluster members in
each direction separately, (Section \ref{ss_direct_calc}~and Table
\ref{tab_dispersions}), the mean velocity dispersions in the three
non-overlapping bins brighter than $K'=18.0$~are
$(\overline{\sigma_x}, \overline{\sigma_y}) = (0.149 \pm 0.011, 0.124
\pm 0.016)$~mas yr$^{-1}$, which translates to velocity dispersion
($5.7 \pm 0.51, 4.68 \pm 0.65$)~km s$^{-1}$ along detector XY
coordinates.

Included in Table \ref{tab_kinparams} is the position-angle (East of
North in the VPD) of the major axis of the cluster component (denoted
$\theta_{cl}$). Comparing independent magnitude-bins, we see that the
orientation of the cluster major axis changes from $70.2\degr \pm
21.7\degr$~to $67.5\degr \pm 48.3\degr$~to $114.0 \pm 66.4\degr$~as
fainter magnitude-bins are considered. Such an extreme variation
indicates the apparent variation in the cluster major axis orientation
is likely a sampling artifact, and does not reflect
underlying variation. The detector X and Y directions therefore sample
a symmetric distribution in two directions that are arbitrary with
respect to the cluster velocity distribution. We can therefore compare
the two measures by averaging the major and minor axes from kinematic
fitting to make a direction-invariant measure from kinematic fitting,
and average the directly-calculated dispersions along the two detector
directions to form a second direction-invariant dispersion
measure. This yields mean velocity dispersions $5.4 \pm 0.3$~km
s$^{-1}$~and $5.5 \pm 0.4$~km s$^{-1}$~respectively for dispersions
estimated from kinematic fitting $(14.0 \le K' < 18.0)$~and those
estimated directly from cluster members ($10.0 \le K' < 18.0$).

\subsection{Velocities for mass estimates}\label{ss_projvels}

Sections \ref{ss_fit} \& \ref{ss_direct_calc}~establish that
mixture-modeling (used to establish membership probabilities) and
direct calculation from cluster members produce the same estimate of
the dispersion. To be compatible with mass estimates based on
kinematic modeling, proper motions expressed in components aligned
with the detector axes, are converted into proper motion components
along the radial vector away from the cluster center, and along the
vector tangential to it. Specifically, the velocities and their errors
used for mass estimates, $v_R, v_T, \delta_R, \delta_T$, are computed
from the following relations:
\begin{eqnarray}
  \theta & \equiv & \arctan{\frac{Y_{\ast} - Y_0}{X_{\ast} - X_0}} \nonumber \\
  v_R & = & v_X \cos \theta  + v_Y \sin \theta \nonumber \\
  v_T & = & -v_X \sin \theta + v_Y \cos \theta \nonumber \\
  \delta^2_R & = & \delta_X^2 \sin^2 \theta + \delta_Y^2 \cos^2 \theta \nonumber \\
  \delta^2_T & = & \delta_X^2 \cos^2 \theta + \delta_Y^2 \sin^2 \theta 
\end{eqnarray}
\nonumber where $v_X, v_Y, \delta_X, \delta_Y$~are the velocities and
their errors along detector-X and detector-Y that were estimated from
proper motions. The positions $X_{\ast}, Y_{\ast}$~denote the position
of the star on the detector, and $X_0, Y_0$~the location of the
cluster center on the detector. The sample is then broken into
concentric annuli, and the velocity dispersion and its error within
each annulus computed by the method of Section
\ref{ss_direct_calc}. To select a sample of cluster member stars,
formal membership probabilities $P_{cluster} > 0.995$~were used (see
Figure \ref{fig_CMD} for an illustration of the VPD and CMD using this
membership probability as a cutoff).

The cluster center itself is not apparent as a peak in individual
motions or surface density within the restricted field of view of our
central-field observations \citep[compare with][]{espinoza09,
  andersonvdm10}, although the dynamical center of the cluster
probably is within the rough center of our field of view. Four
randomly-chosen locations for the cluster center $X_0, Y_0$~are chosen
(Table \ref{table_fieldcenters}), all within 2.6''~from the center of
the field. Mass limits reported in Section \ref{ss_massmodel} are
taken from the ensemble range over all four choices of cluster center;
in practice, the range of derived masses is not strongly dependent on
the location of the cluster center.


\section{Discussion}\label{s_discuss}

A key goal of this work is to compare mass limits set by proper motion
dispersion measurements to literature mass estimates. The set of
literature mass estimates is quite heterogeneous, so some discussion
of notation is in order before proceeding. We use $\IMF$~to represent
the IMF slope where quoted in the cited report, and $\PDMF$~to
represent the present-day mass function; both exponents refer to the
form $dN/d(\log M) \propto M^{-\Gamma}$. Simulations
\citep[e.g.,][]{kim00, kim03} suggest that the Arches may have lost
about half its mass since formation due to dynamical effects; we
therefore distinguish between present-day mass $M_{cl}$~and
initial-mass $M_0$~in the following discussion. We use radius $r$~to
refer to a distance from the cluster center in three dimensions and
$R$~to refer to distance from the cluster center as projected onto the
sky. The term ``half-mass radius'' refers to the radius within which
half the cluster mass is found, but the precise meaning of this term
often depends on the application. In our notation, $r_{hm}$~refers to
the radius of a sphere within which half the cluster mass is contained
\citep[the sense often used by theorists;][]{pz10}~and $R_{hm}$~is the
radius of a cylinder oriented along the line of sight, within which
half the cluster mass (usually the mass of the directly-observed
tracer stars) is observed. This latter quantity is the half-mass
radius most commonly seen in observational estimates\footnote{For
  example, our $R_{hm}$~is the same as the quantity $r_{hm}$~found in
  \citet{figer99} and \citet{stolte02}.}. Where a total cluster mass is reported,
we denote it as $M_{cl}$~or $M_{0}$; where mass within a given
projected radius is reported as \MProjObs~so that the outer radius
limit is clear. Throughout this report the term ``projected
mass''~refers to the mass within a cylinder of radius $R$~on the sky
whose long axis is aligned along the line of sight.

\subsection{Velocity dispersion and mass}\label{ss_massmodel}

The only previous use of stellar motions to estimate the mass of
  the Arches of which we are aware, is that of
\citet{figer02}. Radial velocities of eight emission-line stars within
$R = 0.23$~pc of the cluster center were used to place an upper limit
on the one-dimensional velocity dispersion. Their 22 km s$^{-1}$~was
converted to an upper limit on the enclosed mass using the Virial
theorem; a spherically symmetric, gravitationally-bound cluster with
this velocity dispersion would have enclosed mass $M_{cl}(r <
0.23~{\rm pc}) = 7 \times 10^4 M_{\odot}$. For such massive stars with
strong stellar winds, interpretation of the line profile shape to
estimate systematic velocities is strongly dependent on model
atmospheres for massive stars \citep{figer02}.

A mass estimate based on proper motions is independent of the details
of the line profile of the young, massive stars to which we are
sensitive. We choose {\it not} to employ the moment-based kinematic
mass estimator of \citep[][hereafter LM89]{lm89}, since its power to
estimate the total mass given the projected-radial coverage is
strongly sensitive to the outer radius-limit (e.g. Figure 11 of
\citealt{schoedel09}).

The method used is as follows: we start with a model for the mass
density profile $\rho(r)$~of the cluster, whose parameters are varied
to evaluate the comparison to our proper motion-based dispersion
estimates. Several model choices are possible. We choose to use
  the King (1962) prescription to describe the radial density profile
  $\rho(r)$~of the cluster. Our choice is motivated by three
  observations from the literature. Firstly, the Arches cluster is
  likely already in a post-collapse phase; simulations suggest that
  for the Arches cluster, core collapse takes place only about 0.6 Myr
  after formation \citep{kim99}. This collapse erases the signature of
  substructure in the parent cloud and leads to cluster relaxation on
  a shorter timescale than the initial half-mass relaxation
  time. Secondly, the resulting cluster undergoes dynamical mass
  segregation on a timescale of 1-2 Myr, much shorter than suggested
  by the current crossing-time of most stars in the cluster
  \citep{allison09}. The cluster is thus much older {\it dynamically}
  than its current crossing time would suggest \citep{bastian08,
    allison09}. This suggests that a model assuming a relaxed cluster
  should be a reasonable first approximation to the Arches, even given
  its young age. Thirdly, the {\it observed} surface density profile
  $\Sigma(R)$~(units stars pc$^{-2}$)~of massive stars ($M > 10 M_{\odot}$) is
  indeed reasonably well-fit by a King (1962) profile
  \citep{espinoza09}.

To use the King (1962) model to predict observed velocity dispersions,
we make four further assumptions; we anticipate that the next step in
the analysis will be to move straight to full numerical modeling of
the cluster, but that is beyond the scope of this communication. We
assume that the cluster is (i) not rotating strongly; (ii) can be
characterized as being in equilibrium, (iii) has spherical symmetry,
and (iv) shows isotropic motion. Assumption (ii) allows us to use the
Jeans equation to predict the velocity dispersion profile
$\sigma^2(r)$~corresponding to each model parameter-set. This profile
is then projected onto the sky to predict the observed velocity
dispersion components parallel and perpendicular to the radial
direction away from the cluster center. This prediction is compared to
observations (Section \ref{ss_projvels}), and a figure of merit
($\chi^2$) evaluated for each set of parameters. The set of
$\chi^2$~values mapped out in this way is used to derive confidence
limits for the model parameters. Finally, confidence limits on the
model parameters are converted into confidence limits on parameters of
interest derived from the model - particularly the projected mass
estimate \MProjObs.

We begin with the predictions for intrinsic velocity dispersion by
  substituting the model for $\rho(r)$~into the isotropic Jeans
  equation. Specifically, we have for the mass density $\rho(r)$~and
the surface density (by mass) $\Sigma(R)$:
\begin{eqnarray}
  z^2 & \equiv & \frac{r_c^2 + r^2}{r_c^2 + r_t^2} \nonumber \\
  \rho(r) & = & \frac{K}{\pi r_c \left[1+(r_t/r_c)^2 \right]^{3/2}} \frac{1}{z^2}\left( \frac{1}{z} \arccos z - \sqrt{1-z^2}\right) \nonumber \\
  \Sigma(R) & = & K \left( \frac{1}{\left[ 1 + (R/R_c)^2\right]^{1/2}}   - \frac{1}{\left[1 + (R_t/R_c)^2 \right]^{1/2} } \right)^2 
\end{eqnarray}
\label{eq_kingmodel}
\nonumber where $r_c, R_c$~refer to the core radius, $r_t, R_t$~refer
to the tidal radius, and lowercase/uppercase radii denote the radius
in three dimensions or projected on the sky, respectively. The Jeans
equation takes the form
\begin{eqnarray}
  \sigma^2_{iso}(r) & = & \frac{1}{\rho(r)} \int^{+\infty}_{r} G \frac{\rho(r) M(<r)}{r^2} dr \nonumber \\
  & = & \frac{G}{\rho(r)} \int^{+\infty}_{r} \frac{\rho(r)}{r^2} \int^{r}_{0} 4\pi r'^2 \rho(r') dr' dr 
\end{eqnarray}
\noindent which is readily evaluated numerically. This model
dispersion profile is projected onto the sky using equations (8) and
(10) of LM89 for comparison with observational data.

With this choice of model, three parameters determine the density
profile, the dispersion profile and therefore the derived masses of
interest: the core radius $r_c$, the tidal radius $r_t$~and the total
cluster mass $M_{cl}$. The latter parameter is used to normalize the
model through the condition
\begin{equation}
  M_{cl} = 4\pi \int^{r_t}_0 r^2 \rho(r) dr
\end{equation}
We vary the parameters of the King model (core radius $R_c$, tidal
radius $R_t$~and total cluster mass $M_{cl}$) and map the variation of
$\chi^2$~when the projected dispersion profile is compared with that
obtained from observation. With three model parameters varying we
adopt $\Delta \chi^2 = 3.50,~7.82~\&~13.93$, which correspond to 68\%,
95\% and 99.7\% of probability ($``1\sigma''$,
$``2\sigma''$~\&~$``3\sigma''$) when three model parameters are
allowed to jointly vary (see, for example, Press et al. 2002). The
limits on quantities {\it derived} from these parameters are then
given by the range of values of the derived parameters within each
$\Delta \chi^2$~region of interest.

Our proper motion data do not by themselves constrain the shape of the
cluster, as they are concentrated in its innermost regions (for
example, our data fall entirely within the estimated $R_{hm} \approx
0.4$~pc of Stolte et al. 2005). We therefore incorporate surface
density data \SurfdensN~(units stars pc$^{-2}$)~from the
literature. We have proper motion constraints from five radial annuli;
comparison of these data alone to the velocity dispersion model yields
the figure of merit $\chi^2_{\rm kinem}$. Comparison of the seven
radial estimates of \SurfdensN~from \citet{espinoza09}~to model
prediction then yields the figure of merit $\chi^2_{full}$. The full
figure of merit is then
\begin{eqnarray}
  \chi^2_{kinem} & \equiv & \chi^2_{R} + \chi^2_{T} \nonumber \\
  & = & \sum^{5}_{i=1} \frac{\left[\sigma_R({\rm data}) - \sigma_R(\rm model)\right]^2_i}{\Delta^2_{R,i}} + \sum^{5}_{i=1} \frac{\left[\sigma_T({\rm data}) - \sigma_T(\rm model)\right]^2_i}{\Delta^2_{T,i}} \nonumber \\
  \chi^2_{full} & \equiv & \chi^2_{R} + \chi^2_{T} + \chi^2_{\Sigma} \nonumber \\
  & = & \chi^2_{kinem} + 
  \sum^{7}_{i=1} \frac{\left[\Sigma_N({\rm data}) - \Sigma_N(\rm model)\right]^2_i}{\Delta^2_{\Sigma,i}}
\end{eqnarray}
\label{eq_chiqsuared}
\noindent where $\Delta^2$~represent the squared errors on each
datapoint. 

Since the cluster is in reality mass segregated (e.g. Figer et
al. 1999; Stolte et al. 2005), the underlying mass distribution that
dominates the velocity dispersion is unlikely to be more centrally
concentrated than the massive stars directly amenable to
observation. Espinoza et al. (2009) assess \SurfdensN~for massive
stars in different mass ranges; we report here the mass limits using
the least centrally-concentrated massive-star sample ($10 \le
M_{\ast}/M_{\odot} \le 30$) of Espinoza et al. (2009). Fitted
parameters and the behavior of the $\chi^2$~surface for different
choices of \SurfdensN~(as well as {\it no}~constraint on \SurfdensN,
i.e., fitting with kinematic data only) are discussed in Appendix E.

Figure \ref{fig_chibubble_mtot_10-30}~shows the behaviour of
$\chi^2$~as the model parameters $R_c, R_t, M_{cl}$~are varied,
incuding the \SurfdensN~sample just discussed.  Figure
\ref{fig_chibubble_mproj_10-30}~illustrates the variation of
$\chi^2_{full}$~aganst $R_c, R_t, \MProjObs$. Figure
\ref{fig_radprofs_finer_10-30}~shows radial profiles drawn from within
the $\Delta \chi^2_{full} = 7.82$~surface in parameter space, which
corresponds to 95$\%$~formal significance (or 2$\sigma$). As can be
seen, a wide range of \MKineObs~is consistent with the kinematic and
surface-density data, but a rather narrow range of \MProjObs~is
consistent with the flat plateau and magnitude of the velocity
dispersions we measure. Specifically, we find
\bestmass,~\besttot~and~\bestmcl.

All isotropic models tested yield an upper limit on \MProjObs~of
$1.30\times 10^4~M_{\odot}$~at formal $3\sigma$~confidence (Appendix
E). Inclusion of \SurfdensN~removes the very low estimates of
\MProjObs~from consideration, with the largest lower limit obtained
using the full mass range of \citealt{espinoza09}). With the most
massive stars included in \SurfdensN, core radii $R_c < 0.13$~pc are
rejected at the 3$\sigma$~level; this level is well above the minimum
grid value of $R_c = 0.05$~pc. Therefore the grid boundaries are not
leading us to assume an artificially compact cluster. The total
cluster mass is only very weakly constrained from kinematic data
alone. 

We also attempted to account for a wide range of cluster anisotropies
using the algorithm of LM89. This method differs from the ``forward''
modeling we describe here, in that the isotropic velocity dispersion
profile is modified for anisotropy, the enclosed mass
$M(<r)$~estimated from the full Jeans equation using this dispersion
profile, and the density profile $\rho(r)$~estimated from the form of
$M(<r)$. While we were unsuccessful in reproducing the LM89 approach
for a King (1962) profile, parameterization of the cluster with a
Plummer profile and allowing for anisotropy (following
\citealt{leonard92})~was more fruitful. This yielded a slightly wider
range of compatible values of \MProjObs, though still below 1.5$\times
10^4~M_{\odot}$~at the formal $2\sigma$~confidence level. Appendix E
outlines the mass modeling using anisotropic cluster models following
the method of Leonard \& Merritt (1992); mass limits so produced do
not alter the conclusions of this report.

\subsection{Comparison to literature mass estimates} \label{ss_masscomp}

Unlike photometric mass estimates, which rely heavily on an accurate
completeness and extinction correction to map the observed population
onto the underlying population, a kinematic mass estimate only
requires that the motion of a selection of tracer objects be
well-measured (and of course that the assumptions in the mass modeling
be reasonable). We compare our mass estimates with literature
estimates here.

\subsubsection{Literature mass estimates}

With the exeption of the radial-velocity kinematic estimate of
\citet{figer02}, all {\it observational} mass estimates of the Arches
cluster refer to the projected mass within some radius on the sky,
i.e., \MProjRad. Since this is also the best-constrained of our
kinematic mass estimates, we focus our literature discussion on these
estimates.

\citet{serabyn98} extrapolated the mass estimated from observed
O-stars down to low masses to estimate the cluster stellar mass; they
used $JHK'$~imaging with NIRC on Keck-I to estimate a total of $5,000
\pm 1000~M_{\odot}$~in $100 \pm 50$~massive O-stars in the cluster,
which they extrapolated to the full range of stellar masses using a
mass function exponent $\PDMF = 1.35$. This yielded~ $M_{cl}(R <
0.35~{\rm pc}) = (1.5,6)\times 10^4 M_{\odot}$, for lower stellar-mass
limits (2,0.1)
$M_{\odot}$~respectively.\footnote{\citet{serabyn98}~interpret
  $R=0.35$~pc as the total cluster extent, whereas more recent work
  \citep[e.g.,][]{stolte02}~suggests a cylinder of this radius
  contains only about half the total cluster mass. Thus we refer to
  their estimate as a projected mass estimate.}

\citet{figer99}~used $NICMOS$~on $HST$~to perform a photometric
census~down to a photometric limit corresponding to about $6
M_{\odot}$. Within an annulus $0.12 \le R < 0.37~pc$~they measured
about $0.51\times 10^4 M_{\odot}$~in stars, which, using the PDMF
measured for the same stars, was extrapolated to a total mass for this
outer annulus. The number counts of bright stars were then used to
estimate the scale factor from the outer annulus ($0.12 \le R <
0.37$~pc) to the entire inner cluster ($R < 0.37$~pc), yielding a
total mass of the inner cluster $M_{cl}(R < 0.37~{\rm pc}) =
(1.08-1.20) \times 10^4 M_{\odot}$~depending on the lower mass cutoff
adopted (1.0 - 0.1)~$M_{\odot}$, and a top-heavy PDMF exponent ($\PDMF
\approx 0.6$).

\citet{stolte02} used Gemini NGS/AO photometry and the HST/NICMOS data
of \citet{figer99} to search for variation of $\PDMF$~with projected
radius from the cluster center, using Geneva isochrones
\citep{lejeune01} to convert from magnitude to mass. They obtained a
half-mass radius $R_{hm}=10''= 0.4$~pc. By summing the observed mass
histogram in two projected-radial bins within $R_{hm}$, they estimated
a total mass amongst the stars {\it measured}, of $0.63\times10^4
M_{\odot}$. The authors preferred not to extrapolate the mass function
beneath their photometric limit of about $2 M_{\odot}$~due to the
uncertainties in so doing. They pointed out also that their estimate is
not corrected for incompleteness, and so their estimate for the total
mass within $r_{hm}$~is therefore $M_{cl}(< 0.4~{\rm pc}) \sim 10^4
M_{\odot}$. Note that this is the total stellar mass within a cylinder
of radius $R = 10'' \equiv 0.4$~pc at $\sim$8 kpc, {\it not} the total
stellar mass in the cluster. \citet{stolte02}~found a PDMF exponent
$\PDMF \sim 0.8 \pm 0.2$~as a spatial average, but with considerable
spatial variation as a function of projected radius, though this does
not affect their mass estimate. At $R < 5''$~the PDMF is nearly flat,
developing to $\PDMF = 1.04 \pm 0.29$~at $5'' < R < 9''$~and
consistent with Salpeter at greater radii ($\PDMF = 1.69 \pm 0.66$~for
$10'' < R < 20''$).

\citet{espinoza09} report the use of VLT/NACO photometry to fit mass
function exponents for two annuli in projected radius; $R < 0.2$~pc
and $0.2 < R < 0.4$~pc. Differential reddening-corrections on a
star-by-star basis were used. The authors prefer to quote the initial
masses from Geneva isochrones rather than present-day masses, and
therefore give the {\it initial} mass function exponents $\IMF$; they
find a spatially-averaged $\IMF = 1.1 \pm 0.2$~for $M > 10M_{\odot}$,
consistent with Salpeter, and point out that this index is about
0.1-0.15 dex steeper than PDMF indices reported in the
literature. Integrating the IMF down to a low-mass cut-off of $1
M_{\odot}$, \citet{espinoza09} report $M_{cl}(R < 0.4)~{\rm pc} = (2 \pm
0.6) \times 10^4 M_{\odot}$.\footnote{Most of the mass-loss over the
  Arches' history is probably dynamical; therefore \citet{espinoza09}
  are really reporting the total initial mass of the stars presently
  in the cluster, {\it not} the initial total cluster stellar
  mass. This number is closer to the present-day cluster mass
  $M_{cl}(<R)$~than $M_0(<R)$, so we identify their mass with
  $M_{cl}(<R)$~here.}  Whether a low-mass truncation exists at all in
the Arches mass function is an open question;
\citet{espinoza09}~therefore use a \citet{kroupa02}~mass function to
estimate $M_{cl}(R < 0.4~{\rm pc}) = (3.1 \pm 0.6) \times 10^4
M_{\odot}$ without a lower-mass cut-off. While the extrapolation to
total cluster mass depends on the radial dependence of the PDMF and
the density profile, N-body models suggest \citep{harfst10} that
roughly half the cluster mass is observed between projected radius
$R=0.4$~pc and the tidal radius $\sim 1$~pc, which would suggest that
the total present-day cluster mass indicated by \citet{espinoza09} is
closer to $M_{cl} = (4 \pm 1.2)\times 10^4 M_{\odot}$~and $M_{cl} =
(6.2 \pm 1.2) \times 10^4 M_{\odot}$~for lower mass limits of $1.0
M_{\odot}$~and $0.08 M_{\odot}$,~respectively.

\subsubsection{Additional mass estimates}

In addition to photometric mass estimates discussed above, the Arches
total mass is often used as input to models of the formation and
evolution of massive clusters. To better place our work in context,
all reported mass estimates for the Arches of which we are aware have
been collated into Table \ref{table_all_mass_estimates}. Most of the
total initial mass estimates used in Fokker-Planck (e.g. Kim et
al. 1999) and N-body simulations \citep[e.g.,][]{kim00, pz02,
  harfst10}~lie in the range $M_0 \sim 1-5 \times 10^4
M_{\odot}$. However there are some notable outliers; in particular,
work approaching the Arches formation from the point of view of cloud
fragmentation (generating the IMF) assumes a very high initial cluster
mass \citep[e.g. $M_0 \sim 15 \times 10^4 M_{\odot}$][]{dib07}. The
connection between initial cluster mass for models and observed
present day mass depends on a number of complicated factors that
render direct comparison of $M_0$~to observation highly assumption
dependent; the Arches has probably lost roughly half of its stellar
mass since formation \citep[e.g.][]{kim00, harfst10}.

\subsubsection{Proper motion-derived mass estimate compared to literature mass estimates}

Isotropic King (1962) profiles produce estimates of the present-day
projected mass \MProjObs~that are at least $3\sigma$~below the
\MProjObs = $(3.1 \pm 0.6) \times 10^4~M_{\odot}$~derived by
\citet{espinoza09}~under the assumption of a non-top-heavy mass
function and no lower-mass truncation. In contrast, photometric
estimates assuming either a low-mass truncation or top-heavy mass
function, or both, are more compatible with our dispersion data under
the assumptions of our mass models. In particular, our upper limit of
$1.30 \times 10^4 M_{\odot}$~from isotropic King modeling is
1.5$\sigma$~below the photometric estimate of \citet{espinoza09}~with
a lower-mass cut-off at 1$M_{\odot}$~using a mass function that is not
strongly top-heavy. In addition, the mass ranges of
\citet{figer99}~are highly compatible with our dispersion-based mass
estimate.

Within the limitations of our modeling, then, our velocity dispersion
estimate is compatible with a mass function that is either top-heavy,
truncated at low-mass, or both. A Salpeter PDMF without low-mass
truncation is not indicated by our data. 

\begin{figure}
\begin{center}
\includegraphics[width=12cm]{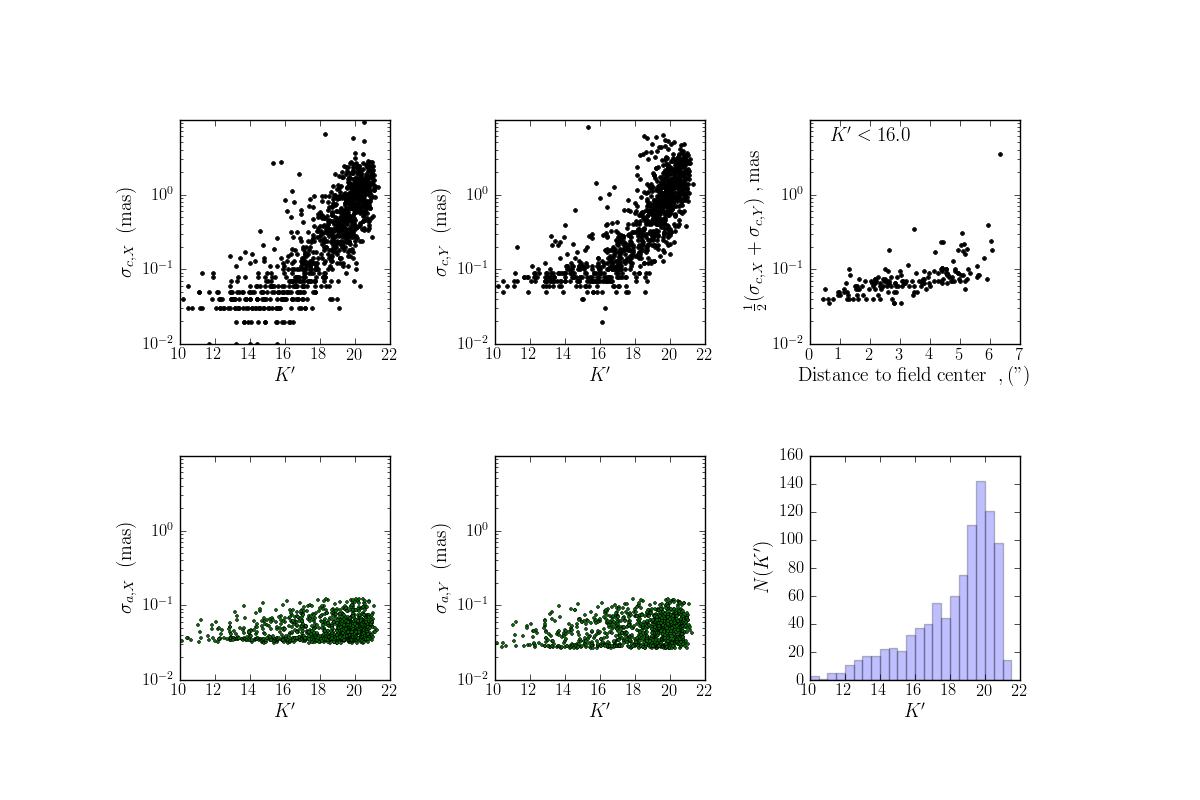}
\caption{Positional errors as measured for an example epoch
  (2008.50). Positions are those in the image-stack with centroiding
  errors assessed as the rms of measurements within an epoch (Section
  \ref{ss_centroiding}). Top row: centroiding errors along detector-X
  and detector-Y (top-left and top-middle respectively), and the
  average of the two as a function of distance from the field center
  (top-right). Bottom row: alignment errors along X and Y (bottom-left
  and bottom-middle; Section \ref{ss_align}). The magnitude histogram
  is given in the bottom-right panel.}
\label{fig_additive_det}
\end{center}
\end{figure}

\begin{figure}
\begin{center}
\includegraphics[height=120mm]{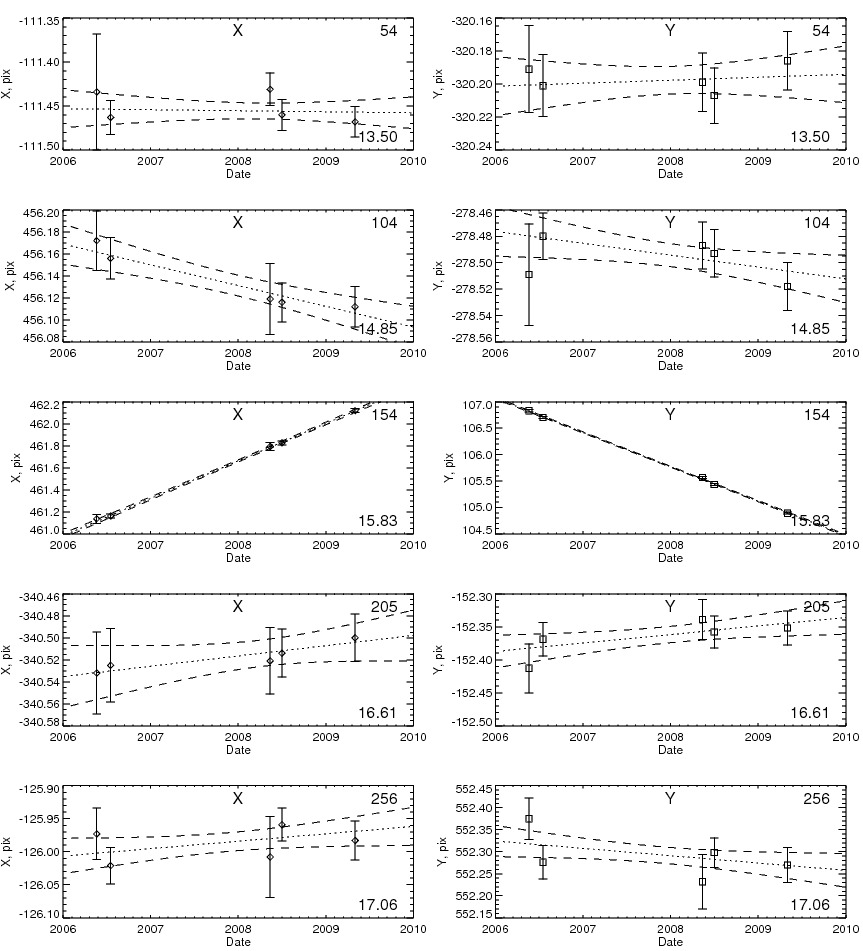}
\caption{Five example positional time-series. Left columns show motion along X, right columns along Y. Object IDs and $K'$~magnitudes are indicated in the right top and bottom corners respectively. Note that each vertical axis is scaled to accommodate the motion of the star and is in units of pixels in the \treference~reference frame. The best-fit straight line to the motions are indicated in each case, as are 1$\sigma$~positional error curves. Object 154 is likely a field object, as indicated by its large proper motion relative to the reference frame.}
\label{f_example_curves}
\end{center}
\end{figure}

\begin{figure}
\begin{center}
\includegraphics[width=15cm]{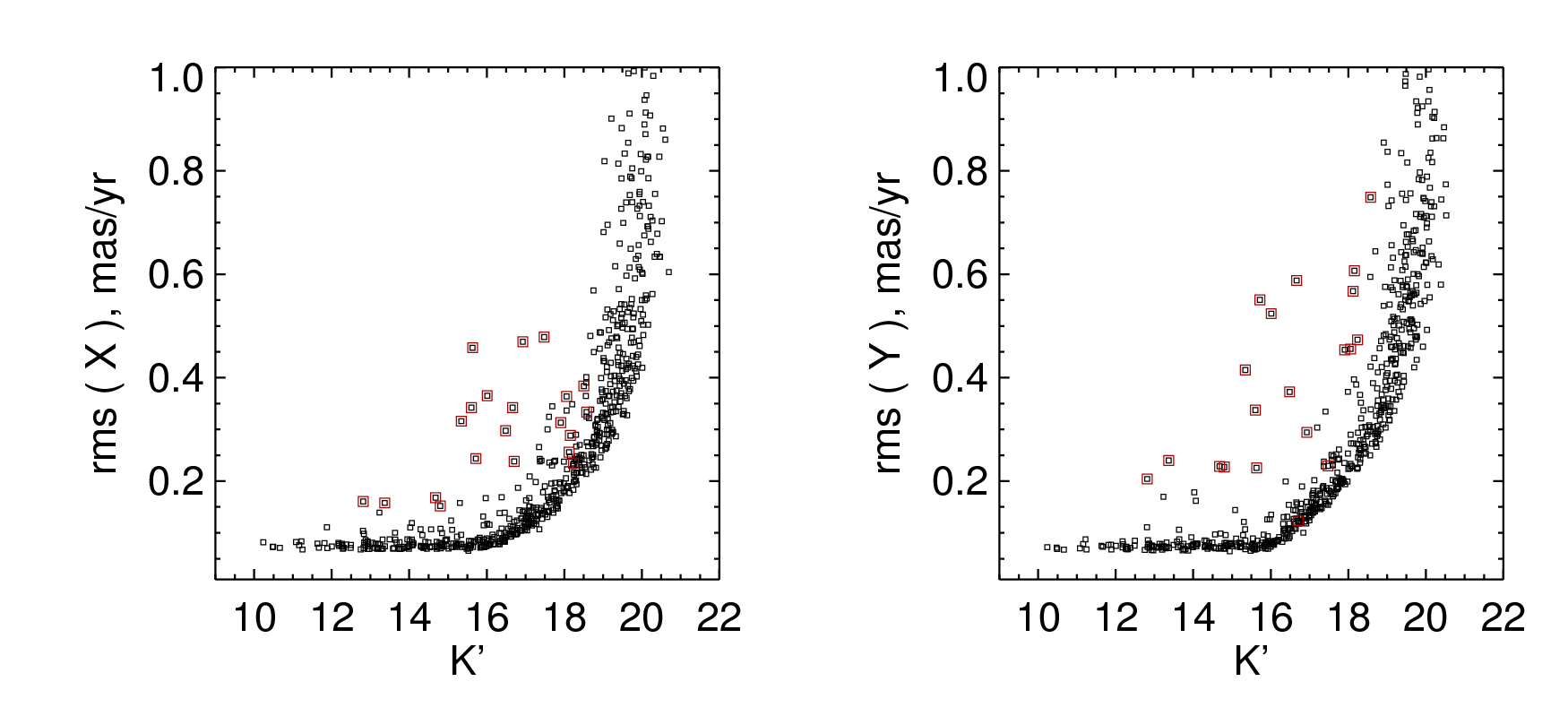}
\caption{The distribution of adopted proper motion precision (Section \ref{ss_errors} and Table \ref{tab_budget}), for all objects with five position-measurements. Outliers due to likely mismatches are indicated by squares and were removed from the analysis. An object qualifies as an outlier if the rms in either co-ordinate falls obviously outside the sequence defined by most of the points.}
\label{f_motion_errors}
\end{center}
\end{figure}

We remind the reader that our mass limit is a first estimate with a
straightforward model, which is likely subject to update when the full
machinery of simulation is brought to bear on the problem using our
dispersions as a constraint. On the modeling side, several factors
complicate the interpretation of velocity dispersion data. Firstly,
the location of the Arches cluster in a strong tidal field suggests
the assumptions of spherical symmetry and negligible rotation may be
violated. Secondly, while the degree of mass segregation in a young
post-collapse cluster is different for stars of different masses,
after $\sim 2$~Myr, the massive stars whose motion we measure ($M \ga
10 M_{\odot}$) are likely to have undergone some mass segregation
\citep{allison09}. Therefore we are measuring tracer stars whose
velocity dispersion may be biased to low values. Velocity dispersions
constructed from proper motion observations of stars solely with
masses $M \la 10 M_{\odot}$~(i.e., $K' \ga 16$) would provide a second
mass estimate from a population less sensitive to mass segregation
\citep[see Figure 2 of][]{allison09}. For the present dataset this
sample is almost entirely on the steeply-rising part of the
error-magnitude curve (Figure \ref{fig_additive_det}). Investigation
of this sample is outside the scope of the present communication. At
the present stage of our investigations, we limit ourselves to
pointing out that our mass estimate may be biased to low values by our
sample selection of massive stars that have likely undergone some
degree of dynamical mass segregation.

On the observational side, we do not yet have sufficiently precise
motions outside 0.2pc to constrain M(R) outside this region, and have
had to resort to projection of models that have significant caveats
when applied to this cluster. Future observations of the outer fields
should remedy this situation.

Note that, unlike with radial-velocity studies, binaries are unlikely
to have an effect on the velocity dispersion measurement we report
here. To produce an effect of the same order of magnitude as the
proper motion dispersion we obtain, a substantial binary population
would be needed in which the binary orbit shifted the center of
near-IR light by $\sim 0.3$~mas over the three-year timebase of our
observations. A binary with components $150~\&~50~M_{\odot}$~in a
1,000d orbit would exhibit semimajor axis and orbital speed of
adequate magnitude ($a \simeq 0.36$~mas, $v \simeq 0.13$~mas
yr$^{-1}$,~respectively) to produce this effect (assuming the near-IR
brightness ratio corresponding to this mass ratio is sufficient for
the center of light to move). However, we expect such systems to be
too rare to produce any effect on the dispersion measurement over the
Arches population. Furthermore, such systems would be confined to the
brightest magnitude-bin in our analysis, leading to a decrease in
measured dispersion with increasing apparent magnitude, which is not
observed. Appealing to high eccentricity introduces an additional
selection-effect (on the orientation of the orbit to produce
measurable motion). Thus we conclude binaries are an insignificant
contributor to the measured velocity dispersion in the Arches cluster
(compare with \citealt{gieles10}).

\subsection{Mean motion of the cluster}\label{ss_meanmotion}

The 2D Gaussian profile of the field component in the VPD shows an
axis ratio that is roughly constant with magnitude, while its
contribution to the sample in the field of view increases as fainter
objects are probed. Its orientation in the VPD is consistent with the
galactic plane, indicating that the velocity dispersion along the
field major-axis in the VPD is strongly affected by differential
rotation.

The bulk motion of the Arches with respect to the field population is
172 $\pm$~15 km s$^{-1}$~(the inverse variance-weighted average of the
$14 < K < 16$ and $16 < K < 18$~bins). Including the $18 < K < 20$~bin
revises this figure downward to 153 $\pm$~11 km s$^{-1}$; however in
this magnitude range the proper motion error curve rises steeply with
magnitude (Figure 5), so objects this faint may be particularly
vulnerable to misclassification bias (Appendix D). This is slightly
smaller than the $212 \pm 29$~km s$^{-1}$~determined previously
\citep{stolte08}. This is probably due to two competing biases in the
previous work that oppose each other: \citet{stolte08}~imposed a hard
membership limit, where all objects within a certain velocity from the
cluster center in the VPD are denoted cluster objects, which tends to
exaggerate the cluster-field separation in the VPD by cutting off one
side of the field component. Conversely, they included objects at all
magnitudes in their estimate of the bulk motion, which tends to reduce
the estimated component separation (see Appendix D).

In their study of the Arches bulk motion, Stolte et al. (2008) found
that, although the Arches is unlikely to be on a circular orbit,
integration of its path through the potential of the inner Milky Way
indicated the cluster was unlikely to pass sufficiently close to the
GC to spiral in towards it and donate its stars to the GC nuclear
cluster. Our revised motion estimate makes the Arches orbit slightly
more compatible with circular motion. Following the arguments of
Stolte et al. (2008), if on a circular orbit, ${\bf v.r} = 0$~then
demands an enclosed mass only $1.5\sigma$~above that measured
photometrically \citep{launhardt02}. At first glance, our new bulk
motion supports the conclusions of Stolte et al. (2008) that a cluster
that is dynamically similar to the Arches is unlikely to be a future
source of young stars for the GC star cluster. Integration of the
Arches motion through the potential of the inner Milky Way using our
new motion determination is required to draw further conclusions about
the formation and subsequent motion of the cluster.

Interpretation of the cluster bulk motion is complicated by three
factors. First, the kinematic parameters of the field component
depends on the distribution of tracer stars along the line of sight as
well as their motion. Differential rotation by field stars
participating in Galactic rotation may therefore vary with tracer-star
brightness (with observations to different depths picking up different
field-tracer populations). Second, the field population (or a
significant component of the field population) may show its own motion
beyond galactic rotation; for example this motion may be dominated by
bar rotation on the nearside of the GC (if the far side of the bar
suffers from higher extinction). Or, rotation of the NSD could impose
a mean motion of the field component with respect to the Arches
cluster. Third, extinction variations along the line of sight coupled
with the low-number statistics ($\sim$~few tens of field objects in
each magnitude bin; Table \ref{tab_kinparams}) may reduce the validity
of a gaussian to represent the field component in the first
place. Thus our quoted error of $15$~km s$^{-1}$~on the Arches bulk
motion is likely an underestimate. \citet{stolte08}~discuss further
some of the difficulties associated with interpreting a bulk motion
against a mean-field population.

\subsection{Properties of the field population}\label{ss_fieldprops}

Within the measurement errors, the orientation of the field ellipse is
entirely consistent with the direction of the vector joining the
cluster and field centers in the VPD, indicating the Arches moves
along the direction of preferential motion for the field (Table
\ref{tab_kinparams}).

To our knowledge, the covariance $\zk$~of the field component provides
the first estimate of the stellar velocity dispersion of the bulge
along such a close sight-line to the Galactic center. This will allow
a direct constraint on the bulge potential along this sight-line,
which itself is a key ingredient in the use of cluster bulk motion to
assess its likely path through the inner Milky Way (Stolte et
al. 2008). Here we restrict ourselves to a comparison of $\zk$~with
the velocity dispersion of the Bulge at higher latitudes.

The bulge is a highly complex stellar structure, with many basic
parameters presently under debate, complicated by shifting
nomenclature in the literature. Several components appear to be
present, with the relationship between them still far from
settled. Chemical evidence suggests most of the stars formed early and
rapidly, as might be expected for a ``classical'' bulge component
\citep[e.g.,][]{mcwill94, lecureur07}, while the spatial arrangement
and motion of the stars suggests a bar structure, driving a
``boxy/peanut''~bulge \citep[e.g.,][]{dwek95, howard09, mcwill10,
  shen10}.

Whatever its formation history, the present-day bulge shows variation
of stellar kinematics with metallicity. \citet{soto07}~report
variation of the shape of the $\sigma_r, \sigma_l$~velocity ellipsoid
with metallicity~(\citealt{zhao94}~provide an early detection of
vertex deviation in the bulge). \citet{babusiaux10} present proper
motion dispersions as a function of metallicity for the Baade's Window
(hereafter, BW) field ($l = 0\degr, b = -0.4\degr$), significantly
farther out from the GC than our sample. For stars of approximately
solar and higher metallicity, they report $\sigma_l, \sigma_b = 107 \pm
6, 94 \pm 6$~km s$^{-1}$. For stars with $[Fe/H] < -0.14$,
\citet{babusiaux10}~report $\sigma_l, \sigma_b = 138 \pm 12, 103 \pm
7$~km s$^{-1}$.

To estimate the velocity dispersion of the Arches field component we
take the variance-weighted mean of the determinations from the three
non-overlapping bins ($14.0 \le K' < 16.0$; $16.0 \le K' < 18.0$~and
$18.0 \le K' < 20.0$). This yields major-axis and
minor-axis~dispersions of $\overline{\sigma_{a,f}} = 2.72 \pm
0.15$~mas yr$^{-1}$~and $\overline{\sigma_{b,f}} = 1.69 \pm 0.10$~mas
yr$^{-1}$, which scale to the distance of the Galactic center ($8.4
\pm 0.4~{\rm kpc}$; Ghez et al. 2008) as $103 \pm 7.7$~km s$^{-1}$~and
$64 \pm 5.0$~km s$^{-1}$. The orientation of the field component in
the vector point diagram (VPD), expressed as a position-angle East of
North, is~$\overline{\theta_f} = +30.5\degr \pm 7.1\degr$~(motion
along the galactic plane corresponds to about $+27.1\degr$~in the
VPD). For the remainder of this section we therefore identify the
field velocity dispersion major axis with the galactic plane, so
$\sigma_{a,f}, \sigma_{b,f}$~represent the field velocity dispersion
in galactic longitude and latitude respectively. Thus our field
velocity dispersion is more consistent with the higher-metallicity BW
stars, which is currently interpreted by \citet{babusiaux10} as a
bar-dominated population.

Some caution is warranted interpreting $\zk$~for the field, as a
number of parameters of the field population are still not fully
understood. First, the NSD imprints its own motion on the field, which
may be coherent and different from the motion of the bulge/bar
system. Second, the mass distribution of the composite bulge along our
line of sight is not fully constrained; for example the range in bar
orientation estimates in the galactic plane still ranges by
$45\degr$~\citep[e.g.,]{benjamin05, robin09, mcwill10}. Third, astrometric
completeness likely biases the bulge motion we observe towards the
near-side of the GC, so the field motion is sampled at some preferred
mean distance from the GC. Fourth, the mixture of bulge and bar
components making up the field population is not yet fully
constrained.

\subsection{Membership probabilities and $L'$-excess sources} \label{ss_excess}

Our refined membership probabilities allow a clean-cluster sample to
be extracted for further work.  Appendix F lists the formal membership
probabilities (equation 2) for well-measured stars in our sample. The
use of this cleaned sample to probe the cluster mass function is
underway by our group \citep{mccrady11}, and will be reported in a
following communication.  For now, we note that the Keck-Keck motions
provide support to the conclusions of \citet{stolte10}, that a
significant number of stars with circumstellar disks are likely
present in the Arches cluster. These stars appear redward of the locus
of most stars in the Arches in the $H-K$~color-magnitude diagram;
since $L'$-excess sources cannot be distinguished from field stars
based on $H-K'$~color alone, kinematic separation is essential. Of
sixteen objects falling within the locus of points redward of the
cluster main sequence within the CMD (Figure \ref{fig_CMD}, six show
proper motions suggestive of cluster membership. More generally,
however, we find that most of the outliers from the main locus of
Arches stars in the $H-K$~CMD are indeed kinematically associated with
the field.

\section{Conclusions}

With uniform observational setup over a sufficient time baseline and
careful accounting for a number of sources of proper motion error, we
have measured the internal velocity dispersion of the Arches cluster
for the first time, finding $\sigma = 0.15 \pm 0.01$~mas yr$^{-1}$,
which corresponds to $5.4 \pm 0.4$~km s$^{-1}$~at a distance of 8.4
kpc.

We have used this dispersion to test the photometric estimates of the
present-day mass function (PDMF) of the Arches cluster. The total mass
is likely in the range \besttot, but this is only weakly constrained
by kinematic data and is consistent with nearly all suggestions of
total cluster mass from the modeling literature. The projected mass
(i.e., mass contained within a cylinder of radius $R$~on the sky) is
rather better constrained; we find \bestmass~at formal
$3\sigma$~confidence. The upper bound of this range is $3\sigma$~below
the photometric estimate for \MProjObs~estimated by
\citet{espinoza09}~under the assumptions of a non-top-heavy mass
function without low-mass truncation. If a substantial contribution
from massive binaries were unknowingly included in our measurement,
this would strengthen our conclusion because the upper cluster mass
bound would accordingly be reduced. This is the first mass estimate
for the Arches based on proper motion velocity-dispersion.

We have also revised the bulk motion of the Arches slightly
downward. Our updated motion of $172 \pm 15$~km s$^{-1}$~is only
slightly lower than the $212 \pm 29$~km s$^{-1}$~determined previously
(Stolte et al. 2008). Taken at face value, this supports the previous
conclusion that the Arches Cluster is unlikely to pass within 10~pc of
the GC.

Finally, we have provided the first estimate (to our knowledge) of the
velocity dispersion of the Bulge along such a close sight-line to the
Galactic center; this estimate is (103, 64)~$\pm$~(7.7,5.0)~km
s$^{-1}$, with the major axis coincident with the Galactic plane, to
within the uncertainties.


\begin{figure}
\begin{center}
\includegraphics[height=18cm]{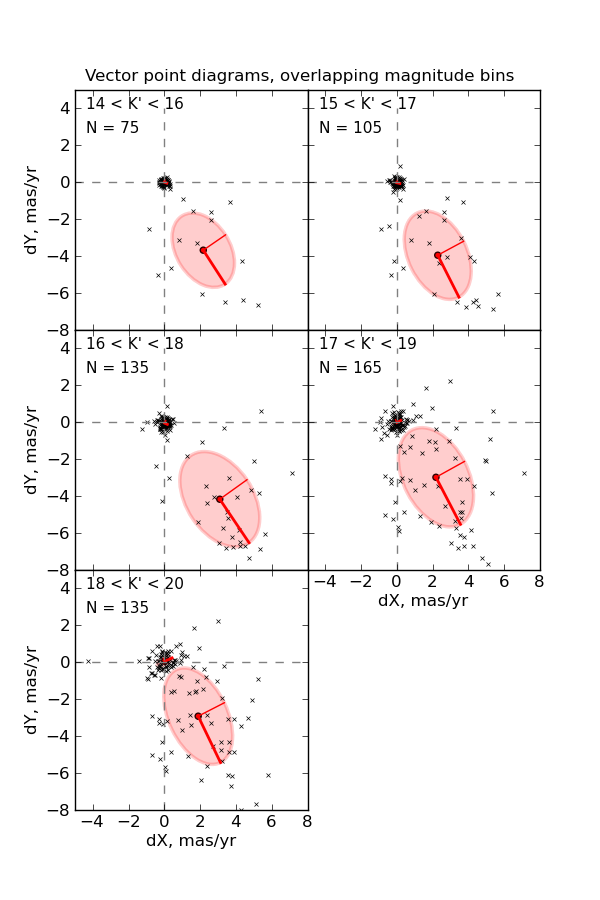}
\caption{Vector Point Diagrams for the overlapping magnitude-ranges of Section \ref{ss_fit} and Table \ref{tab_kinparams}. Shaded ellipses give the $1\sigma$~contours for the two-dimensional gaussian components fit to the field and cluster components. Within each ellipse, the lines indicate the length and direction of the semimajor (thick red line) and semiminor (thin red line) axes.}
\label{f_VPD_compilation}
\end{center}
\end{figure}

\begin{figure}
\begin{center}
\includegraphics[height=10cm]{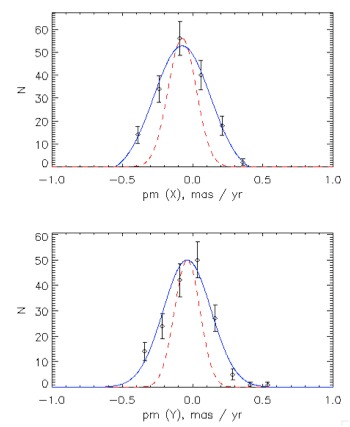}
\caption{Observed velocity dispersion in each coordinate for stars
  with $14.0 \le K' < 17.0$~compared to a gaussian of width equal to the mean measurement error over this range (Section \ref{ss_errors} and Table \ref{tab_budget}). Panels correspond to detector-X (Top) and detector-Y (bottom). This figure was constructed after removing likely field objects (Section \ref{ss_membprob})}
\label{f_compare_erro}
\end{center}
\end{figure}

\begin{figure}
\begin{center}
\includegraphics[width=15cm]{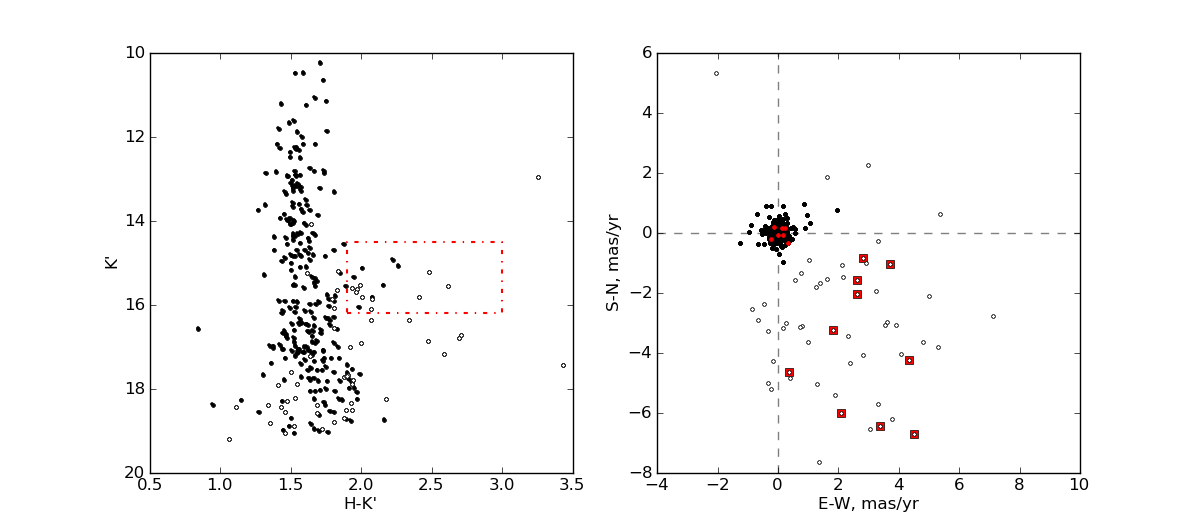}
  \caption{Color-magnitude diagram (CMD) and Vector Point Diagram
    (VPD) for all objects with proper motion error $< 0.5$~mas
    yr$^{-1}$~and five epochs of measurement. The CMD presented here
    was constructed by matching $K'$~measurements to photometry taken
    in $H$-band in 2006 May with Keck-2/NIRC2-LGS \citep{mccrady11},
    which limits the depth in the CMD. Objects with $P_{\rm{cluster}}
    > 0.995$~are shown in black, all other objects denoted with open
    circles. Red objects in the VPD correspond to the stars within the
    red dot-dashed box in the CMD, and represent well-measured objects
    with a possible $H-K'$~excess. Of these objects, those with
    $P_{\rm cluster} > 0.995$~are shown with a red circle; their field
    counterparts are shown with red squares. See Section
    \ref{ss_excess}~and \citet{stolte10}~for more information on these
    objects.}
\label{fig_CMD}
\end{center}
\end{figure}

\begin{figure}
\includegraphics[width=8cm]{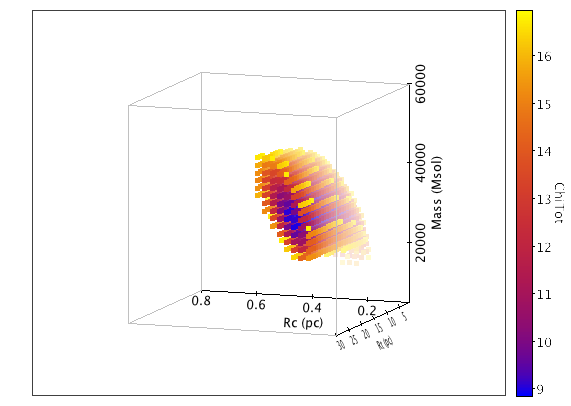}
\includegraphics[width=8cm]{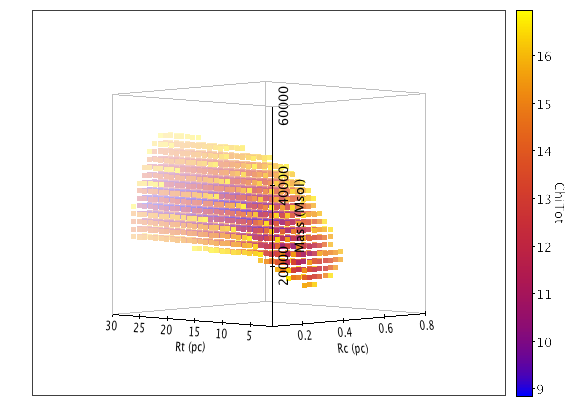}
\caption{Views of the $\Delta \chi^2_{full} < 7.82$~region when both
  kinematic and surface density data (for stars of mass $10 \le
  M/M_{\odot} \le 30$; Espinoza et al. 2009) are included in the
  assessment. Axes are: $R_c, R_t, M_{cl}$, with total cluster mass
  $M_{cl}$~vertical in each case. Limits shown are: $0.05 \le R_c \le
  0.8$~pc; $1.0 \le R_t \le 30$~pc; $0.5 \le M_{cl} \le 6.0 \times
  10^4 M_{\odot}$.}
\label{fig_chibubble_mtot_10-30}
\end{figure}

\begin{figure}
\includegraphics[width=8cm]{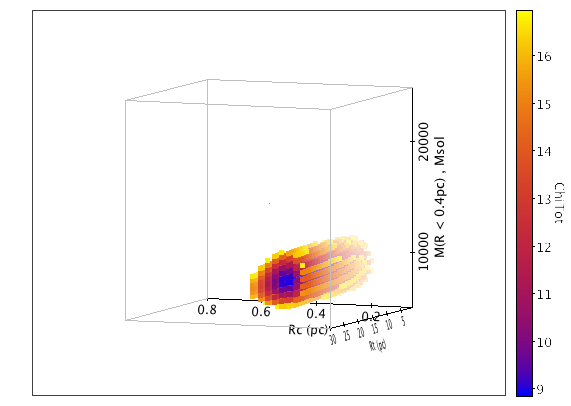}
\includegraphics[width=8cm]{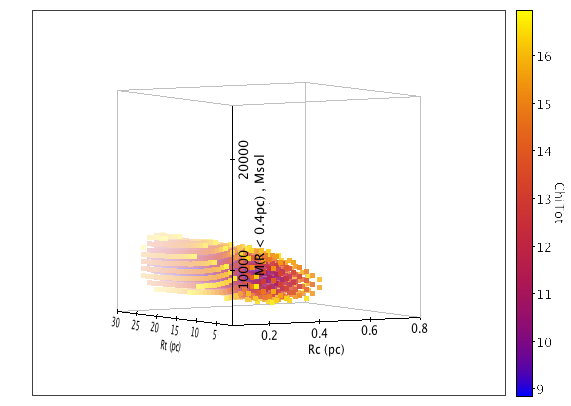}
\caption{As Figure \ref{fig_chibubble_mtot_10-30}, but with \MProjObs~along the vertical axis. Limits shown are: $0.05 \le R_c \le 0.8$~pc; $1.0 \le R_t \le 30$~pc; $0.5 \le \MProjObs \le 2.5 \times 10^4 M_{\odot}$.}
\label{fig_chibubble_mproj_10-30}
\end{figure}

\begin{figure}
\includegraphics[width=16cm]{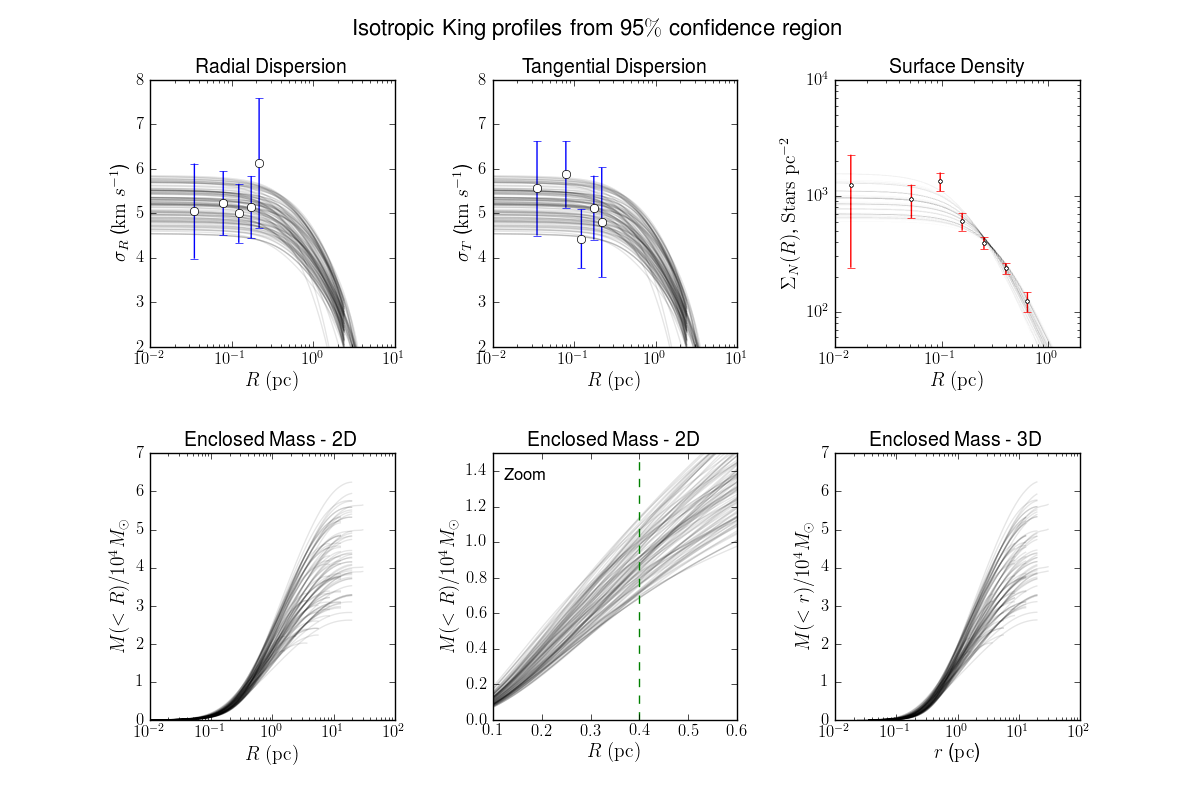}
\caption{Radial profiles corresponding to parameter-sets within the
  $\Delta \chi^2_{full} < 7.82$~surface; both our own kinematic data
  and surface density (by number) \SurfdensN~data are used
  (corresponding to stars of mass $10 \le M/M_{\odot} \le 30$;
  Espinoza et al. 2009). Top-left and top-middle panels show radial
  and tangential velocity dispersions from proper motions (points)
  over the projected profiles corresponding to model parameters
  (lines). Top-right panel shows the \SurfdensN~dataset with model
  predictions. Bottom-left and bottom-middle panels show the total
  mass within cylindrical radius $R$~on the sky, with $R=0.4$~pc
  indicated by the vertical dashed line. Bottom-right panel shows the
  mass enclosed within a sphere of radius $r$~pc from the cluster
  center. See also Table \ref{table_massest_10-30_full}.}
\label{fig_radprofs_finer_10-30}
\end{figure}

\begin{figure}
\begin{center}
\includegraphics[width=16cm]{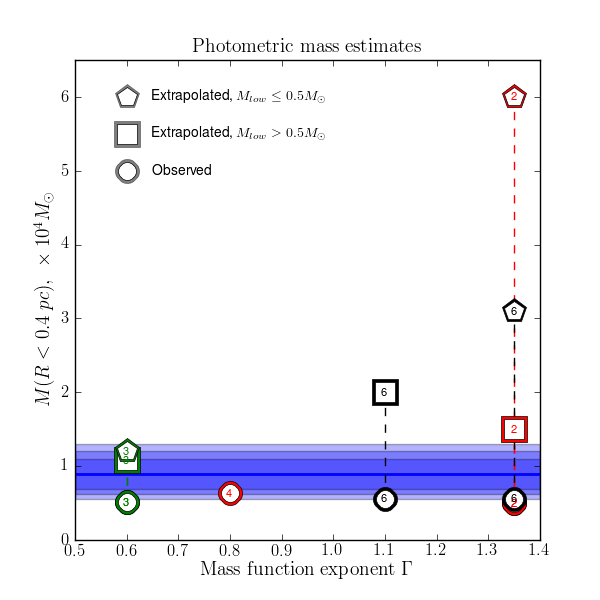}
\caption{Photometric mass estimates for the Arches cluster from the literature. Symbols give the directly-observed photometric mass (circles) and extrapolated mass (where reported; squares and pentagons) depending on the low-mass cut-off assumed. The citation for each estimate is shown inside the symbols (using the numbering of Table \ref{table_all_mass_estimates}). The horizontal bands show our model-dependent mass estimate using our dispersion data (at 1$\sigma$, 2$\sigma$~and 3$\sigma$), and the \SurfdensN~dataset of \citet{espinoza09}; see Section \ref{ss_massmodel} for more detail on the mass modeling used. All masses are reported as \MProjObs.}
\label{fig_literature_estimates}
\end{center}
\end{figure}



\begin{table}
\small
\begin{center}
 \begin{tabular}{lcccccccc} 
   Epoch & (t$_{int} \times N_{coadd}$) & $N_{images}$ & $N_{use}$ & FWHM & Strehl & $N_{\ast}$ & $N_{\ast, {\rm uncrowd}}$ & $K'_{lim}$ \\ 
        &         (s)                &             &           & (mas) &        &       &      & (mag)    \\ 
   \hline
2006 May 21 &   3.00 $\times$ 10 & 15 & 15 &  61.05 &  0.261 & 660 & 649 & 19.43 \\ 
2006 Jul 18 &   3.00 $\times$ 10 & 52 & 38 &  56.95 &  0.349 & 657 & 642 & 19.89 \\ 
2008 May 13 &   3.00 $\times$ 10 & 146 & 72 &  66.66 &  0.219 & 556 & 536 & 19.74 \\ 
2008 Jun 01 &   3.00 $\times$ 10 & 89 & 83 &  54.96 &  0.373 & 845 & 810 & 20.42 \\ 
2009 May 02 &   2.80 $\times$ 10 & 119 & 108 &  51.47 &  0.442 & 968 & 917 & 20.59 \\ 
   \end{tabular}
\end{center}
\normalsize
\caption{Summary of observations. Reading left-right, the columns are: Epoch of observation, the total integration time for each image, the number of images observed, the number of images used, the median FWHM and Strehl ratio over the set of accepted images $N_{use}$, the number of stars measured within the mean image stack in each epoch; the number surviving the cut on proximity to a known neighbour, and finally the magnitude $K'_{lim}$~at which the cumulative distribution function of the observed $K'$~magnitudes reaches 90$\%$~of the total number of stars in the sample at each epoch.}
\label{tab_obsns}
\end{table}

\scriptsize 
\begin{deluxetable}{lcccccccc} 
\tablewidth{0pt} 
\tablecaption{PSF stars. Reading left-right, columns are: sequential star number in the master table of membership probabilities, estimated brightness, and finally the position of the star expressed as an offset in arcseconds from the reference star along the (E-W) and (S-N) directions. See Figure 1 for the locations of these stars on the field of view.} 
\tablehead{ 
Row & $K'$ & $\Delta x$ & $\Delta y$ \\ 
 & (mag) & ($"$) & ($"$)  \\ 
} 
\startdata 
   1 &   10.24 &    2.736 &   -3.943 \\
   2 &   10.48 &    2.063 &   -1.193 \\
   3 &   10.49 &    0.791 &    0.755 \\
   4 &   10.66 &    3.150 &   -2.899 \\
   5 &   11.08 &   -0.633 &   -4.252 \\
   7 &   11.22 &   -1.650 &    1.730 \\
   8 &   11.25 &    1.385 &   -2.334 \\
  16 &   12.18 &    1.012 &   -5.199 \\
  24 &   12.49 &  0 &  0 \\
  25 &   12.50 &    5.407 &   -0.218 \\
  32 &   12.88 &    2.499 &   -5.402 \\
\enddata 
\label{tab_psfstars}
\end{deluxetable} 
\normalsize 


\begin{sidewaystable} 
\centering
\scriptsize 
\begin{tabular}{l|r||l|l|l|l|l||} 
\multicolumn{2}{l||}{Parameter} & 2006.38 & 2006.54 & 2008.37 & 2008.50 & 2009.33 \\ 
\hline 
\hline 
\multicolumn{2}{l||}{$N_{ref}$} & 238  & 239  & 235  & 241  & 233  \\ 
\multicolumn{2}{l||}{$N_{<4}$} & 9     & 10    & 21    & 11    & 6     \\ 
\hline 
\hline 
$\Delta$ & $x'$ & ~-83.94 $\pm$~$6.03 \times 10^{-3}$ & ~9.26 $\pm$~$4.01 \times 10^{-3}$ & ~4.16 $\pm$~$3.71 \times 10^{-3}$ & (-4.06 $\pm$~17.716)$\times 10^{-4}$ & ~6.53 $\pm$~$3.52 \times 10^{-3}$\\(pix)& $y'$ & ~-16.88 $\pm$~$4.74 \times 10^{-3}$ & ~5.69 $\pm$~$4.39 \times 10^{-3}$ & ~3.35 $\pm$~$5.64 \times 10^{-3}$ & (3.66 $\pm$~25.421)$\times 10^{-4}$ & ~3.51 $\pm$~$3.66 \times 10^{-3}$\\ 
\hline 
$x$ & $x'$ & ~1.00005 $\pm$~$2.11 \times 10^{-5}$ & ~0.99994 $\pm$~$9.12 \times 10^{-6}$ & ~1.00025 $\pm$~$1.34 \times 10^{-5}$ & ~0.99999 $\pm$~$6.18 \times 10^{-6}$ & ~1.00002 $\pm$~$8.24 \times 10^{-6}$\\()& $y'$ & (6.64 $\pm$~0.185)$\times 10^{-4}$ & (7.73 $\pm$~0.079)$\times 10^{-4}$ & (1.33 $\pm$~1.425)$\times 10^{-5}$ & (2.32 $\pm$~5.858)$\times 10^{-6}$ & (2.75 $\pm$~0.090)$\times 10^{-4}$\\ 
\hline 
$y$ & $x'$ & (-3.15 $\pm$~0.177)$\times 10^{-4}$ & (-3.53 $\pm$~0.089)$\times 10^{-4}$ & (1.23 $\pm$~0.105)$\times 10^{-4}$ & (-2.40 $\pm$~4.766)$\times 10^{-6}$ & (-2.54 $\pm$~0.072)$\times 10^{-4}$\\()& $y'$ & ~1.00046 $\pm$~$1.97 \times 10^{-5}$ & ~1.00028 $\pm$~$9.62 \times 10^{-6}$ & ~1.00014 $\pm$~$1.34 \times 10^{-5}$ & ~1.00000 $\pm$~$6.66 \times 10^{-6}$ & ~1.00009 $\pm$~$7.99 \times 10^{-6}$\\ 
\hline 
$x^2$ & $x'$ & ~28.90 $\pm$~5.946 & ~20.61 $\pm$~3.463 & ~-18.24 $\pm$~3.989 & ~-0.27 $\pm$~1.986 & ~4.63 $\pm$~2.818\\($\times 10^{-8}$ pix$^{-1}$)& $y'$ & ~11.78 $\pm$~4.751 & ~15.66 $\pm$~3.266 & ~23.10 $\pm$~4.404 & ~0.48 $\pm$~1.937 & ~20.65 $\pm$~2.938\\ 
\hline 
$xy$ & $x'$ & ~1.80 $\pm$~5.818 & ~20.39 $\pm$~3.772 & ~58.96 $\pm$~4.594 & ~-0.28 $\pm$~2.040 & ~-0.58 $\pm$~2.973\\($\times 10^{-8}$ pix$^{-1}$)& $y'$ & ~-5.05 $\pm$~6.935 & ~-10.19 $\pm$~3.985 & ~-28.61 $\pm$~5.203 & ~-1.52 $\pm$~2.680 & ~-20.03 $\pm$~3.335\\ 
\hline 
$y^2$ & $x'$ & ~-3.04 $\pm$~5.917 & ~5.55 $\pm$~3.048 & ~1.09 $\pm$~3.540 & ~1.71 $\pm$~1.686 & ~-0.59 $\pm$~2.959\\($\times 10^{-8}$ pix$^{-1}$)& $y'$ & ~33.22 $\pm$~6.228 & ~52.07 $\pm$~3.664 & ~75.11 $\pm$~5.451 & ~-1.13 $\pm$~2.488 & ~-0.47 $\pm$~2.633\\ 
\hline 
\hline 
\multicolumn{2}{l||}{$1.0$-$M$} & ~-2.55 $\pm$~0.150 & ~-1.14 $\pm$~0.064 & ~-1.95 $\pm$~0.095 & (1.97 $\pm$~4.303)$\times 10^{-2}$ & ~-0.58 $\pm$~0.062\\ 
\multicolumn{2}{l||}{($\times 10^{-4}$ )} & & & & &\\ 
\hline 
\multicolumn{2}{l||}{$1.0$-$M_y/M_x$} & ~-4.08 $\pm$~0.275 & ~-3.34 $\pm$~0.135 & ~1.10 $\pm$~0.189 & (-5.72 $\pm$~9.564)$\times 10^{-2}$ & ~-0.71 $\pm$~0.104\\ 
\multicolumn{2}{l||}{($\times 10^{-4}$ )} & & & & &\\ 
\hline 
\multicolumn{2}{l||}{$\theta_{rot}$} & ~-10.11 $\pm$~0.250 & ~-11.62 $\pm$~0.129 & ~1.13 $\pm$~0.182 & (-4.88 $\pm$~7.925)$\times 10^{-2}$ & ~-5.46 $\pm$~0.116\\ 
\multicolumn{2}{l||}{($''$)} & & & & &\\ 
\hline 
\multicolumn{2}{l||}{$\theta_{skew}$} & ~7.20 $\pm$~0.553 & ~8.66 $\pm$~0.234 & ~2.82 $\pm$~0.364 & -0.002 $\pm$~0.153 & ~0.43 $\pm$~0.245\\ 
\multicolumn{2}{l||}{($''$)} & & & & &\\ 
\hline 
\end{tabular} 
\normalsize 
\caption{Transformation parameters taking the starlist in each epoch into the reference frame $t_{ref}$. {\it Top row}: Number of reference stars $N_{ref}$~used in the mapping, along with the number of reference stars $N_{<4}$~that appear in fewer than four epochs. {\it Next six rows:} Coefficients of the polynomial fits $x' = f(x,y)$~and $y' = g(x,y)$~(top and bottom rows respectively in each pair). For reference, a quadratic term of size $10.0\times10^{-8}$~pix$^{-1}$~would introduce displacement 0.25 mas at the edges of the detector, comparable to the centroiding error for bright ($K' < 16$)~objects (Table \ref{tab_budget}). {\it Bottom four rows:} The linear parts of the transformations re-expressed as a global scaling $M$, nonuniform magnification $M_y/M_x$, rotation $\theta_{rot}$~and departure from perpendicular axes $\theta_{skew}$. (Global shifts $\Delta$~appear in the polynomial fits and are not repeated.)} 
\label{tab_transform}
\end{sidewaystable}


\begin{table} 
\scriptsize 
\begin{center} 
\begin{tabular}{l|l||l|l|l|l|l||c||l||l||} 
 \multicolumn{2}{l||}{$K'$} & \multicolumn{5}{|l||}{Centroiding,  Alignment (mas)} & Additive & Confusion & Motion \\ 
\multicolumn{2}{l||}{}  & 2006.39  & 2006.54  & 2008.37  & 2008.50  & 2009.33  & (mas) & bias (mas) & (mas/y)   \\ 
\hline 
\hline 
\hline 
\multirow{2}{*}{$10-16$}  & $x$  & 0.25, 0.09 & 0.10, 0.05 & 0.25, 0.06 & 0.06, 0.04 & 0.08, 0.03 & 0.16 $\pm$~0.02& 0.0 - 0.03 & {\bf 0.076 } \\ & $y$  & 0.23, 0.08 & 0.07, 0.05 & 0.19, 0.08 & 0.11, 0.04 & 0.09, 0.05 & 0.15 $\pm$~0.02 &   & {\bf 0.074 } \\ \hline 
\multirow{2}{*}{$16-18$}  & $x$  & 0.41, 0.11 & 0.23, 0.05 & 0.40, 0.08 & 0.14, 0.05 & 0.14, 0.04 & 0.24 $\pm$~0.02& 0.03 - 0.10 & {\bf 0.130 } \\ & $y$  & 0.42, 0.10 & 0.20, 0.05 & 0.43, 0.09 & 0.17, 0.05 & 0.18, 0.05 & 0.30 $\pm$~0.03 &   & {\bf 0.153 } \\ \hline 
\multirow{2}{*}{$18-20$}  & $x$  & 1.10, 0.11 & 0.92, 0.05 & 1.03, 0.08 & 0.60, 0.05 & 0.59, 0.04 & 0.59 $\pm$~0.06& 0.1- 1.0 & {\bf 0.378 } \\ & $y$  & 1.38, 0.10 & 1.05, 0.05 & 1.35, 0.09 & 0.81, 0.05 & 0.77, 0.06 & 0.71 $\pm$~0.08 &   & {\bf 0.478 } \\  
\end{tabular} 
\caption{Astrometric error budget. For each magnitude bin, the top (bottom) row gives errors in X (Y). For each star, centroiding, alignment and additive error describe random variation between epochs. The effect of confusion bias on motions depends on its variation between epochs; random variation is already included in the additive error, while linear trends masquerading as spurious motions are expected to be $\lesssim 10\%$~of the confusion bias across the epochs for all objects (Section \ref{ss_additive}).} 
\label{tab_budget} 
\end{center} 
\normalsize\end{table} 


\scriptsize 
\begin{deluxetable}{lcccccccc} 
\tablewidth{0pt} 
\tablecaption{Membership table for objects in the Arches Central field. Reading left-right, columns are: Sequential star number, estimated brightness, offset from reference star (E-W and S-N), proper motion and error, and the formal probability that the object is associated with the cluster and field, respectively.$^\dagger$}
\tablehead{ 
Row & $K'$ & $\Delta x$ & $\Delta y$ & $\mu_x$ & $\mu_y$ & $P({\rm cluster})$ & $P({\rm field})$ \\ 
 & (mag) & ($"$) & ($"$) & (mas yr$^{-1}$) & (mas yr$^{-1}$) &  &   \\ 
} 
\startdata 
   1$^{\ast}$ &   10.24 &    2.736 &   -3.943 &    0.14 $\pm$  0.07 &    0.13 $\pm$  0.08 &  0.999 &  $8.54 \times 10^{-4}$ \\
   2$^{\ast}$ &   10.48 &    2.063 &   -1.193 &    0.08 $\pm$  0.07 &    0.08 $\pm$  0.07 &  0.999 &  $5.85 \times 10^{-4}$ \\
   3$^{\ast}$ &   10.49 &    0.791 &    0.755 &    0.15 $\pm$  0.07 &   -0.09 $\pm$  0.07 &  0.999 &  $1.06 \times 10^{-3}$ \\
   4$^{\ast}$ &   10.66 &    3.150 &   -2.899 &   -0.01 $\pm$  0.07 &    0.20 $\pm$  0.07 &  0.999 &  $8.91 \times 10^{-4}$ \\
   5$^{\ast}$ &   11.08 &   -0.633 &   -4.252 &    0.03 $\pm$  0.07 &   -0.11 $\pm$  0.08 &  0.999 &  $6.86 \times 10^{-4}$ \\
   6 &   11.16 &    4.603 &    1.092 &    0.00 $\pm$  0.08 &    0.07 $\pm$  0.08 &  1.000 &  $4.88 \times 10^{-4}$ \\
   7$^{\ast}$ &   11.22 &   -1.650 &    1.730 &    0.21 $\pm$  0.09 &    0.05 $\pm$  0.08 &  0.999 &  $1.11 \times 10^{-3}$ \\
   8$^{\ast}$ &   11.25 &    1.385 &   -2.334 &   -0.28 $\pm$  0.07 &    0.06 $\pm$  0.07 &  0.999 &  $1.34 \times 10^{-3}$ \\
   9 &   11.63 &   -1.758 &   -1.287 &   -0.00 $\pm$  0.07 &    0.05 $\pm$  0.08 &  1.000 &  $4.72 \times 10^{-4}$ \\
  10 &   11.67 &    2.038 &    0.445 &    0.05 $\pm$  0.07 &    0.03 $\pm$  0.07 &  0.999 &  $5.04 \times 10^{-4}$ \\
  11 &   11.81 &   -2.337 &   -0.540 &   -0.28 $\pm$  0.08 &   -0.20 $\pm$  0.08 &  0.997 &  $2.59 \times 10^{-3}$ \\
  12 &   11.88 &    5.528 &   -3.874 &   -0.04 $\pm$  0.08 &    0.04 $\pm$  0.11 &  1.000 &  $4.66 \times 10^{-4}$ \\
  13 &   11.89 &    0.285 &   -1.191 &    0.04 $\pm$  0.07 &    0.01 $\pm$  0.07 &  1.000 &  $4.98 \times 10^{-4}$ \\
  14 &   12.00 &   -0.158 &   -3.382 &   -0.06 $\pm$  0.07 &   -0.08 $\pm$  0.07 &  0.999 &  $5.54 \times 10^{-4}$ \\
  15 &   12.18 &    5.362 &    1.667 &   -0.02 $\pm$  0.08 &    0.09 $\pm$  0.08 &  0.999 &  $5.02 \times 10^{-4}$ \\
  16$^{\ast}$ &   12.18 &    1.012 &   -5.199 &    0.02 $\pm$  0.07 &   -0.13 $\pm$  0.08 &  0.999 &  $7.41 \times 10^{-4}$ \\
  17 &   12.19 &   -1.490 &    0.681 &    0.06 $\pm$  0.08 &   -0.07 $\pm$  0.07 &  0.999 &  $6.26 \times 10^{-4}$ \\
\enddata 
\tablenotetext{\ast}{PSF Star}
\tablenotetext{\dagger}{This Table will be published in its entirety in the electronic edition of the Astrophysical Journal, A portion is shown here for guidance regarding its form and content. Until publication, an electronic copy of this table is available from the first author.} 
\label{table_memprob}
\end{deluxetable} 
\normalsize 


\begin{table} 
\scriptsize 
\begin{center} 
\begin{tabular}{c||c|c|c|c|c} 
$K'$  & 14.0-16.0  & 15.0-17.0  & 16.0-18.0  & 17.0-19.0  & 18.0-20.0  \\ 
  &  &  &  & \\ 
\hline 
\hline 
 $N$   &   75  &  105  &  135  &  165  &  135  \\ 
   &  &  &  & \\ 
\hline 
 $\pi_{cl}$  &  0.80 $\pm$~0.036  &  0.72 $\pm$~0.035  &  0.72 $\pm$~0.030  &  0.59 $\pm$~0.029  &  0.52 $\pm$~0.032  \\ 
   &  &  &  & \\ 
\hline 
 $\Delta \mu$  &  4.28 $\pm$~0.526  &  4.54 $\pm$~0.430  &  5.15 $\pm$~0.356  &  3.68 $\pm$~0.293  &  3.51 $\pm$~0.320  \\ 
 (mas yr$^{-1}$)  &  &  &  & \\ 
\hline 
 $\phi_{f}$  &  30.9 $\pm$~4.8  &  30.5 $\pm$~3.9  &  37.1 $\pm$~3.1  &  36.6 $\pm$~2.5  &  32.5 $\pm$~2.7  \\ 
 ($^o$)  &  &  &  & \\ 
\hline 
 $\sigma_{a,f}$  &  2.21 $\pm$~0.338  &  2.56 $\pm$~0.280  &  2.87 $\pm$~0.240  &  2.89 $\pm$~0.198  &  2.80 $\pm$~0.216  \\ 
 (mas yr$^{-1}$)  &  &  &  & \\ 
\hline 
 $\sigma_{b,f}$  &  1.50 $\pm$~0.231  &  1.64 $\pm$~0.193  &  1.85 $\pm$~0.159  &  1.81 $\pm$~0.130  &  1.64 $\pm$~0.137  \\ 
 (mas yr$^{-1}$)  &  &  &  & \\ 
\hline 
 $\sigma_{a,cl}$  &  0.15 $\pm$~0.013  &  0.17 $\pm$~0.012  &  0.16 $\pm$~0.014  &  0.24 $\pm$~0.022  &  0.45 $\pm$~0.034  \\ 
 (mas yr$^{-1}$)  &  &  &  & \\ 
\hline 
 $\sigma_{b,cl}$  &  0.12 $\pm$~0.010  &  0.16 $\pm$~0.010  &  0.16 $\pm$~0.012  &  0.16 $\pm$~0.019  &  0.17 $\pm$~0.029  \\ 
 (mas yr$^{-1}$)  &  &  &  & \\ 
\hline 
 $\theta_{f}$  &  33.9 $\pm$~17.3  &  27.8 $\pm$~14.2  &  35.0 $\pm$~11.8  &  28.5 $\pm$~8.8  &  26.7 $\pm$~10.3  \\ 
 ($^o$)  &  &  &  & \\ 
\hline 
 $\theta_{cl}$  &  70.2 $\pm$~21.7  &  78.6 $\pm$~30.3  &  67.5 $\pm$~48.3  &  117.1 $\pm$~63.5  &  114.0 $\pm$~66.4  \\ 
 ($^o$)  &  &  &  & \\ 
\hline 
 $\sigma_{b,cl} / \sigma_{a,cl}$  &  0.83 $\pm$~0.086  &  0.91 $\pm$~0.078  &  0.96 $\pm$~0.069  &  0.69 $\pm$~0.075  &  0.37 $\pm$~0.085  \\ 
   &  &  &  & \\ 
\hline 
 $\sigma_{b,f} / \sigma_{a,f}$ &  0.68 $\pm$~0.130  &  0.64 $\pm$~0.113  &  0.65 $\pm$~0.099  &  0.63 $\pm$~0.083  &  0.58 $\pm$~0.090  \\ 
  &  &  &  & \\ 
\hline 
\end{tabular} 
\end{center} 
\caption{Fitted kinematic parameters of cluster and field. For each magnitude range, rows give: the cluster fraction, the separation between cluster and field centers in the Vector Point Diagram (VPD), the orientation of the separation vector from the cluster center to the field center in the VPD, the semimajor and minor axes of the field component, the semimajor and minor axes of the cluster component, the orientation of the semimajor axis of the field component, the orientation of the semimajor axis of the cluster component, and finally the axis ratio (minor/major) of the cluster and field components. Errors are estimated from Monte Carlo simulations: populations in the VPD are simulated under the intrinsic kinematic parameters estimated from observation, perturbed by the measured errors for stars in each magnitude range, and re-fitted. Orientations are position-angles reported in degrees East of North.} 
\label{tab_kinparams}
\normalsize 
\end{table} 


\begin{table} 
\begin{center} 
\begin{tabular}{l|c||c|c||c|c} 
$K'$ & $N$ & $\sigma_x$ & $\sigma_y$ & $\sigma_x$ & $\sigma_y$ \\ 
 & & (mas yr$^{-1}$) & (mas yr$^{-1}$) & (km s$^{-1})$ & (km s$^{-1}$) \\ 
\hline 
10.0-14.0 & 67 & 0.130 $\pm$~0.017 & 0.123 $\pm$~0.016  & 4.912 $\pm$~0.639 & 4.680 $\pm$~0.593 \\ 
14.0-16.0 & 72 & 0.161 $\pm$~0.019 & 0.129 $\pm$~0.016  & 6.088 $\pm$~0.739 & 4.878 $\pm$~0.606 \\ 
16.0-18.0 & 107 & 0.177 $\pm$~0.027 & 0.180 $\pm$~0.030  & 6.721 $\pm$~1.034 & 6.839 $\pm$~1.142 \\ 
18.0-20.0 & 97 & 0.224 $\pm$~0.039 & 0.148 $\pm$~0.046  & 8.508 $\pm$~1.498 & 5.629 $\pm$~1.753 \\ 
\end{tabular} 
\end{center} 
\caption{Arches velocity dispersion in each co-ordinate. Reading left-right, columns are: Magnitude range of interest, number of cluster stars in this magnitude range, intrinsic velocity dispersion and error in each coordinate, first in mas yr$^{-1}$~and then km s$^{-1}$~assuming the Arches is at 8.4~kpc.} 
\label{tab_dispersions}
\end{table} 


\begin{table}
\begin{center}
\begin{tabular}{cc|cc}
\multicolumn{2}{c|}{Offset from star 24} & \multicolumn{2}{|c}{Offset from field center} \\
$\Delta_{E-W}$ & $\Delta_{S-N}$ & $\Delta_{E-W}$ & $\Delta_{S-N}$ \\
\tableline
+1.5'' & -1.5'' & +0.23'' & +0.22'' \\
+1.5'' & -2.5'' & +0.23'' & -0.79'' \\
+2.5'' & -2.5'' & +1.23'' & -0.79'' \\
-0.5'' & -3.5'' & -1.77'' & -1.79'' \\
\tableline
\end{tabular}
\end{center}
\caption{Location of the field centers chosen to evaluate the radial and transverse components of the velocity dispersion from proper motions (Section \ref{ss_projvels}; see also Section \ref{ss_massmodel}). The left column gives positions as offsets (E-W) and (S-N) from the reference star marked in Figure 1 (see also Tables 2 \& 7). The right column gives offsets from the center of the field of view.}
\label{table_fieldcenters}
\end{table}


\begin{table} 
\begin{tabular}{r|c|c|c} 
$\Delta \chi^2_{full}$ & 3.50  & 7.82  & 13.93  \\ 
Confidence & 68\% & 95\% & 99.7\% \\ 
& "$1\sigma$" & "$2\sigma$" & "$3\sigma$" \\ 
\hline 
$M(R< 0.40~{\rm pc})$ & 0.69 - 1.10 & 0.62 - 1.20 & 0.55 - 1.30\\ 
($10^4~M_{\odot}$) &   &   &   \\ 
\hline 
$M(r < 1.0~{\rm pc})$ & 1.16 - 1.88 & 1.04 - 2.06 & 0.91 - 2.24\\ 
($10^4~M_{\odot}$) &   &   &   \\ 
\hline 
$\rho_0$ & 0.45 - 1.66 & 0.30 - 2.34 & 0.20 - 3.19\\ 
($10^5 M_{\odot}~{\rm pc^{-3}}$) &   &   &   \\ 
\hline 
$R_c$ & 0.18 - 0.31 & 0.15 - 0.44 & 0.13 - 0.80\\ 
(pc) &   &   &   \\ 
\hline 
$R_t$ & 2.00 - 30.00 & 2.00 - 30.00 & 1.00 - 30.00\\ 
(pc) &   &   &   \\ 
\hline 
$M_{cluster}$ & 1.64 - 4.29 & 1.45 - 4.86 & 1.07 - 5.62\\ 
($10^4~M_{\odot}$) &   &   &   \\ 
\hline 
$1000 \times \Sigma_{N,0} / \rho_0$ & 0.00 - 0.07 & 0.00 - 0.18 & 0.00 - 15.62 \\ 
(stars pc$^{-2} / M_{\odot}~{\rm pc}^{-3}$) &   &   &   \\ 
\hline 
\end{tabular} 
\caption{Significance regions for isotropic King modeling of the Arches cluster. Ranges of each parameter corresponding to the stated significance level are given, when $R_c, R_t, M_{cluster}$~are all allowed to vary. The quantity $\chi^2_{full}$~denotes the badness-of-fit when comparing model predictions to both the Arches kinematic dataset and the surface density dataset of Espinoza et al. (2009), over the mass range ($10 \le M \le 30$)~$M_{\odot}$.} 
\label{table_massest_10-30_full}
\end{table} 


\begin{sidewaystable}
\scriptsize
\begin{tabular}{lcccccccccl}
Ref& $R_{in}$& $R_{out}$& $R_{ext}$&  $\Gamma$& $M_{low}$& $M_{obs}$& $M_{calc}$& $\delta (M_{obs})$& $\delta(M_{calc})$& Notes \\
     &      (pc) &    (pc)     &  (pc)    &           & $(M_{\odot})$ &  $(\times 10^4 M_{\odot})$ &  $(\times 10^4 M_{\odot})$ &  $(\times 10^4 M_{\odot})$ & & \\
\tableline
\tableline
1 & -   &  1.15 & -   & -     & -     & 0.08  &  0.24 &   -  &  -    &  Lower limit on total mass \\
2 & -   &  0.35 & -   &  1.35 &  2.0  & 0.50  &  1.5  &  0.1 &  -    &  PDMF Salpeter; \MProjRad \\
2 & -   &  0.35 & -   &  1.35 &  0.1  & 0.50  &  6.0  &  0.1 &  -    &  PDMF Salpeter; \MProjRad \\
3 & 0.12 & 0.35 & -   &  0.6  &	  1.0 & 0.51  &  1.08 &    - &  -    &  PDMF Top-heavy; \MProjRad \\
3 & 0.12 & 0.35 & -   &  0.6  &	  0.1 & 0.51  &  1.20 &    - &  -    &  PDMF Top-heavy; \MProjRad \\
4 & -    & 0.40 & -   &  -    &   2.0 & 0.63  &  -    &	   - &  -    &  Rough limit on \MProjRad~reported \\
5 & -	 & 0.23 & -   &  -    &  -    & 7.0   &  -    &    - &  -    &  Upper limit on \MProjRad~from radial velocities \\
6 & - 	 & 0.40 & -   &  1.1  &	  1.0 & 0.557 & 2.0 &    - &  0.6 &  low-mass truncation; \MProjRad \\
6 & -	 & 0.40 & -   &  1.35 &  0.08 & 0.557 & 3.1  &    - &  0.6 &  Kroupa PDMF; no low-mass truncation; \MProjRad \\
\tableline
7 & -   & 2.50 & -    & 0.5   &	 1.0  & -     &  1.60 &    - & 	-    & N-body; IMF top-heavy; $M_0$~reported \\
7 & -   & 2.50 & -    & 0.75  & 1.0   & -     & 2.00  &    - &	-    &	  ``        `` \\ 
7 & -  	& 2.50 & -    & 1.0   &	 1.0  & -     & 2.80  &    - &	-    &	  ``        `` \\
8 & -	& 1.26 & 3.0  & 2.8   & -     & -     & 4.00  &    - &	-    & N-body; Multi-component IMF \\
9 & -	& 0.35 & 2.5  & 0.9   &	 1.3  & -     & 4.00  &    - &	-    & N-body; Representative $M_0$~reported in Figure caption \\
10 & -	& 0.35 & 2.5  & 0.9   &	 0.1  & -     & 14.50 &    - &	-    & Turbulent-fragmentation calculation \\
11 & -	& 0.35 & 2.4  & 1.1  & 0.9    &	-     & 5.90  &    - &	-    & N-body; Observations reported in Kim et al. (2006) \\
12  & -	& 0.40 & -    & 1.35 & 0.5    & -     & 1.80  &    - &	-    & N-body, Salpeter IMF;  Present-day \MProjObs \\
12  & -	& 0.40 & 1.0  & 1.35 & 0.5    & -     & 3.60  &    - &	-    & Present-day simulated mass within projected radius $R = 1.0$~pc \\
12  & -	& 0.40 & 2.8  & 1.35 & 0.5    & -     & 4.90  &    - &	0.8  & Total initial cluster mass $M_0$~with low-mass truncation \\ 
12  & -	& 0.40 & 2.8  & 1.35 & 1.0    & -     & 3.60  &    - &	0.6  & Total initial cluster mass $M_0$~with moderate-mass truncation \\
12  & -	& 0.40 & 2.8  & 1.35 & 4.0    & -     & 1.90  &    - &	0.3  & Total initial cluster mass $M_0$~with moderate-mass truncation \\
\tableline
\tableline
\end{tabular}
\caption{All Arches literature mass estimates of which the authors are aware, current as of September 2011. Observational estimates are listed first, followed by cluster mass estimates from models. Reading left-right, columns give: 1: Reference cited. 2,3: Inner and outer radii within which stars were observed $R_{in}, R_{out}$. 4: Radius to which mass function has been extrapolated $R_{ext}$. 5: Mass function slope $\Gamma$. 6, lower stellar mass used for IMF. 7:Total mass of stars directly observed, $M_{obs}$. 8: Extrapolated mass $M_{calc}$~where appropriate. 9,10: errors (where given) in the observed and extrapolated masses; 11. Brief description. References are: 1. \citet{cotera96}; 2. \citet{serabyn98}; 3. \citet{figer99}; 4. \citet{stolte02}; 5. \citet{figer02}; 6. \citet{espinoza09}. 7. \citet{kim00}; 8. \citet{pz02}; 9. \citet{kim06}; 10. \citet{dib07}; 11. \citet{chatterjee09}; 12. \citet{harfst10}}
\label{table_all_mass_estimates}
\end{sidewaystable}


\acknowledgements

Support for this work was provided by NSF grants AST 04-06816 and AST
09-09218, and the NSF Science and Technology Center for Adaptive
Optics, managed by the University of California, Santa Cruz (AST
98-76783), and the Levine- Leichtman Family Foundation. AS is
supported by a DFG Emmy Noether grant under ID STO 469/3-1. The
W. M. Keck Observatory is operated as a scientific partnership among
the California Institute of Technology, the University of California
and the National Aeronautics and Space Administration. The Observatory
was made possible by the generous financial support of the W. M. Keck
Foundation. The authors wish to recognize and acknowledge the very
significant cultural role and reverence that the summit of Mauna Kea
has always had within the indigenous Hawaiian community. We are most
fortunate to have the opportunity to conduct observations from this
mountain. The authors thank Sungsoo Kim for helpful discussion
  and for bringing a very useful paper to our attention. We also thank
  the anonymous referee for insightful comments which clarified the
  presentation of some of the points in this paper.


\appendix

\section{Production of motions from star-lists}

Here we provide details of the procedures used to produce proper
motions from the star-lists. The steps are:

{\it 1. Choice of initial reference stars, and initial mapping onto
  \tzero:} An initial list was constructed of eleven bright stars that
were well-measured in all epochs and cover the full area of the
detector, and used for the initial registration of each epoch onto
\tzero = 2008.5. This epoch was chosen because its starlist is of high
quality (Table \ref{tab_obsns}) and the epoch itself will be near the
pivot point of the straight line fits to the positional time-series
for most of the stars. A 6-term linear transformation was used for
this mapping for each epoch, accounting for positional shift, global
scaling, rotation, a difference in scale factors in X and Y (``on-axis
skew'') and differences in the angle between axes (``off-axis skew'').

{\it 2. Matching of all stars within \tzero:} Using this initial
mapping, all stars were matched to their counterparts in \tzero (where
present) by proximity in \tzero and magnitude. Matching radius 5 pix
(approximately the PSF core FWHM) and a broad magnitude tolerance 3
mag were used. This yields positional differences between predicted
and observed positions (hereafter ``deltas'') in \tzero~for matched
pairs using the first-guess transformation.

\begin{figure}
\begin{center}
\includegraphics[height=80mm]{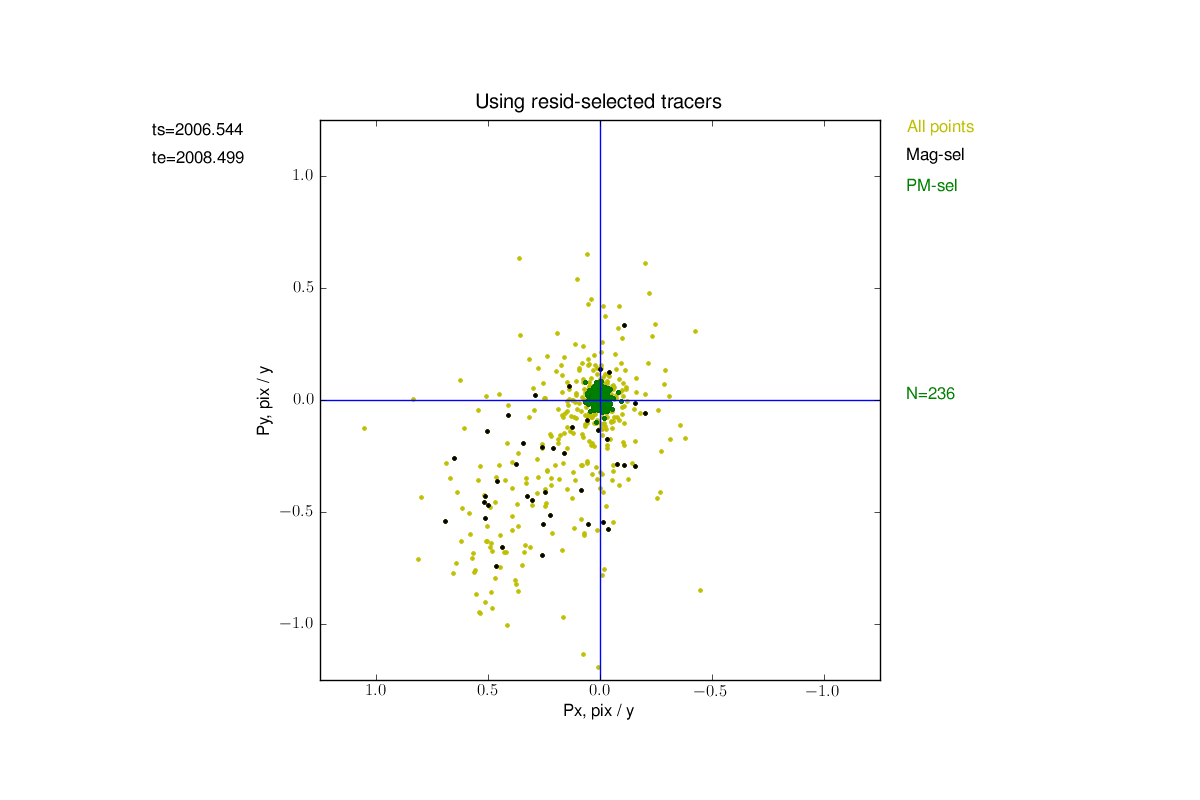}
\caption{Culling of reference stars during the frame mapping process, in this case mapping epoch 2006.54 onto the frame of epoch \tzero=2008.50. {\it Yellow points:} all matches. {\it Black points:} selected for magnitude ($K' \le 17.5$). {\it Green points:} reference stars selected by position residual from the center of mass of the magnitude-selected sample and with outliers clipped during the mapping (in this plot the motions have been shifted to the center of mass of the selected objects).}
\label{fig_cull_refstars}
\end{center}
\end{figure}

{\it 3. Fitting of reference-frame mapping for matching:} The previous
step typically produces deltas for $\sim 300$~stars at
$K'<17.5$~across each pair of epochs ($t$-\tzero). The field
population displays significant motion dispersion in a preferential
direction close to the galactic plane. Field objects must therefore be
removed from the sample of reference stars to avoid biasing the
offsets and magnification factors when mapping the reference
frames. From the positional deltas of reference stars, the center of
the distribution in the vector point diagram is estimated and the
standard deviation of motions in each direction from this center of
mass estimated. Objects farther than 2$\sigma$~from this estimated
center of mass are removed. This process is repeated twice to produce
a cleaned list of reference stars; typically 260 objects survive this
process. These objects are used to re-map the epochs onto \tzero using
a full 6-term linear transformation. Clipping of outliers in this
epoch mapping typically removes a further 30 stars in each epoch
(Figure \ref{fig_cull_refstars}).

{\it 4: Trim coincident close pairs of stars: } At this stage we have
the master-list of measurements of each object, in the frame in which
the object was originally measured. To mitigate confusion by known
objects as much as possible, all coincident pairs with separation $<
75$~mas are removed from consideration for each epoch. This typically
removes 20-40 objects from the position-lists at each epoch (column
$N_{\ast, uncrowd}$~in Table 1). The result is a matched catalogue of
1114 objects present in at least two epoochs.

{\it 5. Reference frame-mapping for motions:} Armed with the matched
list of objects and their measurements at each epoch, likely-cluster
members (at $K<17.5$) are used as reference stars to map each epoch
onto \tzero, using the same weighting and clipping as step 3. We find
(Appendix \ref{ap_frameorder}) that a second-order transformation in X and Y
is sufficient to capture most of the residual higher-order effective
distortions between epochs without falling prey to overfitting of few
stars with a high-order transformation. 

{\it 6. Motion extraction in \tzero:} A first pass at stellar motions
is estimated by fitting a linear trend to the positional time-series
$x(t), y(t)$~of each star in the reference frame \tzero. For each
star, the weighted mean time $\tmeanm = \sum_{i} t_i w_i / \sum_i
w_i$~is evaluated so that the fit becomes $x(t) = a +
b(t-\tmeanm)$. Weights $w_i = 1/\sigma^2_i$~are the inverse of the
variance of each measurement due to positional uncertainty. This
removes correlation between errors in the parameters (e.g. Press et
al. 1992); the center of mass of the data is first determined then the
slope pivoted about this point to find the best-fit positional
gradient. The proper motion error is the formal error on the best-fit
slope: $\sigma^2_b = 1/\left( \sum_{i=1}^N
\frac{(t-\tmeanm)^2}{\sigma_i^2} \right)$. To mitigate sensitivity to
short-term excursions in position, for objects measured in $\ge
4$~epochs, two passes of sigma-clipping at 3$\sigma$~are applied. Note
that \tmean is a property of $\sigma_i(t)$~and thus is estimated
separately for each star and for each co-ordinate. Motions are
estimated for all 805 stars with $\ge 3$~epochs of measurement.

\begin{figure}
\begin{center}
\includegraphics[height=80mm]{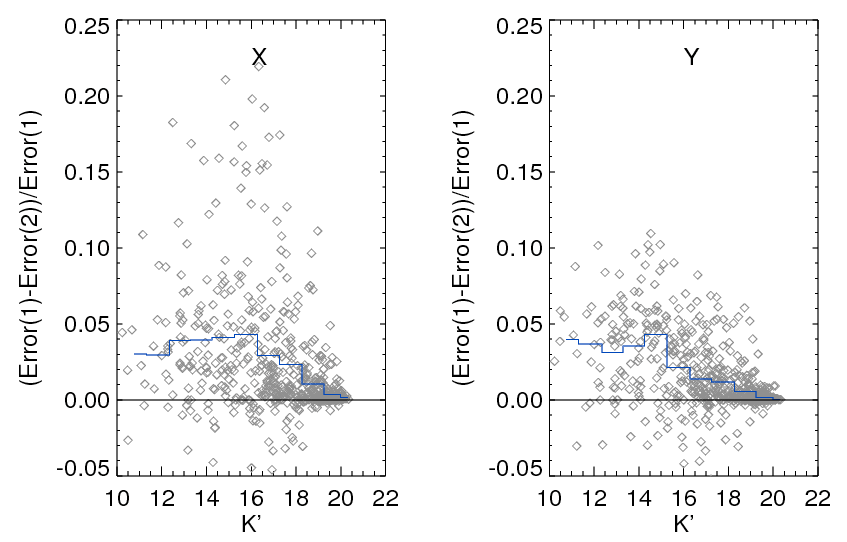}
\caption{Reduction in velocity error when frames are mapped onto a
  reference list constructed from a first pass at proper motion
  fitting (2), over frame mapping onto a single starlist at epoch
  \tzero (1). The change in error is expressed as the ratio of the
  difference between (1) and (2) to the original error (1). The blue
  line reports the median improvement within each magnitude
  bin. Reading left-right,panels indicate errors along X and Y.}
\label{f_motion_improvement}
\end{center}
\end{figure}

{\it 7. Refinement of the reference position-list:} When choosing a
reference frame in which to evaluate motions, our goal is a
reference-list onto which cluster members can be mapped with as little
scatter as possible due to measurement and fitting error. We generate
a reference frame by evaluating at some time \treference the
straight-line fits to the positional time-series of cluster reference
stars. Positional errors in this predicted frame (``predictive
errors'') are evaluated by propagating the errors on the fit
coefficients $a,b$~for each star. By choosing \treference to be near
the pivot point \tmean of the greatest number of reference stars, we
aim to minimize the error of the predicted positions in the
constructed reference frame. The distribution of \tmean is nearly
Gaussian with \tmean = 2008.0 $\pm$~0.4 (1$\sigma$); we therefore
adopt \treference = 2008.0 to evaluate the reference epoch. To
evaluate the degree to which this mean reference frame improves the
mapping, motions and their errors for each star were evaluated using
the quad sum (centroiding + alignment) errors when mapped to \tzero and
(centroiding + alignment + predictive) when mapped to
\treference. Motion errors are improved by up to 20\% for some bright
objects, with median improvement up to $4\%$~for well-measured objects
(Figure \ref{f_motion_improvement}).

{\it 8. Re-mapping and re-extraction of motions:} Finally, the
star-list from each epoch is mapped onto the constructed
reference-frame \treference and motions in this frame evaluated in the
manner of Step 7. Table \ref{tab_transform} gives the fitted
parameters and the number of reference stars used in the mapping from
each epoch to the \treference frame. Provided the motions of cluster
stars in the field of view do not themselves describe a second-order
or lower transformation (e.g. rotation or contraction of the cluster)
to within our ability to determine, then the parameters taking
reference frame 2008.50 to \tmean should be consistent with zero, as
is observed (Table \ref{tab_transform}). We find that, when
applied to stars near the edges of the detector, the size of the
positional shifts due to the quadratic terms in the mapping are 1-few
times the centroiding error for bright ($K' < 16$)~objects (Table
\ref{tab_transform}).

{\it 9. Evaluation and incorporation of additional error sources:}
The distribution of fits to the velocities thus produced were examined
for additional sources of random error. It became readily apparent
that a significant source of error along the time-series was not taken
into account by the steps above. When characterised (Section
\ref{ss_additive}), steps 6-9 were repeated with this error term
included.

\section{Additional errors beyond centroiding and alignment errors}

\subsection{Estimating the magnitude of additive random errors}\label{ss_additivedet}

When frames are mapped together and motions estimated using the
quadrature-sum of the centroiding and additive errors, the resulting
distribution of $\chi^2_{\nu}$~from the velocity-fits is significantly
different from the canonical $\chi^2$~distribution, indicating that
the random errors characterized in Sections \ref{ss_centroiding} \&
\ref{ss_align} is not sufficient to account for the random variation
about the best-fit actually observed. We detail here the estimate of
the ``additive'' error $\addx, \addy$~that must be added in
quadrature to rectify this situation.

\begin{figure}
\begin{center}
\includegraphics[width=15cm]{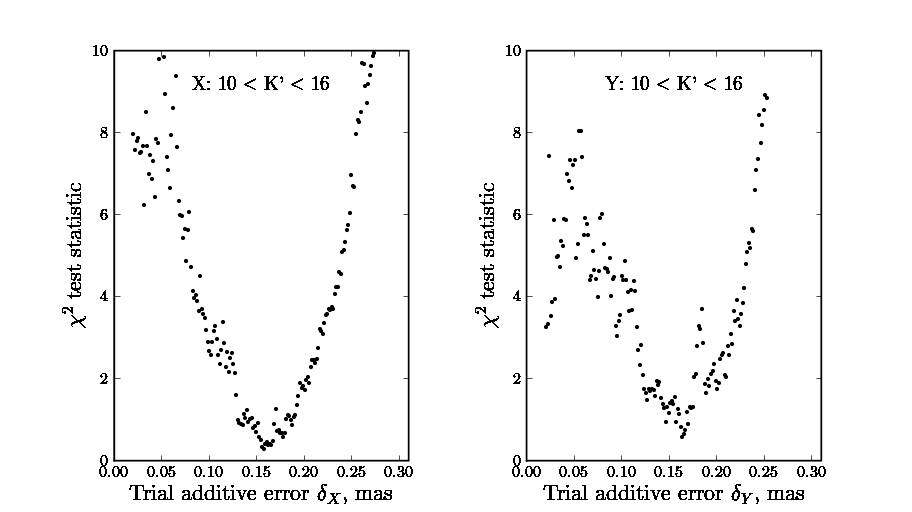}
\caption{Determination of additive errors $\addx$~(left) and $\addy$~(right), for objects in the brightness range ($10.0 \le K' \le 16$.). For each trial additive error, the distribution of chi-squared values from the velocity-fits to each star is compared to that expected under statistical error, using the $\chi^2$~test. The statistic of this comparison is plotted here.}
\label{f_test_adderrs}
\end{center}
\end{figure}

Once stars are aligned into the \treference reference frame,
velocities are re-fit for trial values of $\addx, \addy$, and the
resulting distribution of $\chi^2$~values from the velocity fits are
compared to expectations. Two tests were evaluated to make the
comparison. First, the chi-squared test was evaluated for the
difference between the $\chi^2$~histogram and the theoretical
expectation at each trial additive error. Because of the binning
required, this statistic does not vary smoothly with the trial error;
to estimate the minimum, a second-order polynomial was fit to the
trough in the fit statistic. This yields estimates for the best-fit
additive error in each co-ordinate (denoted $\addxbest,
\addybest$). Second, the two-sided Kolmogorov-Smirnov test was used as
a fit statistic to obviate the need for binning. While the minima
returned by the two measures are broadly consistent with each other,
we adopt the chi-squared test since it appears to provide a more
sensitive determination of the best-fit additive error (Figure
\ref{f_test_adderrs}). Errors on this determination of
$\addxbest$~\&~$\addybest$~are estimated by simulation; sets of
positional time-series are constructed under gaussian noise with
amplitudes as in the real data and perturbed by additional spatially
uniform error $\delta_{x,in}$,$\delta_{y,in}$~(while keeping the error
used to re-determine the additive errors as the quadrature-sum of
alignment and centroiding errors). The rms of ($\delta_{x,in} -
\addxbest$), ($\delta_{y,in} - \addybest$) are then adopted as the
error in the additive errors. Only stars with 5 measurements are used
to estimate $\addx$~\&~$\addy$.

\begin{figure}
\begin{center}
\includegraphics[width=8cm]{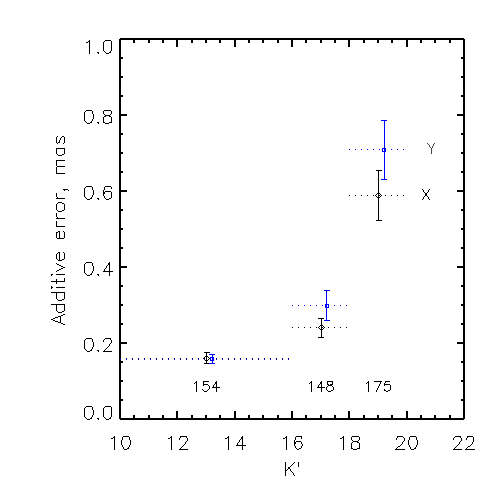}
\caption{Additive error as a function of magnitude. Black diamonds: X; blue squares: Y, offset along the horizontal axis slightly for clarity. The sample size in each magnitude bin are indicated within the frame. Horizontal dashed lines indicate the magnitude-ranges in each bin.}
\label{f_aderr_mag}
\end{center}
\end{figure}

As the balance of dominant error terms evolves with magnitude
\citep[e.g.,][]{fritz10}, we might expect $\overline{\adde}$~to also
vary with magnitude. We therefore break the sample into three
non-overlapping magnitude bins such that the number of stars with 5
good measurements is approximately uniform across the bins. The
additive error and its uncertainty is then determined for each
magnitude bin following the above prescription in the previous
paragraph (Figure \ref{f_aderr_mag}).

\begin{figure}
\begin{center}
\includegraphics[width=15cm]{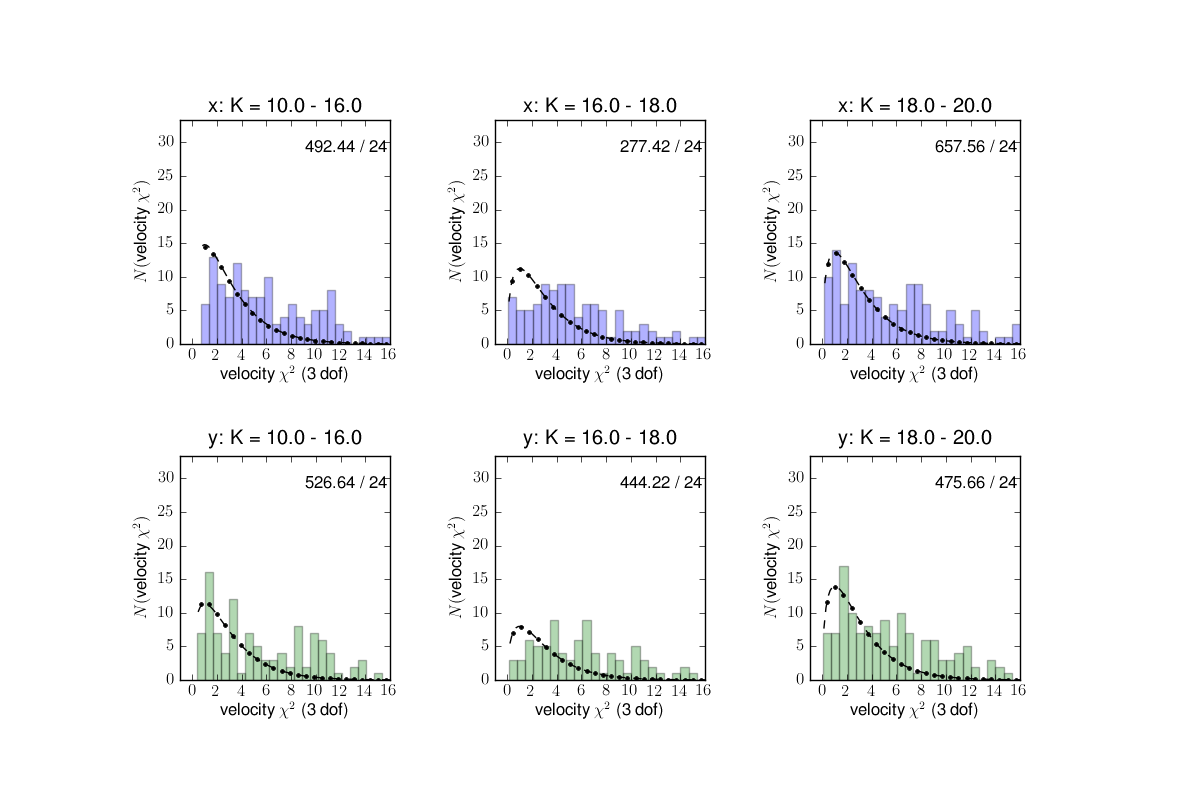}
\caption{Assessment of the distribution of $\chi^2$~from the velocity fits when motions are fit using only centroiding and alignment errors. Columns break the sample of stars into bright, medium and faint magnitiude bins. Histograms show the distribution of velocity-fit $\chi^2$~values in X (blue; top-row) and Y (green;bottom row). The numbers inset give the values of the chi-square test statistic per degree of freedom for the comparison of the observed histogram to the predicted distribution (dashed curve).}
\label{f_aderr_test1}
\end{center}
\end{figure}

\begin{figure}
\includegraphics[width=15cm]{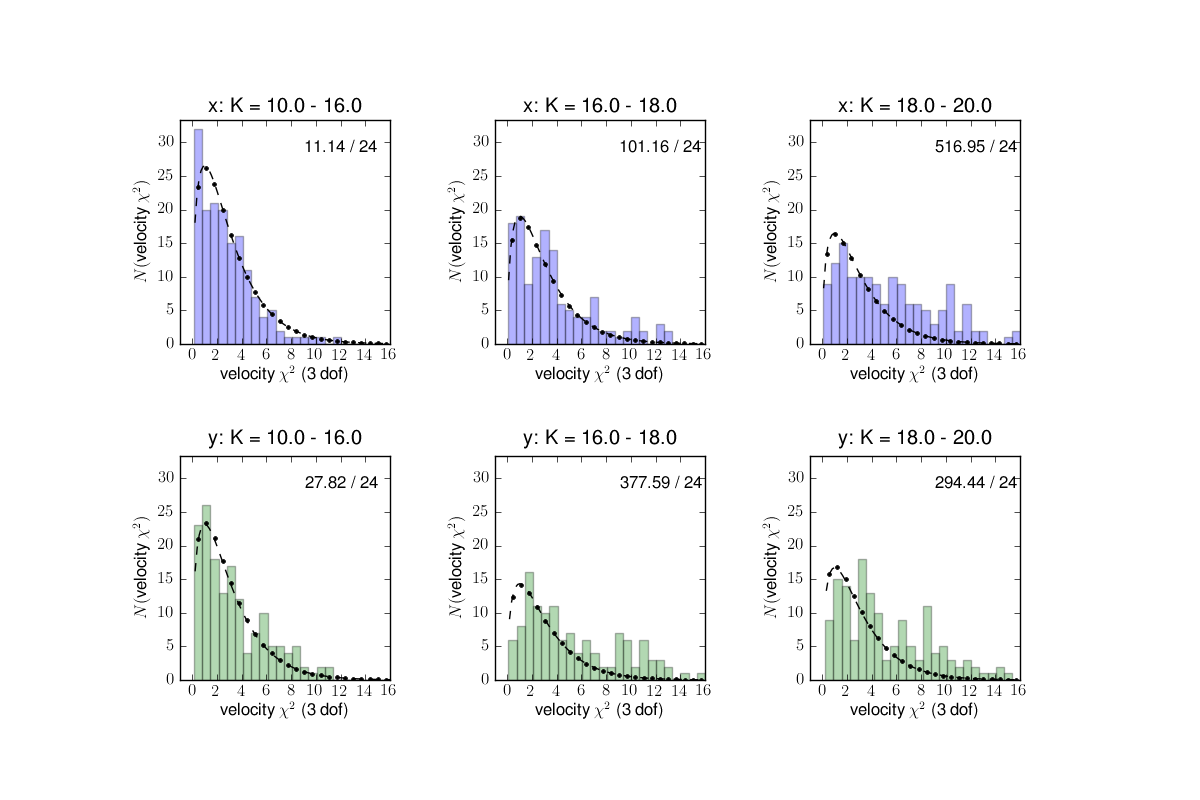}
\includegraphics[width=15cm]{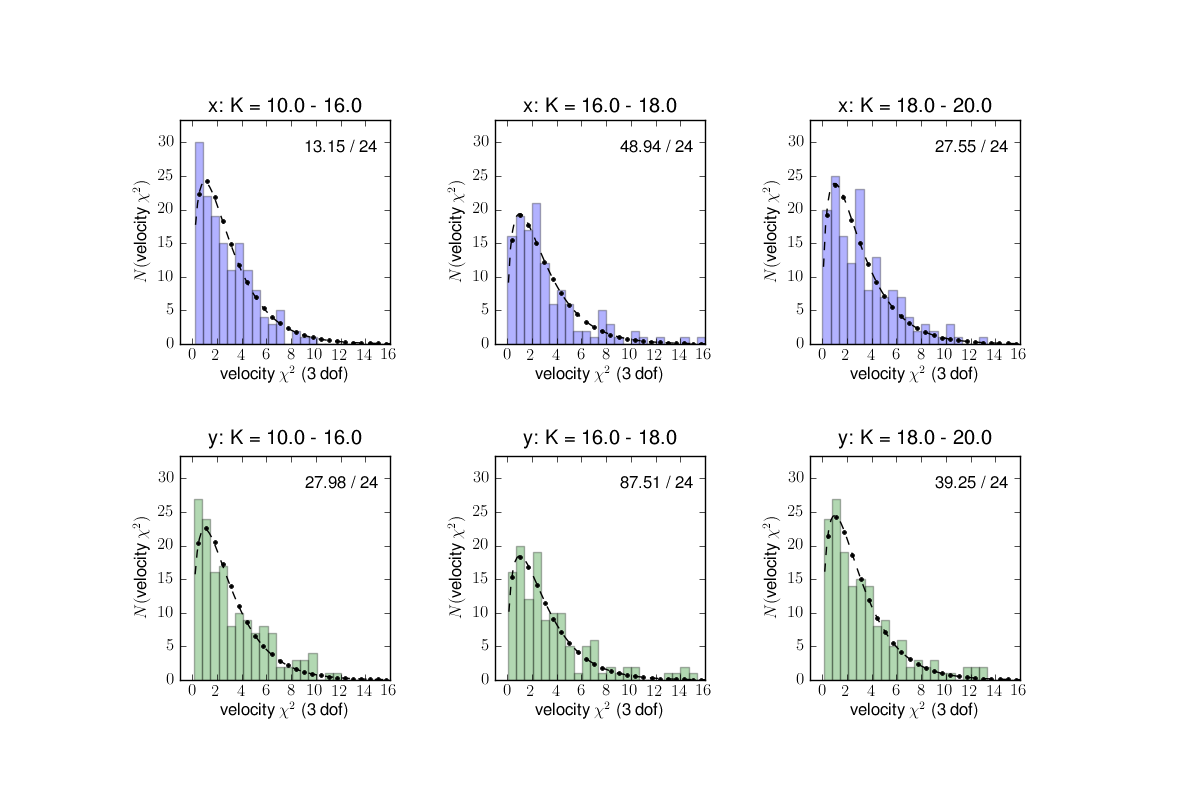}
\caption{As Figure \ref{f_aderr_test1}, for constant additive error (Top) and additive error computed from a fit to separate determinations for each magnitude-range (Figure \ref{f_aderr_mag})}
\label{f_aderr_test2}
\end{figure}

Because they describe the mean additional statistical scatter required
between epochs, the additive errors $\addxbest, \addybest$~are applied
to the position lists at the stage of frame mapping. The distribution
of $\chi^2$~values from the velocity fits after re-mapping and
re-fitting including additive error was evaluated for three cases: (1)
no additive error; (2) a flat additive error (as determined from the
$10 < K' < 16$~sample), and (3) additive error allowed to vary with
magnitude. We find that a flat distribution of additive error with
magnitude produces a velocity $\chi^2$~distribution significantly more
discrepant from statistical expectation than a magnitude-dependent
additive error (Figures \ref{f_aderr_test1}~\&~\ref{f_aderr_test2}). We
therefore adopt the magnitude-dependent additive error estimate.

\subsection{Confusion bias}\label{ap_confusion}

When a sufficiently bright star passes within $\sim 1$~FWHM of the PSF
of a star of interest, the shape of its PSF can be sufficiently
altered that its position measurement is biased, but not so altered
that the measurement is rejected. In some cases this bias can be much
larger than the positional measurement error \citep[$\gtrsim 2$~mas
  for $\Delta K' < 2$;][]{ghez08}. The distribution of this confusion
bias across the sample of stars depends on the spatial crowding and
magnitude distribution of stars in the field of view. To estimate its
order of magnitude for the Arches central field, we use the
simulations of Fritz et al. (2010), which model the distribution of
astrometric bias as a function of magnitude, for a $K'$~distribution
appropriate for the nuclear star cluster near the Galactic center. The
rms of the confusion bias (denoted here as $\sigma_{bias}$)~follows a
power law whose normalization depends on the stellar density within
the field. Of their three regions of interest, the stellar density
within the Arches field matches most closely that of their
3.5$''$~sample. This then predicts positional bias $\sigma_{bias}$~of
order 20\% of the additive errors $\adde$~for $K' < 18$~and comparable
to $\adde$~at $K' > 18$~(Table \ref{tab_budget}).

Relative motion across the PSF of the two components of a confused
pair would imprint a spurious motion due to the resulting
time-variation of the confusion bias. Inter-epoch variation in the PSF
would thus cause varying positional bias between epochs even in the
case of components that are perfectly stationary with respect to each
other. Under the expectation that PSF variation between epochs is
random, this error is subsumed within the additive random error
(Section \ref{ss_additive}).

Linear trends in the relative separation of the star and its
unrecognized confusing counterpart are in principle more problematic,
as the spurious motion thus induced would be impossible to separate
from the desired intrinsic motion. Indeed, for some of the
rapidly-moving S-stars near the Galactic center, apparent deviations
from the orbital path on a timescale of up to a few years are clearly
visible as the star of interest crosses the region of influence of the
confusing source entirely during the timebase of the observations
\citep{ghez08, gillessen09}. Measurements confined to the
time-interval of confusion would therefore detect linear motion in the
wrong direction entirely. However, for the Arches stars of interest
here, relative motions of members of a confused pair are too slow to
have a significant impact on the motions we measure. We assume that
the bias changes by $1\times \sigma_{bias}$~in the time taken for the
relative separation of confused components to change by the FWHM of
the PSF. With expected velocity dispersion $\sim 0.2$~mas
yr$^{-1}$~(Stolte et al. 2008), confused pairs of cluster objects
change their separation by $\la 3\%$~of the FWHM per year, so that the
astrometric bias will essentially be static for confused
cluster-pairs. Cluster objects confused with field stars may be
subject to relative motions $\sim 5$~mas yr$^{-1}$; in this case the
proper motion bias may approach $\sim 0.1\sigma_{bias}$~yr$^{-1}$. We
conclude that, for our measurements of motions in the Arches central
field, proper motion bias due to confusion trends is a very small
effect compared to other sources of error (Table \ref{tab_budget}) and
can safely be ignored in our analysis.

\section{Transformation order during frame-mapping}\label{ap_frameorder}

\begin{figure}
\begin{center}
\includegraphics[width=17cm]{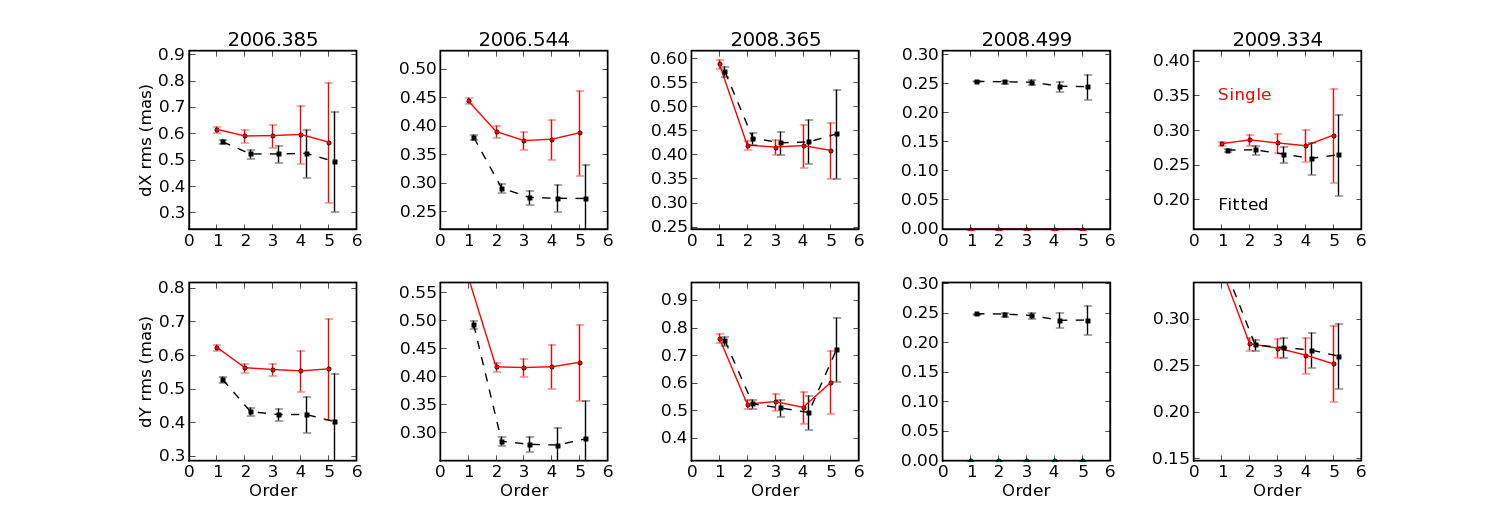}
\caption{Positional residuals after mapping of each epoch onto the
  adopted reference frame, as a function of the polynomial
  transformation order adopted. Red solid line: reference frame used
  is the star-list in 2008.5. Black dashed line: reference frame
  constructed by using the linear fits to the positional time-series
  for likely cluster members. Errorbars were estimated by monte carlo
  simulation using random half-samples. Residuals are evaluated along
  detector X (top row) and detector Y (bottom row)}
\label{f_order}
\end{center}
\end{figure}

At the level of a few percent of a pixel (comparable to the velocity
dispersion we wish to measure), variations in distortion may be
present between epochs. These variations might consist of both a
spatially random and a spatially correlated part, and might consist of
temporally random and/or correlated parts. To quantify spatially
correlated time-variation, mappings between reference frames were
re-fit separately across each (t-\tzero) pair using polynomials of
order $0 \le M \le 5$~using the same set of likely cluster members at
each order (Appendix A; step 6). The rms of cluster members in X and Y
as transformed to \tzero was evaluated for each order for each epoch
(Figure \ref{f_order}), with errors on the rms evaluated from monte
carlo resampling and re-fitting in a similar manner to the estimation
of alignment errors (Section \ref{ss_align}). Visual inspection
suggests that for each epoch, a significant improvement is gained by
using a second-order polynomial; order 3 is sometimes indicated along
$Y$, and 4th or higher orders rarely bring about significant
improvement.

\begin{figure}
\centerline{\hbox{
\includegraphics[width=8cm]{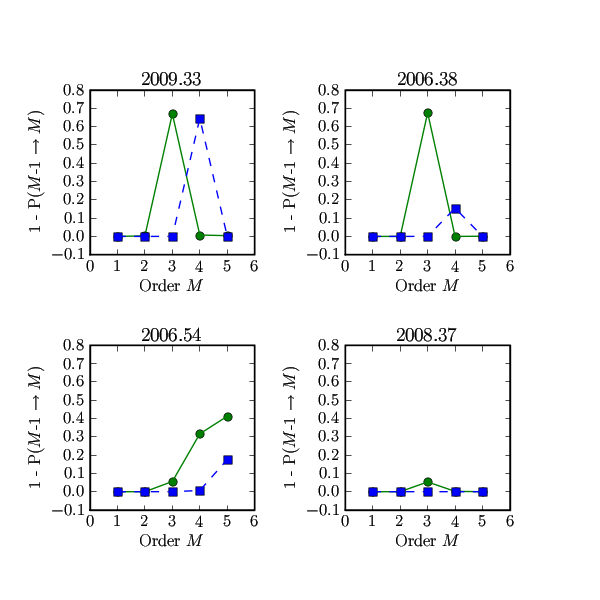}
\includegraphics[width=8cm]{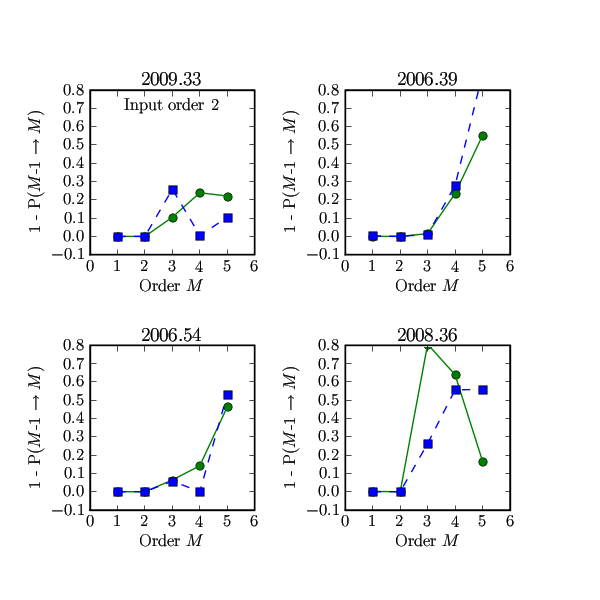}
}}
\caption{Formal significance of the fit improvement when
  transformations of increasing order are used to map starlists
  between epochs. For each step up in order $M$-$1$~to $M$, the
  false-alarm probability is shown that corresponds to random chance
  producing a decrease in badness-of-fit at least as great as that
  observed (Appendix \ref{ap_frameorder}). This statistic is evaluated
  separately for residuals in X (green circles, solid line) and Y
  (blue squares, dashed line). {\it Left $2\times2$~panels:} measured
  positions. {\it Right $2\times2$~panels:} the same test applied to a
  control experiment where the observed positions are perturbed under
  the cluster and field motion distributions, and a second-order
  polynomial of similar magnitude to that fit from the real data is
  added to simulate epoch-to-epoch distortion variations. A polynomial
  of order 2 produces a formally significant improvement in the
  fitting.}
\label{f_ftest}
\end{figure}

The formal significance of the improvement of the fit when stepping up
from order $M$-$1$~to $M$~was estimated by evaluating the ratio
$(\chi_{\nu}^2(M$-$1)$-$\chi^2_\nu(M))/\chi^2_{\nu}(M)$~for order $1 <
M < 5$; assuming the residuals after mapping are indeed
$\chi^2$~distributed, this ratio should follow the $F$-distribution
for the corresponding pairs of degrees of freedom for $M-1$~and
$M$~\citep[e.g., Chapter 11 of][]{bevington03}. This produces a formal
false-alarm probability that a difference in badness-of-fit of
$\chi_{\nu}^2(M$-$1)$-$\chi^2_\nu(M)$~or greater could arise from
random chance alone. This suggests that order $M > 3$~is not warranted
for fits to either coordinate (Figure \ref{f_ftest}; Left). The
apparent improvement in fit significance at order $M=5$~is probably an
artefact of overfitting to the $\sim 235$~reference objects
($M=5$~corresponds to only $\sim 10$~points per term in the
polynomial).

A control test was conducted where stars at the observed positions
were moved randomly under the expected velocity distribution of the
cluster or (for 15\% of objects) the field, perturbed by measurement
error, and subjected to a 2nd order polynomial of similar amplitude to
the parameters fitted to the real stellar positions. This indicated
that the formal fit statistic is indeed sensitive to the polynomial
order, provided the number of reference stars is sufficient. A 5th
order polynomial (21 terms, or $\sim$~11 stars per term) is often
spuriously indicated (Figure \ref{f_ftest}; Right). We therefore adopt
a second-order polynomial for the frame mapping when extracting
motions. In principle, relative distortions between epochs might
require a more complicated description, but this cannot be determined
from the sample at hand.

\section{Misclassification bias in kinematic fitting}\label{ss_misclass}

Section \ref{ss_fit} details the steps taken to estimate membership
probabilities by simultaneously fitting cluster and field kinematic
properties (denoted \mukm, \sigkm for each component)~and membership
fraction (\pikm). Because of the inter-relation between \pikm~and
(\mukm, \pikm), any biases in fitting the kinematic components
translate into biases in the membership probabilities, and vice
versa. To mitigate the effect of magnitude-dependent error on \pikm,
as well as allow for any intrinsic changes in $\zk$~with magnitude,
the maximum likelihood fitting was carried out in a
magnitude-dependent way as described in Section \ref{ss_fit}.

\begin{figure}
\begin{center}
\includegraphics[width=16cm]{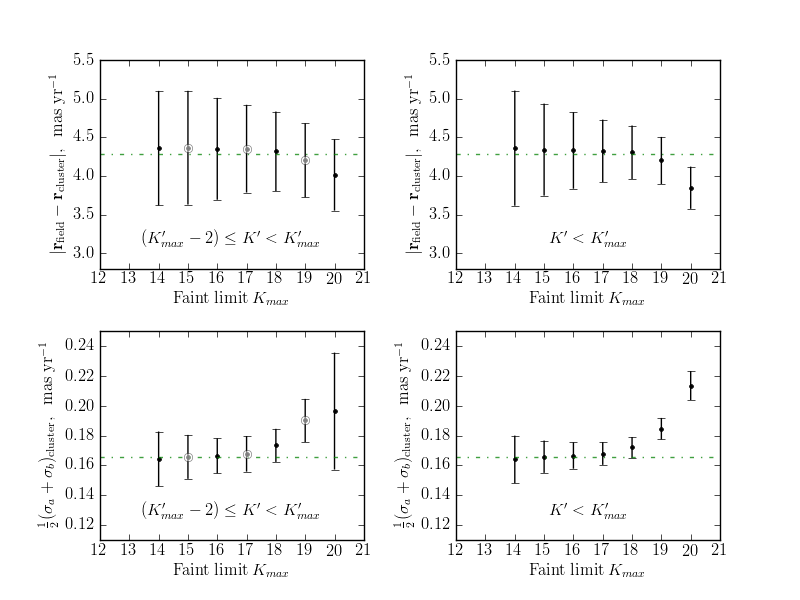}
\caption{Dependence of fitted kinematic parameters on star
  brightness. Magnitudes and errors were sampled from the observed
  magnitude and error distribution (bottom-right panel of Figure
  \ref{fig_additive_det}). Two measures are shown: the separation
  between components in the VPD (top row) and the average of the
  cluster major and minor axes (bottom row). The same underlying
  kinematic parameters were used for all simluations in this figure
  (green broken lines). The left column shows fits evaluated over
  two-magnitude-wide magnitude-bins (with non-overlapping bins
  indicated using the symbols), the right column shows fits evaluated
  over all stars brighter than $K'$~in each bin. If all stars are fit
  together to estimate kinematic parameters for the cluster as a whole
  (corresponding to the faintest bin in the right column) then the
  fitted parameters are biased.}
\label{f_VPD_bias}
\end{center}
\end{figure}

To investigate the size of any misclassification biases, synthetic
datasets were simulated using the same parameters ($\muk, \sigk,
\pik$) at all magnitudes and perturbed by proper motion errors sampled
from the magnitude-error curves observed (Figure 2). The fitting
process in Section \ref{ss_fit}~was performed for a large number of
trials and the recovered parameters observed as a function of
magnitude. For this set of tests, errors on the recovered parameters
are the standard deviation of the parameters recovered over the
trials. For each trial dataset, the fitting process was carried out
using tracer stars selected according to two schemes: (1). using all
stars regardless of brightness; and (2). the magnitude-local scheme
described in Section \ref{ss_membprob}. Figure \ref{f_VPD_bias}~shows
the result. Both techniques show some degree of bias at fainter
magnitudes (higher errors), though as the sample is usually dominated
by the faintest stars in each sample, the biases are comparable.  

\section{Dynamical mass estimate}

Many proper motion datasets (including that reported here) cover only
the inner region of the cluster, over which the velocity anisotropy
varies too slowly with radius to be well-constrained by the proper
motions. In this case the popular moment-based estimator of Leonard \&
Merritt (1989 Equation 19) should not be used (as pointed out by
LM89). A full non-parametric modeling of the dataset (see Schoedel,
Merritt \& Eckart 2009) is not appropriate without data spanning a
wider radius range than we have at present. Instead we use a
prescription for the mass density $\rho(r)$~and evaluate the predicted
velocity dispersion profile for comparison to our dispersion data.

\subsection{Model and Method}

For this first examination we assume the cluster can be adequately
parameterised by an isotropic, spherical King (1962) model. In this
limit, the model is completely described by three parameters; the core
radius $R_c$, the tidal radius $R_t$~and the total cluster mass
$M_{cl}$. Model parameters were varied over a grid of values, with the velocity
dispersion profile projected onto the sky and compared to our dataset
in each case. The variation of $\chi^2$~with parameter values was then
used to estimate confidence limits on the model parameters, as
discussed in the main text. 

The kinematic dataset covers the innermost region of the cluster,
within $\sim 1-2 \times R_c$. For this reason the shape parameters of
the King profile are poorly constrained by the kinematic data
alone. We therefore evaluate $\chi^2$~in two ways and present
confidence limits derived from both. Firstly, ranges are estimated
comparing model predictions to kinematic data alone. Secondly, ranges
are estimated by comparing predictions to the kinematic dataset and
also to the surface density by number $\Sigma_N(R)$~for massive
stars. \citet{espinoza09}~report \SurfdensN~profiles for three mass
ranges: ($10 \le M < 30$)~$M_{\odot}$, ($30 \le M \le
120$~$M_{\odot}$) and the union of the two, ($10 \le M \le
120$~$M_{\odot}$). We examined both the full mass range and the ($10
\le M \le 30$~$M_{\odot}$)~mass range when comparing \SurfdensN~to
data, to gain insight into the dependence of the derived mass on the
shape parameters of the cluster. Two grids were used to explore
parameter-space, one coarse: \\ $\bullet$ ($0.05 \le R_c \le 0.8$)~pc
in 40 steps \\ $\bullet$ ($1.0 \le R_t \le 50.0$)~pc in 40 steps
\\ $\bullet$ ($0.5 \le \MCl \le 10.0$)~$\times 10^4 M_{\odot}$~in 40
steps \\
\noindent the other somewhat more finely-spaced near the apparent $\chi^2$~minimum: \\
$\bullet$ ($0.05 \le R_c \le 0.8$)~pc in 50 steps \\
$\bullet$ ($1.0 \le R_t \le 30.0$)~pc in 50 steps \\
$\bullet$ ($0.5 \le \MCl \le 6.0$)~$\times 10^4 M_{\odot}$~in 50 steps \\

Each $R_c, R_t, M_{Cl}$~combination predicts a pattern of surface
density by mass \SurfdensM~(units $M_{\odot}~{\rm pc}^{-2}$), which is
optimally scaled to the surface density by number \SurfdensN~(units
stars pc$^{-2}$).\footnote{As this scale factor is optimized to fit
  the data for each trial-set of the other three parameters, the
  appropriate $\Delta \chi^2$~regions for significance ranges are
  unchanged from the kinematic-only comparison; three parameters are
  allowed to vary.}. This scale factor \Surfdensfac~relates to the
mass function of the cluster, and so we include it in the reported
quantities derived from the model parameters.

With a few exceptions, observational mass estimates report the total
mass in stars in a cylinder of radius $R=0.35-0.4$~pc on-sky (Section
5.4). Hereafter we refer to this quantity as the ``projected mass''
$M(<R)$~to distinguish it from the total mass enclosed within a sphere
of radius $r$, i.e., $M(<r)$. We report confidence limits on
\MProjObs~to provide a direct comparison with the observational
literature. We also report limits on \MKineObs, as well as the total
cluster mass \MCl~(when all three model parameters are allowed to
vary). For interest we also report the central volume density
$\rho_0$~in the table of confidence regions. 

The form for \MProjObs~can be analytically derived from the form for
$\Sigma_{mass}(R)$. It is reported in King (1962); for convenience we
repeat the form here:
\begin{eqnarray}
  M(< R) & = & \pi R_c^2 K \left[ \ln(1 + x) - 4\frac{\sqrt{1+x}-1}{\sqrt{1+x_t}} + \frac{x}{1+x_t}  \right]; \nonumber \\
  x \equiv (R/R_c)^2 \nonumber \\
  x_t \equiv (R_t/R_c)^2 
\end{eqnarray}
\noindent while \MKineObs~is estimated from
\begin{equation}
  M(<r)  =  4 \pi \int^{r = 1{\rm pc}}_0 r^2 \rho(r) dr
\end{equation}
\noindent and $\rho_0$~is evaluated by setting $r=0$~in
(\ref{eq_kingmodel}).

\subsection{Results}

\begin{figure}
  \includegraphics[width=8cm]{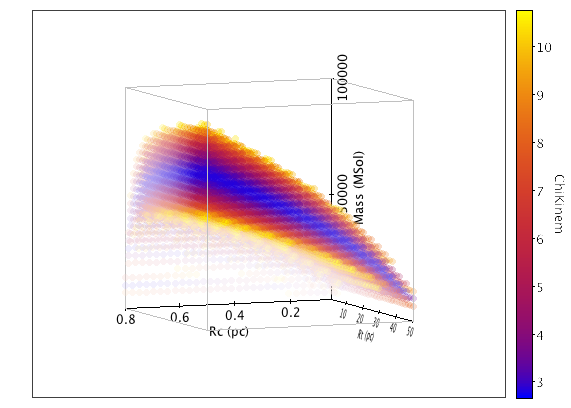}
  \includegraphics[width=8cm]{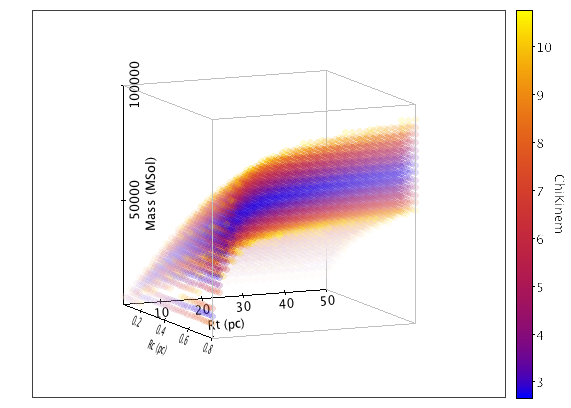}
  \caption{Views of the $\Delta \chi^2_{kinem} < 7.82$~region when only the kinematic dataset is used to assess the mass model. Axes are: $R_c, R_t, M_{cl}$, with total cluster mass $M_{cl}$~vertical in each case. Limits shown are: $0.05 \le R_c \le 0.8$~pc; $1.0 \le R_t \le 50$~pc; $0.5 \le M_{cl} \le 10.0 \times 10^4 M_{\odot}$.}
\label{fig_chibubble_mtot_kinem}
\end{figure}

\begin{figure}
  \includegraphics[width=8cm]{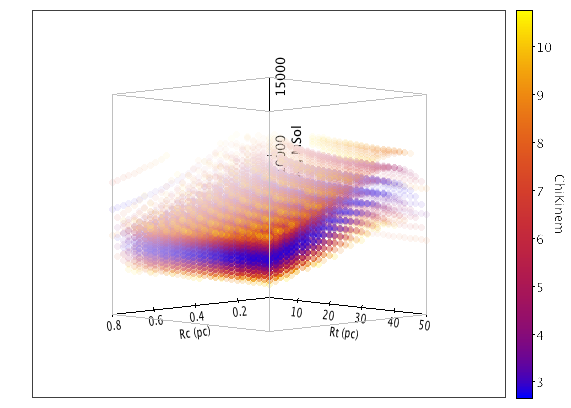}
  \includegraphics[width=8cm]{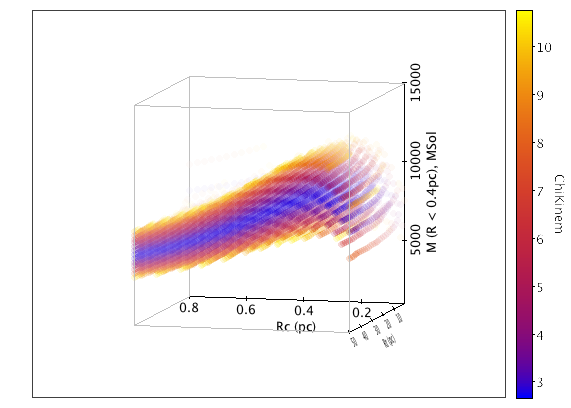}
  \caption{As Figure \ref{fig_chibubble_mtot_kinem}, but with \MProjObs~along the vertical axis. Limits shown are: $0.05 \le R_c \le 0.8$~pc; $1.0 \le R_t \le 50$~pc; $0.5 \le \MProjObs \le 1.5 \times 10^4 M_{\odot}$.}
\label{fig_chibubble_mproj_kinem}
\end{figure}

\begin{figure}
  \includegraphics[width=16cm]{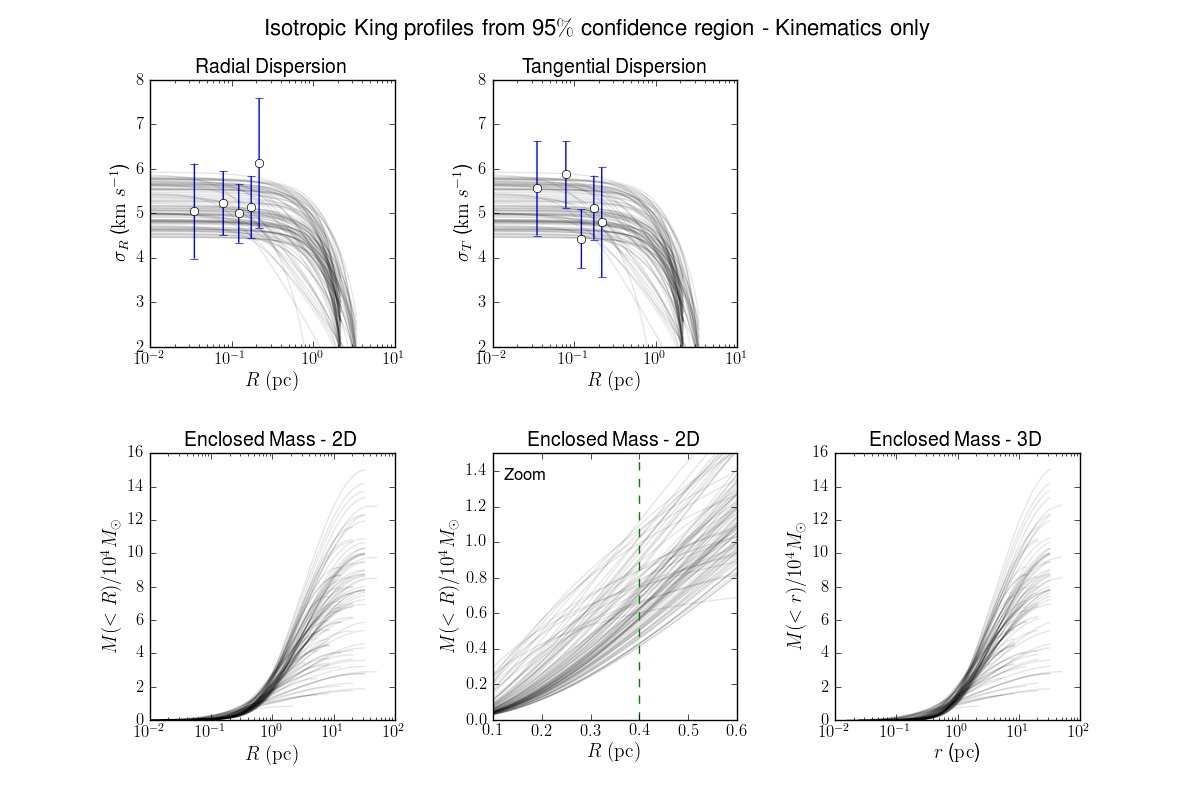}
  \caption{Radial profiles corresponding to parameter-sets within the $\Delta \chi^2_{kinem} < 7.82$~surface. Top-left and top-middle panels show radial and tangential velocity dispersions from proper motions (points) over the projected profiles corresponding to model parameters (lines). Bottom-left and bottom-middle panels show the total mass within cylindrical radius $R$~on the sky, with $R=0.4$~pc indicated by the vertical dashed line. Bottom-right panel shows the mass enclosed within a sphere of radius $r$~pc from the cluster center. See also Table \ref{table_massest_kinem}.}
\label{fig_radprofs_coarse_kinem}
\end{figure}

Figures
\ref{fig_chibubble_mtot_kinem}~\&~\ref{fig_chibubble_mproj_kinem} show
the behaviour of the 95\% significance region in $R_c, R_t,
M_{cl}$~and $R_c, R_t, \MProjObs$~space respectively, for the coarse
grid of parameter values and {\it without}~using any constraints on
the surface density \SurfdensN. Our maximum tidal radius $R_t = 50$~pc
already would suggest a very extended cluster, and it is likely that
increasing $R_t$~still further would decrease \MProjObs. Although the
total mass and shape parameters of the King (1962) model are poorly
constrained by kinematic data alone, the requirement that the cluster
arrange itself in order to yield the low velocity dispersion that we
measure, imposes an upper limit to \MProjObs~of about 1.30$\times
10^4~M_{\odot}$. With this dataset only, the total cluster mass is
essentially unconstrained (Table \ref{table_massest_kinem}). Figure
\ref{fig_radprofs_coarse_kinem}~shows radial profiles drawn from
within the $\Delta \chi^2_{kinem} = 7.82$~surface in parameter space.

The inter-dependence of the model parameters when representing the
observed dispersions can be intuitively understood in the following
way: under a King model, for constant tidal radius $R_t$, a more
centrally-concentrated cluster (i.e., smaller $R_c$) produces a larger
projected velocity dispersion for the same cluster mass
$M_{cl}$. Thus, to match the observed dispersion plateau we measure,
without \SurfdensN~information the total cluster mass $M_{cl}$~should
decrease as $R_c$~decreases at constant $\chi^2_{kinem}$, as is
observed (Figure \ref{fig_chibubble_mtot_kinem}). However when
evaluating \MProjObs, we see that as $R_c$~is increased, a smaller
fraction of the total cluster mass is observed within a cylinder of
radius $R=0.4$~pc on the sky, while $M_{cl}$~increases as
$R_c$~increases. The trade-off between these two trends leads to the
shape of the $\chi^2_{kinem}$~surface for \MProjObs.

\begin{figure}
  \includegraphics[width=8cm]{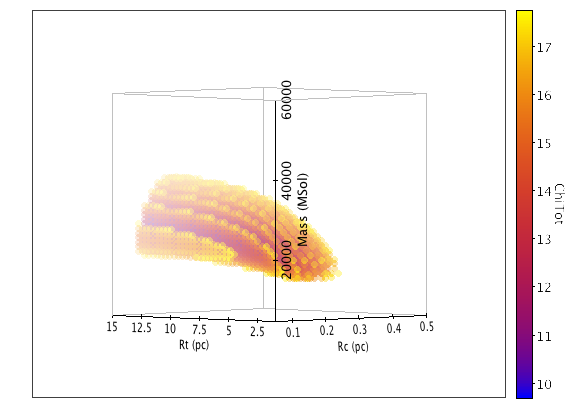}
  \includegraphics[width=8cm]{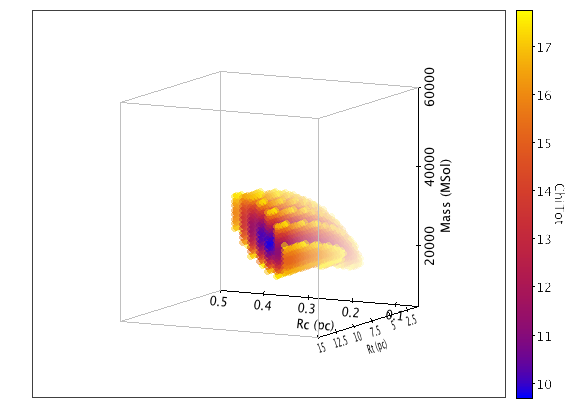}
  \caption{Views of the $\Delta \chi^2_{full} < 7.82$~region when both
    kinematic and surface density data (using the full mass range of
    Espinoza et al. 2009) are included in the assessment. Axes are:
    $R_c, R_t, M_{cl}$, with total cluster mass $M_{cl}$~vertical in
    each case. Limits shown are: $0.05 \le R_c \le 0.5$~pc; $1.0 \le R_t \le 25$~pc; $0.5 \le M_{cl} \le 6.0 \times 10^4 M_{\odot}$.}
\label{fig_chibubble_mtot_full}
\end{figure}

\begin{figure}
  \includegraphics[width=8cm]{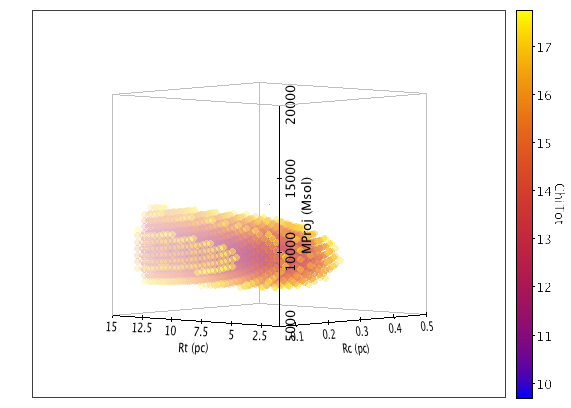}
  \includegraphics[width=8cm]{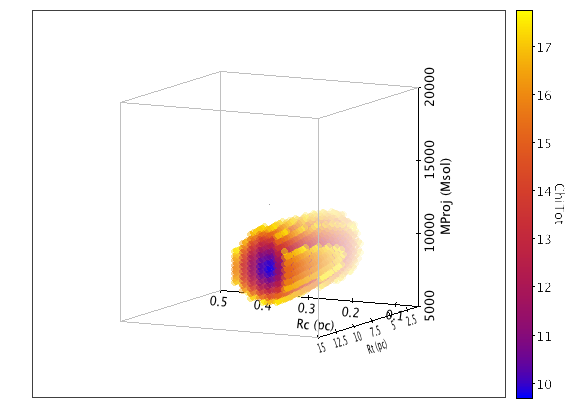}
  \caption{As Figure \ref{fig_chibubble_mtot_full}, but with \MProjObs~along the vertical axis. Limits shown are: $0.05 \le R_c \le 0.5$~pc; $1.0 \le R_t \le 25$~pc; $0.5 \le \MProjObs \le 2.0 \times 10^4 M_{\odot}$.}
\label{fig_chibubble_mproj_full}
\end{figure}

\begin{figure}
  \includegraphics[width=16cm]{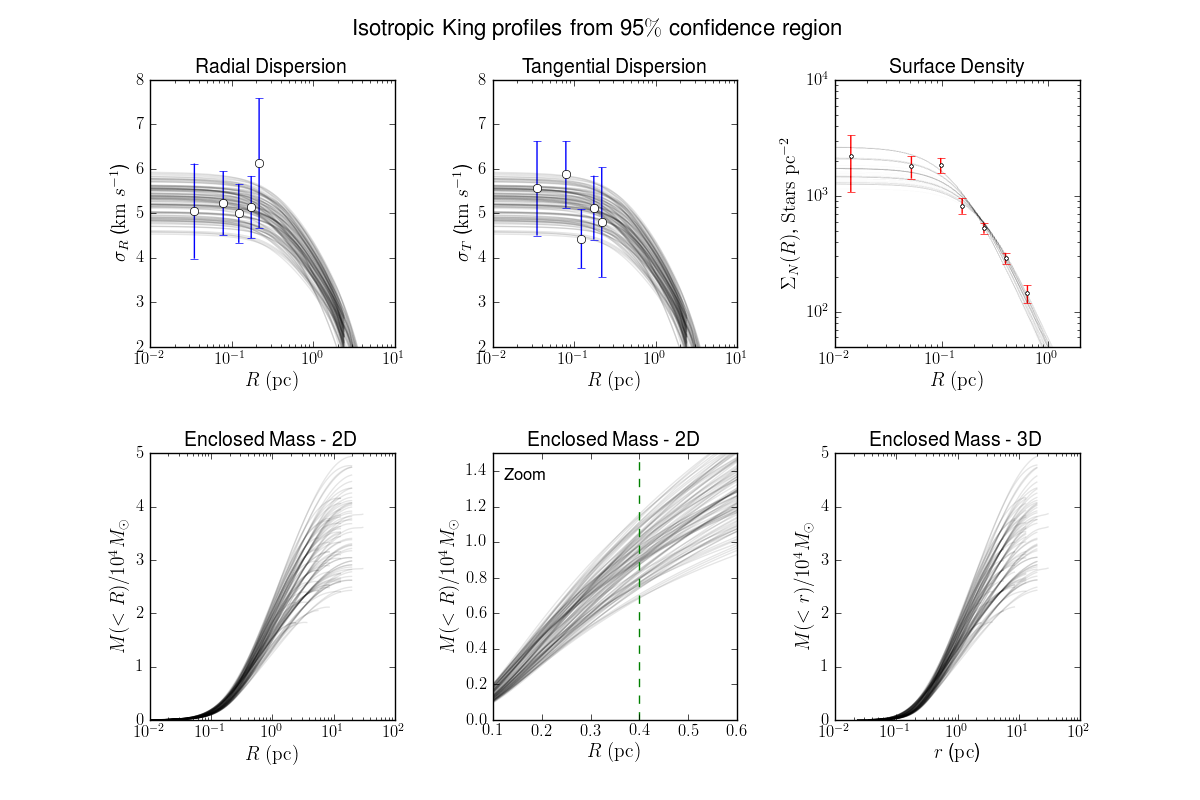}
  \caption{As Figure \ref{fig_radprofs_finer_10-30}, but this time \SurfdensN~corresponds to stars in the range ($10 \le M \le 120$)~$M_{\odot}$. See Table \ref{table_massest_full}.}
\label{fig_radprofs_finer_full}
\end{figure}

Inclusion of the \SurfdensN~data produces tighter constraints on the
derived parameters because the cluster shape is now more tightly
constrained (Tables
\ref{table_massest_10-30_full}~\&~\ref{table_massest_full} and Figures
\ref{fig_chibubble_mtot_full} \& \ref{fig_chibubble_mproj_full};
Figure \ref{fig_radprofs_finer_full}~shows radial profiles in this
case). In this latter case, the upper limit on \MProjObs~is still
about 1.30$\times 10^4~M_{\odot}$, but now the lower limit is
increased and the total cluster mass is itself constrained, to
\besttot~at 99.7\% confidence. 


\subsection{Beyond the Isotropic King model}


The isotropic King (1962) profile assumes that the cluster has
achieved dynamical relaxation through multiple collisions between its
constituent stars. While the Arches cluster is young compared to
  a typical crossing time of most of its stars, the cluster is likely
  to be significantly older {\it dynamically}~than this consideration
  would suggest, since it has likely already undergone gravitational
  collapse \citep{allison09}. Thus, we selected a King (1962) model as
  a reasonable first-order approximation.

Full constraints await realistic N-body simulations using our
dispersion measurements as a constraint. In the meantime, we attempted
to follow the practical algorithm of LM89 for a (possibly strongly)
anisotropic cluster. In their approach, a velocity dispersion profile
$\sigma^2_{iso}(r)$ is computed under the assumption that the cluster
is isotropic and follows some density model $n(r)$~that is fit to the
subset of stars directly observed. The isotropic dispersion profile
$\sigma^2_{iso}$~is then modified following some prescription for
anisotropic velocity dispersions to produce radial and tangential
dispersions $\sigma^2_r, \sigma^2_t$. The enclosed mass $M(<r)$~and
underlying mass density $\rho(r)$~are then derived from the Jeans
equation for dispersions $\sigma^2_r, \sigma^2_t$~of tracer stars
whose number density distribution follows $n(r)$. Finally, the
quantities of interest (particularly \MProjObs)~can be calculated from
the model, for example by projecting $\rho(r)$~onto the sky and
integrating over projected radius $R$.

\begin{figure}
  \includegraphics[width=10cm]{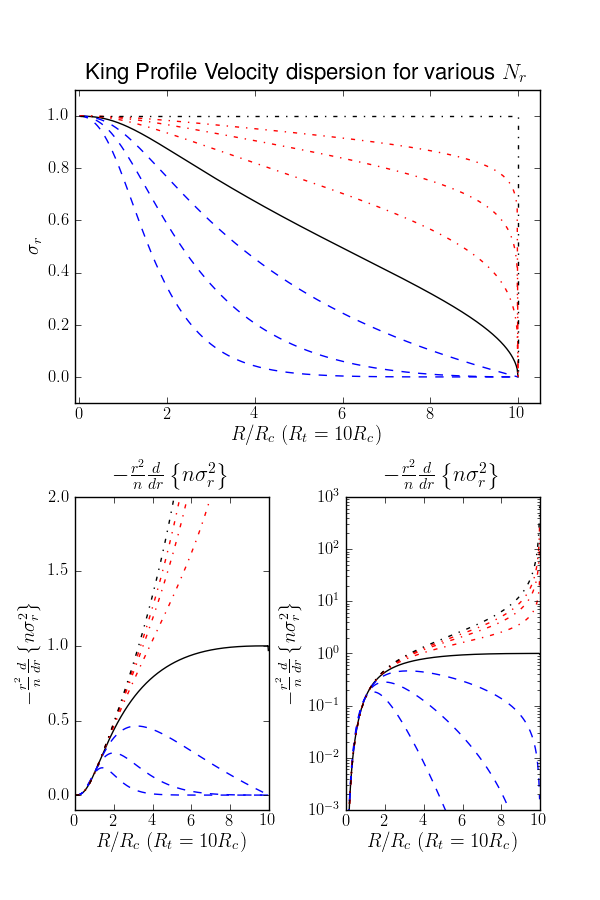}
  \caption{Velocity dispersion profiles following Leonard \& Merritt
    (1989; LM89). Top panel shows the radial velocity dispersion profile under the same range of velocity anisotropy parameters $N_r$~used in Figure 1 of LM89. Dashed lines show (left-right): $N_r = 8, 4, 2$. Dot-dashed lines show (left-right): $N_r = 0.5, 0.25, 0.125, 0.0$. Bottom panels show the behavior of the first term in the Jeans equation under the same values of $N_r$, showing divergence for $N_r < 1$. }
\label{fig_vdisp_LM89}
\end{figure}

LM89 use an isotropic King (1962) profile to generate $n(r)$~and
$\sigma^2_{iso}$. To modify the dispersion for an anisotropic cluster,
they raise $\sigma_{iso}$~to some power $N_r, N_t$~for radial and
tangential dispersions respectively. This allows for an extreme range
of anisotropies (provided only $N_r, N_t$~combinations that yield
positive $\rho(r)$~everywhere are used). When we attempted this
procedure, we found that the first term in the Jeans equation diverges
strongly for significant regions of parameter-space, although we
reproduce the shapes of the velocity dispersion profiles of LM89
exactly (Figure \ref{fig_vdisp_LM89}).

\begin{figure}
  \includegraphics[width=16cm]{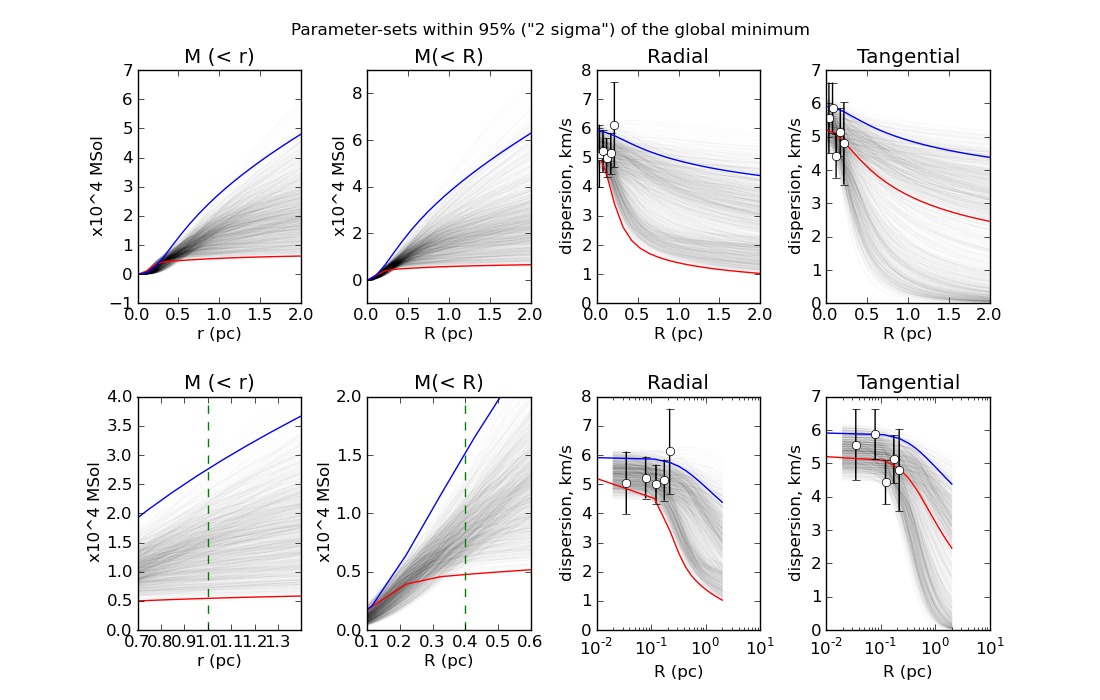}
  \caption{Radial profiles drawn from the 95\% confidence region for Plummer model parameters. Reading left-right: mass within a sphere of radius $r$~pc; projected mass within cylindrical radius $R$~pc; projected radial and tangential velocity dispersion components. Dashed green lines in the lower left two panels indicate 1.0 and 0.4 parsecs. The solid blue and red curves indicate the upper and lower bounds for \MProjObs.}
\label{fig_radprofs_plummer92}
\end{figure}

We also attempted the approach of \citet{leonard92}, in which the
general method of LM89 is modified by using a Plummer profile to
describe $n(r)$~and~$\sigma^2_{iso}(r)$. In this case we were able to
reproduce the \citet{leonard92}~dispersion curves exactly for
anisotropic models without the divergence problems we encountered
estimating $M(<r)$~following LM89. Under a Plummer model the Arches
cluster is more centrally concentrated than the relaxed \citet{king62}
profile. Figure \ref{fig_radprofs_plummer92}~shows radial profiles
drawn from the $\Delta \chi^2_{full} = 7.82$~regions that
result. Because the radial profile of the Plummer model provides a
poorer fit to \SurfdensN~than the King (1962) profile, a wide range of
radial profiles are consistent with the data. In this case the
best-fitting values for \MProjObs~are slightly broader than for the
King (1962) profile, but still below 1.5$\times 10^4~M_{\odot}$~for
all combinations within the formal-$2\sigma$~confidence
region.

Further work along these lines would use a more realistic prescription
for the anisotropy. \citet{leonard92}~never justify their approach to
modify $\sigma^2_{iso}$~for anisotropy; we note here that it allows
for a very large range of velocity dispersion anisotropies yet still
produces \MProjObs~$\la 1.7 \times10^4~M_{\odot}$. Since the King (1962)
profile provides a better fit the \SurfdensN~dataset than anisotropic
Plummer profiles - and is used to model very young clusters regardless
of their evolutionary status (e.g. \citealt{harfst10}) - we report
mass ranges corresponding to isotropic King profiles in this
communication. We expect that more realistic N-body simulations will
yield constraints not bound to spherical models.

\subsection{Functional forms in the Leonard et al. (1992) approach}

\citet{leonard92}~choose not to provide the functional forms of
several of the steps in their analysis. To aid the reader who might be
interested to try this approach, we show the relevant relationships
here. \citet{leonard92}~assume that the measured stars are distributed with
an isotropic spatial distribution $n(r)$, but the same stars move in a
way that reflects the true underlying mass distribution and thus may
show anisotropic motion. This motion is parameterised as a modified
form of the velocity dispersion that would be obtained if the motions
were isotropic. The flow from model to derived quantities is: \\

1. Start with a prescription for the number density of tracers
$n(r)$~for an isotropic distribution, and derive the velocity
dispersion profile that would be obtained if motions were
isotropic. For the Plummer model of \citet{leonard92} this gives
\begin{eqnarray}
  u & \equiv & 1 + ( r / r_0 )^2 \nonumber \\
  n(r) & = & n_0 u^{-5/2} \nonumber \\
  \sigma_r(r) & = & \sigma_0 u^{-N_r/4} \nonumber \\
  \sigma_t(r) & = & \sigma_0 u^{-N_t/4} \nonumber \\
  \Sigma(R) & = & \Sigma_0 u^{-2} 
\end{eqnarray}
\label{eq_obsmodel_params}
\noindent where $N_r, Nt$~are positive indices that are used to modify
the dispersion profile for velocity anisotropy.

Note that while $\Sigma_0 = \frac{4}{3}r_0 \rho_0$~depends on $(r_0,
Nr, Nt, \sigma^2_0)$, this refers to the surface density by mass, not
by the number of stars observed per pc$^{2}$. The relationship between
the surface density (mass) and surface density (number of tracers per
pc$^2$) depends on the mass function and the depth of observations,
and for the purpose of the modeling is treated as a free parameter to
be optimized out when evaluating the fit of the model to surface
density data.

2. Use the Jeans equation to evaluate the enclosed mass distribution
corresponding to the anisotropic velocity dispersions just derived,
but assume the mass density of the measured stars (as opposed to the
underlying mass distribution) is distributed isotropically as $n(r)$:
\begin{equation}
  G \Mrad = - \frac{r^2}{\ntracer} \frac{d}{dr}\left\{ \ntracer \sigr \right\} - 2r(\sigr - \sigt)
\label{eq_jeans_full}
\end{equation}
\noindent which for the Plummer model
(\ref{eq_obsmodel_params})~becomes
\begin{eqnarray}
  M(< r) & = & \frac{r \sigma_0^2}{G} \left[ \left( \frac{r}{r_0} \right)^2 (5 + N_r)u^{-(Nr + 2)/2} - 2u^{N_r/2} + 2u^{-Nt/2}\right] \nonumber \\
  \left(\frac{\Mrad}{M_{\odot}}\right) & = & 231.3 \times  \left(\frac{\sigma_0}{\rm km~s^{-1}}\right)^2 \left(  \frac{r}{\rm pc}\right) ... \nonumber \\
  & ... & \times \left[ \left( \frac{r}{r_0} \right)^2 (5 + N_r)u^{-(Nr + 2)/2} - 2u^{N_r/2} + 2u^{-Nt/2}\right]
\label{eq_mass_enclosed}
\end{eqnarray}

3. Use the enclosed mass to estimate the volume density profile
$\rho(r)$~of the underlying mass distribution:
\begin{equation}
  \rho = \frac{1}{4 \pi r^2} \frac{d}{dr} \left\{ \Mrad \right\}
\end{equation}
\noindent which for the Plummer model is
\begin{eqnarray}
  \rho(r) & = & \frac{\sigma_0^2}{4\pi G r_0^2} \left\{ u^{-(N_r + 2)/2} 
    \left( 2N_r + (5 + N_r)\left[ 3 - \left( \frac{r}{r_0} \right)^2 \left( \frac{N_r + 2}{u}\right)\right]\right) ... \right. \nonumber \\ 
    & & \left. ~~~~ ... - 2N_t u^{-(N_t + 2)/2}
    + 2\left(\frac{r_0}{r}\right)^2 \left( u^{-N_t/2} - u^{-N_r / 2} \right) \right\} \nonumber \\
    \left(\frac{\rho}{M_{\odot} {\rm pc}^{-3}}\right) & = & 18.4 \left( \frac{\sigma_0}{\rm km ~s^{-1}} \right)^2 \left( \frac{r_0}{pc}\right)^{-2} \times \left\{ ... \frac{}{}\right\}
\end{eqnarray}
\noindent with the lower form giving $\rho(r)$~in units
$M_{\odot}$~pc$^{-3}$~with $r_0$~in pc and $\sigma_0$~in km s$^{-1}$.
As $r \rightarrow 0$~then $\{\} \rightarrow 15 + 5N_r - 2N_t$, so the central volume density becomes:
\begin{equation}
  \rho_0 = \frac{\sigma_0^2}{4\pi G r_0^2} \times \left\{ 15 + 5N_r - 2N_t\right\}
\end{equation}

4. Evaluate the mass observed within projected radius $R$~on-sky to
predict \MProjObs. This is given by 
\begin{eqnarray}
  M(< R_f) & = & 4 \pi \int^{R_f}_{0} \int^{+\infty}_{R}  \frac{rR \rho(r)}{\sqrt{r^2 - R^2}} ~dr~dR \nonumber \\
         & = & \frac{\sigma_0^2}{G r_0^2} \int^{R_f}_{0} \int^{+\infty}_{R} 
  \frac{rR}{\sqrt{r^2 - R^2}}\left\{ \right\}~dr~dR
\end{eqnarray}
\noindent where the double integral is evaluated numerically.

5. To compare the anisotropic dispersion profile to data, project the
model dispersions onto the sky. The projection is the same as for the
King model:
\begin{eqnarray}
  \sigma^2_R(R) & = & \frac{\int^{+\infty}_R  \frac{r\ntracer\sigt}{\sqrt{r^2 - R^2}} \left[ \frac{R^2}{r^2}\sigr + \left( 1 - \frac{R^2}{r^2}\right) \sigt \right] dr  }  {\int^{+\infty}_{R} \frac{r \ntracer}{\sqrt{r^2 - R^2}} dr } \nonumber \\       
  \sigma^2_T(T) & = & \frac{\int^{+\infty}_{R} \frac{r\ntracer\sigt}{\sqrt{r^2 - R^2}} dr } {\int^{+\infty}_{R} \frac{r \ntracer}{\sqrt{r^2 - R^2}} dr } 
\end{eqnarray}
\noindent (note that $n(r)$~should be used in this step not $\rho(r)$).


\begin{table} 
\begin{tabular}{r|c|c|c} 
$\Delta \chi^2_{kinem}$ & 3.50  & 7.82  & 13.93  \\ 
Confidence & 68\% & 95\% & 99.7\% \\ 
& "$1\sigma$" & "$2\sigma$" & "$3\sigma$" \\ 
\hline 
$M(R< 0.40~{\rm pc})$ & 0.40 - 1.10 & 0.36 - 1.19 & 0.32 - 1.30\\ 
($10^4~M_{\odot}$) &   &   &   \\ 
\hline 
$M(r < 1.0~{\rm pc})$ & 0.74 - 1.98 & 0.50 - 2.14 & 0.50 - 2.33\\ 
($10^4~M_{\odot}$) &   &   &   \\ 
\hline 
$\rho_0$ & 0.06 - 30.92 & 0.05 - 30.92 & 0.05 - 31.40\\ 
($10^5 M_{\odot}~{\rm pc^{-3}}$) &   &   &   \\ 
\hline 
$R_c$ & 0.05 - 0.80 & 0.05 - 0.80 & 0.05 - 0.80\\ 
(pc) &   &   &   \\ 
\hline 
$R_t$ & 1.00 - 50.00 & 1.00 - 50.00 & 1.00 - 50.00\\ 
(pc) &   &   &   \\ 
\hline 
$M_{cluster}$ & 0.83 - 8.36 & 0.50 - 9.34 & 0.50 - 10.00\\ 
($10^4~M_{\odot}$) &   &   &   \\ 
\hline 
\end{tabular} 
\caption{Significance regions for isotropic King modeling of the Arches cluster. Ranges of each parameter corresponding to the stated significance level are given, when $R_c, R_t, M_{cluster}$~are all allowed to vary. The quantity $\chi^2_{kinem}$~denotes the badness-of-fit when comparing model predictions to the Arches kinematic dataset only.} 
\label{table_massest_kinem}
\end{table} 


\begin{table} 
\begin{tabular}{r|c|c|c} 
$\Delta \chi^2_{full}$ & 3.50  & 7.82  & 13.93  \\ 
Confidence & 68\% & 95\% & 99.7\% \\ 
& "$1\sigma$" & "$2\sigma$" & "$3\sigma$" \\ 
\hline 
$M(R< 0.40~{\rm pc})$ & 0.70 - 1.10 & 0.62 - 1.20 & 0.56 - 1.30\\ 
($10^4~M_{\odot}$) &   &   &   \\ 
\hline 
$M(r < 1.0~{\rm pc})$ & 1.07 - 1.77 & 0.96 - 1.92 & 0.85 - 2.11\\ 
($10^4~M_{\odot}$) &   &   &   \\ 
\hline 
$\rho_0$ & 0.89 - 2.37 & 0.62 - 3.47 & 0.44 - 3.91\\ 
($10^5 M_{\odot}~{\rm pc^{-3}}$) &   &   &   \\ 
\hline 
$R_c$ & 0.15 - 0.21 & 0.13 - 0.26 & 0.13 - 0.33\\ 
(pc) &   &   &   \\ 
\hline 
$R_t$ & 3.00 - 30.00 & 2.00 - 30.00 & 2.00 - 30.00\\ 
(pc) &   &   &   \\ 
\hline 
$M_{cluster}$ & 1.64 - 3.53 & 1.26 - 3.91 & 1.07 - 4.48\\ 
($10^4~M_{\odot}$) &   &   &   \\ 
\hline 
$1000 \times \Sigma_{N,0} / \rho_0$ & 0.00 - 0.01 & 0.00 - 0.06 & 0.00 - 0.12 \\ 
(stars pc$^{-2} / M_{\odot}~{\rm pc}^{-3}$) &   &   &   \\ 
\hline 
\end{tabular} 
\caption{Significance regions for isotropic King modeling of the Arches cluster. Ranges of each parameter corresponding to the stated significance level are given, when $R_c, R_t, M_{cluster}$~are all allowed to vary. The quantity $\chi^2_{full}$~denotes the badness-of-fit when comparing model predictions to both the Arches kinematic dataset and the surface density dataset of Espinoza et al. (2009), over the mass range ($10 \le M \le 120$)~$M_{\odot}$.} 
\label{table_massest_full}
\end{table} 


\end{document}